\documentclass[twocolumn]{aastex701}
\usepackage{amsmath}
\usepackage{amsfonts}
\usepackage{subcaption}
\usepackage{booktabs}
\usepackage{tabularx}
\usepackage{tikz}
\usepackage{enumitem}
\usepackage{xspace}
\usepackage{silence}  \newcommand{\bsym}{\boldsymbol}
\renewcommand{\colhead}[1]{\textbf{#1}}
\newcommand{\MR}{$\{M_{\star}, R_{\star}\}$\xspace} \newcommand{\UQ}{\textit{DQ}\xspace}
\newcommand{\UO}{\textit{DQO}\xspace}
\defcitealias{kalapotharakos_multipolar_2021}{CK21}
\defcitealias{petri_multipolar_2015}{JP15}
\defcitealias{olmschenk_pioneering_2025}{GO25}
\begin{document}
\title{The swept-back multipolar magnetic field of neutron stars: Application to NICER MSP J0030+0451}
\author[0000-0003-2128-1414]{Anu Kundu}
\altaffiliation{NASA Postdoctoral Program Fellow}
\affiliation{Astrophysics Science Division, NASA Goddard Space Flight Center, Greenbelt, MD 20771, USA}
\affiliation{Centre for Space Research, North-West University, Private Bag X6001, Potchefstroom 2520, South Africa}
\affiliation{Center for Space Sciences and Technology, University of Maryland, Baltimore County, Baltimore, MD 21250, USA}
\affiliation{Center for Research and Exploration in Space Science and Technology, NASA/GSFC, Greenbelt, MD 20771, USA}
\email[show]{anu.kundu@nasa.gov}
\author[0000-0003-1080-5286]{Constantinos Kalapotharakos}
\affiliation{Astrophysics Science Division, NASA Goddard Space Flight Center, Greenbelt, MD 20771, USA}
\email{some_emails@email.com}
\author[0000-0002-9249-0515]{Zorawar Wadiasingh}
\affiliation{Department of Astronomy, University of Maryland, College Park, Maryland 20742, USA}
\affiliation{Astrophysics Science Division, NASA Goddard Space Flight Center, Greenbelt, MD 20771, USA}
\affiliation{Center for Research and Exploration in Space Science and Technology, NASA/GSFC, Greenbelt, MD 20771, USA}
\email{some_emails@email.com}
\author[0000-0001-8472-2219]{Greg Olmschenk}
\affiliation{Department of Astronomy, University of Maryland, College Park, Maryland 20742, USA}
\affiliation{Astrophysics Science Division, NASA Goddard Space Flight Center, Greenbelt, MD 20771, USA}
\affiliation{Center for Research and Exploration in Space Science and Technology, NASA/GSFC, Greenbelt, MD 20771, USA}
\email{some_emails@email.com}
\author[0009-0002-9936-4041]{Wendy F. Wallace}
\affiliation{Southeastern Universities Research Association, Washington, DC 20005, USA}
\affiliation{Astrophysics Science Division, NASA Goddard Space Flight Center, Greenbelt, MD 20771, USA}
\affiliation{Center for Research and Exploration in Space Science and Technology, NASA/GSFC, Greenbelt, MD 20771, USA}
\email{some_emails@email.com}
\author[0000-0001-6119-859X]{Alice K. Harding}
\affiliation{Theoretical Division, Los Alamos National Laboratory, Los Alamos, NM 58545, USA}
\email{some_emails@email.com}
\author[0000-0002-2666-4812]{Christo Venter}
\affiliation{Centre for Space Research, North-West University, Private Bag X6001, Potchefstroom 2520, South Africa}
\affiliation{National Institute for Theoretical and Computational Sciences (NITheCS), Potchefstroom, South Africa}
\email{some_emails@email.com}
\author[0000-0002-7435-7809]{Demosthenes Kazanas}
\affiliation{Astrophysics Science Division, NASA Goddard Space Flight Center, Greenbelt, MD 20771, USA}
\email{some_emails@email.com}
\begin{abstract}
NICER observations of millisecond pulsars (MSPs) suggest that non-dipolar magnetic fields are required to explain their surface X-ray hotspots. \citet{kalapotharakos_multipolar_2021} modeled the NICER light curve of MSP J$0030+0451$ (J$0030$) using a static vacuum offset dipole-plus-quadrupole field and corresponding force-free (FF) solutions to jointly reproduce the X-ray and Fermi-LAT $\gamma$-ray emission. We substitute their static vacuum field model with a more realistic swept-back configuration that accounts for rotational effects. This field more closely resembles the corresponding FF solutions, making it a more physically motivated choice for future multiwavelength modeling. We adopt a centered swept-back vacuum multipolar magnetic field (SVM2F; \citealt{petri_multipolar_2015}), expressed as a complete expansion in vector spherical harmonics, enabling flexible descriptions of arbitrary magnetic field geometries. We introduce a metric to quantify the complexity among different field prescriptions, illustrated for the static offset vacuum field. To efficiently explore parameter space, we train a neural network surrogate \citep{olmschenk_pioneering_2025} on SVM2F light curves including components up to the octupole, accelerating Markov chain Monte Carlo sampling by $\sim 10^3$ compared to direct physical model evaluations. Applying this framework to J$0030$, we constrain the field parameter space and find that a centered swept-back multipolar field including terms up to the octupole adequately reproduces the bolometric thermal X-ray light curve. Our study highlights the importance and inherent complexity of prescribing different multipolar magnetic field models for rotating stars, and can be extended to other MSPs to ultimately constrain the masses and radii of neutron stars, and hence their equation of state.
\end{abstract}

\section{Introduction}\label{sec:intro}
Neutron stars (NSs) are the densest and most strongly magnetized known compact objects. Their internal structure is governed by the equation of state (EoS) of dense matter (see reviews by \citealt{lattimer_neutron_2007, lattimer_equation_2016}), which can be directly constrained through observational measurements of the stellar mass, $M_{\star}$, and radius, $R_{\star}$. Recent observations of millisecond pulsars (MSPs), NSs with rotation periods $< 16$~ms \citep{halder_defining_2023}, together with measurements of masses up to $\sim 2~M_{\odot}$ have provided a powerful means of ruling out EoS models that are unable to support such high masses. Most MSPs are in binaries (attaining rapid rotation via accretion from a companion star; \citealt{alpar_new_1982, bhattacharya_formation_1991}), where precise constraints on $M_{\star}$ can be obtained from orbital dynamics or radio Shapiro delay measurements  \citep{shapiro_fourth_1964, shapiro_fourth_1971, demorest_two-solar-mass_2010}. For isolated NSs, like PSR~J$0030+0451$ (hereafter J$0030$), stringent independent constraints are generally unavailable.

Thermal X-ray emission from MSPs originates from heated polar caps, defined by the footpoints of open magnetic field lines\footnote{Open magnetic field lines cross the light cylinder, where the corotation speed of the magnetosphere reaches the speed of light, $c$.} at the stellar surface. These polar cap regions (also referred to as hotspots) are heated to X-ray-emitting temperatures by magnetospheric electron/positron pair cascades or return currents \citep{harding_pulsar_2001, harding_pulsar_2002}. Modeling their geometry and thermal X-ray waveforms constrains the \MR relation via gravitational light bending and provides insight into particle heating mechanisms, atmospheric composition, and magnetic field structure. Early studies with ROSAT and XMM-Newton demonstrated the feasibility of this approach albeit with large uncertainties \citep{zavlin_soft_1998, bogdanov_constraints_2007, bogdanov_thermal_2008, bogdanov_nearest_2013}.

NS Interior Composition ExploreR (NICER; \citealt{gendreau_neutron_2016}) has transformed this field through its exceptional soft X-ray (0.2-12 keV) sensitivity, large effective area, and excellent timing resolution. For J0030 \citep{bogdanov_constraining_2019}, independent NICER analyses by \citet{miller_psr_2019} and \citet{riley_nicer_2019} inferred consistent \MR constraints while revealing a striking hotspot geometry: both emitting regions lie in the same rotational hemisphere (see \citealt{bilous_nicer_2019} for the implications). This configuration disfavors a simple centered dipolar magnetic field, which would produce antipodal polar caps, and instead indicates a more complex surface magnetic topology. Similar evidence of non-antipodal emission regions has emerged from analyzing PSR~J0740+6620 \citep{miller_radius_2021, riley_nicer_2021}.

These studies adopt phenomenological hotspot shapes to reproduce observed pulse profiles, and while effective for constraining \MR, they do not directly probe the underlying magnetic field structure. To address this, several works have implemented physically motivated magnetic field prescriptions for J0030. \cite{chen_numerical_2020} applied a centered dipole plus offset quadrupole constraining the dipole with $\gamma$-ray data. \citet{kalapotharakos_multipolar_2021}, hereafter \citetalias{kalapotharakos_multipolar_2021}, assumed a static offset dipole-plus-quadrupole field to fit X-ray and $\gamma$-ray light curves (discussed in more detail below). In contrast, \cite{carrasco_relativistic_2023} modeled J0030 with a centered dipole under a general-relativistic force-free (FF\footnote{FF field assumes that only the electric field perpendicular to the magnetic field survives ($\boldsymbol{E} \cdot \boldsymbol{B} = 0$) and that the magnetic field dominates ($\boldsymbol{E} < \boldsymbol{B}$).}) approximation, finding it difficult to reproduce the X-ray data, though the method succeeded for other MSPs. \citet{petri_constraining_2023} employed an off-centered dipole to fit $\gamma$-ray, radio, and X-ray light curves, showing that multiwavelength diagnostics are critical to constrain field geometry. Similarly, \citet{huang_physics_2025} used an off-centered dipole and highlighted the need for higher-order multipoles. More recently, \citet{CaoYang_J0030_2026} modeled J0030 with a dissipative dipole plus offset quadrupole, self-consistently reproducing both X-ray and $\gamma$-ray pulse profiles. The importance of these diverse approaches is emphasized by \citet{vinciguerra_updated_2024}, who note that constraints from independent studies help to mitigate the computational costs of \MR analyses.

The need for complex magnetic configurations is physically well motivated. During the accretion-driven recycling phase, MSPs may undergo magnetic burial, which can reduce the large-scale dipole while preserving or enhancing higher-order multipoles near the surface \citep{melatos_hydromagnetic_2001, payne_burial_2004, suvorov_recycled_2020}. Because higher-order components dominate near the stellar surface, they strongly influence the polar cap geometry, pair production efficiency, and, consequently, the current distribution in the outer magnetosphere \citep{timokhin_polar_2015}. MSPs have smaller spin periods ($P$) than normal pulsars, resulting in compact magnetospheres with smaller light-cylinder radii, $R_{\rm LC} = c P/ 2 \pi$. Consequently, emission across all wavebands is strongly shaped by the near-surface magnetic field topology.

We summarize the study by \citetalias{kalapotharakos_multipolar_2021}, which serves as the foundation for the present work. In their static offset dipole-plus-quadrupole model, each component’s orientation and offset varied within a Markov chain Monte Carlo (MCMC) framework, revealing multiple families of vacuum solutions (``field degeneracies") capable of reproducing J0030’s NICER bolometric thermal X-ray light curve. Using the best-fit vacuum parameters as MCMC starting points, they incorporated FF simulations into the pipeline, obtaining FF configurations consistent with the NICER data, while avoiding a prohibitively expensive full FF MCMC. These FF solutions were then combined with the FIDO model (FF inside, dissipative outside the light cylinder; \citealt{kalapotharakos_toward_2012, kalapotharakos_gamma-ray_2014, brambilla_testing_2015, kalapotharakos_fermi_2017}) to compute $\gamma$-ray light curves via curvature radiation in the equatorial current sheet beyond the light cylinder. Comparison with Fermi-LAT $\gamma$-ray data showed that simultaneous X-ray and $\gamma$-ray modeling can partially lift field degeneracies, robustly constraining the magnetic parameter space. However, computational costs limited a thorough exploration of the parameter space using MCMC. To address this, \citet{olmschenk_pioneering_2025}, hereafter \citetalias{olmschenk_pioneering_2025}, developed a neural network (NN) surrogate for the vacuum model, trained on thermal X-ray light curves generated from the physical model. This approach achieved a $> 400$ times speed-up, enabling more complete exploration of the parameter space and revealing that the solutions of \citetalias{kalapotharakos_multipolar_2021} had not fully converged. Additionally, when $\gamma$-ray light-curve fitting is incorporated using the physical model, the complexity increases substantially, and the computational demands become even more severe due to the expensive FF simulations required. While physical models using FF configurations are more than $10^3$ slower than the corresponding vacuum models, the trained NNs maintain the same computational efficiency (T. Lechien et al., in preparation). The NN approach thus proves particularly valuable for efficient multiwavelength modeling of combined NICER and Fermi data, providing an accelerated pathway towards equilibrium distribution that would otherwise be impractical.

The current work aims to address three key objectives. First, we quantify the complexity of MSP magnetic fields by defining a metric based on a complete multipolar magnetic field basis \citep[][hereafter \citetalias{petri_multipolar_2015}]{petri_multipolar_2015}, establishing a common framework for comparing different field prescriptions. Second, we extend \citetalias{kalapotharakos_multipolar_2021} by replacing their static vacuum field with a swept-back multipolar field \citepalias{petri_multipolar_2015}\footnote{Throughout this paper, ``swept-back'' refers to the standard delayed-time vacuum solution, historically known as the retarded vacuum field. The sweepback arises from the finite propagation speed of electromagnetic fields, which makes the exterior solution depend on delayed-time potentials and therefore introduces time-dependent structure. For non-axisymmetric modes, this also gives rise to a low-frequency electromagnetic-wave component driven by stellar rotation.} to fit J0030’s bolometric thermal X-ray light curve. The swept-back fields are expected to better approximate FF configurations, providing a realistic baseline for future FF-based modeling. Third, we obtain converged MCMC solutions for J0030’s X-ray light curve using an NN surrogate \citepalias{olmschenk_pioneering_2025} trained on swept-back vacuum configurations, achieving large computational speed-ups and enabling future NN applications to FF models in a global pulsar magnetosphere framework for multiwavelength modeling.

The paper is organized as follows. Section~\ref{sec:mf_expansion} introduces the multipolar magnetic field basis and addresses the complexity associated with field prescriptions. Section~\ref{sec:application_SVM2F} describes the implementation of the expansion and the magnetic field parameter space. Section~\ref{sec:methodology} summarizes the methods, Section~\ref{sec:results} presents the MCMC results, and Section~\ref{sec:discussionsAndConclusions} discusses the findings and future directions.

\section{Magnetic field expansion and completeness} \label{sec:mf_expansion}
Using a complete vector spherical harmonics (VSH) basis, $\{ \bsym{Y}_{lm}, \bsym{\Psi}_{lm}, \bsym{\Phi}_{lm} \}$ (see Appendix~\ref{appsub:VSHBasis}), \citetalias{petri_multipolar_2015} presented the exact analytical vacuum magnetic field solution \emph{interior} to an NS as,
\begin{eqnarray}
    \boldsymbol{B}_{\rm in} = \sum_{l=1}^{\infty}  \sum_{m=-l}^{l} \nabla \times \left[ f_{lm}^{B,{\rm in}}(r, t) \boldsymbol{\Phi}_{lm}  \right],
    \label{eqn:magfieldBinside}
\end{eqnarray}
where the expansion is taken about the stellar center, i.e., all multipolar components are assumed to originate at the same point.
Here, $l$ denotes the multipolar order and $m$ the azimuthal mode. The temporal dependence of each mode enters as $e^{-im\Omega t}$, where $\Omega=2\pi/P$ is the stellar angular frequency. Since the vector spherical harmonic $\boldsymbol{\Phi}_{lm}$ inherits the azimuthal dependence $e^{im\phi}$ from the scalar spherical harmonic $Y_{lm}$, each non-axisymmetric mode varies as $e^{im(\phi-\Omega t)}$ up to a constant phase offset. In what follows, $f_{lm}$ (and $g_{lm}$ below) denote only the corresponding radial coefficients.

Thus, for this centered interior expansion, the radial dependence is the static, non-propagating form
\begin{equation}
    f_{lm}^{B,{\rm in}}(r) = a_{lm}^{B,{\rm in}}\,r^{-(l+1)},
    \label{eqn:flmInside}
\end{equation}
where $a_{lm}^{B,{\rm in}}$ are complex-valued coefficients determined by the surface field. For $m\neq 0$, these coefficients encode both the amplitude and any constant phase offset of the corresponding $(l,m)$ mode (see Section~\ref{sec:application_SVM2F}).

The corresponding \emph{exterior} swept-back vacuum multipolar magnetic field (hereafter SVM2F) and electric field solutions \citepalias{petri_multipolar_2015} are
\begin{eqnarray}
    &&\boldsymbol{B}_{\rm out}(r, \theta, \phi, t) = \nonumber \\ 
    &&\sum_{l=1}^{\infty}  \sum_{m=-l}^{l} \left( \nabla \times \left[ f_{lm}^{B,{\rm out}}(r, t) \boldsymbol{\Phi}_{lm}  \right] + g_{lm}^{B,{\rm out}}(r, t) \boldsymbol{\Phi}_{lm} \right),
    \label{eqn:magfieldB}
\end{eqnarray}
\begin{eqnarray}
    &&\boldsymbol{D}_{\rm out}(r, \theta, \phi, t) = \nonumber \\ 
    &&\sum_{l=1}^{\infty}  \sum_{m=-l}^{l} \left( \nabla \times \left[ f_{lm}^{D,{\rm out}}(r, t) \boldsymbol{\Phi}_{lm}  \right] + g_{lm}^{D,{\rm out}}(r, t) \boldsymbol{\Phi}_{lm} \right),
    \label{eqn:elecfieldD}
\end{eqnarray}
where $B$ and $D$ represent the magnetic and electric parts, respectively. The exterior solution is described by the complex-valued scalar coefficients $f_{lm}^{\rm out}(r,t)$ and $g_{lm}^{\rm out}(r,t)$, while the temporal dependence of each mode remains $e^{-im\Omega t}$, as
introduced above.
The radial dependence of the coefficients $f_{lm}^{\rm out}(r,t)$ is defined in terms of the complex-valued integration constants, $a_{lm}^{\rm out}$, as \citepalias{petri_multipolar_2015}
\begin{equation}
    f_{lm}^{(B,D),\rm out}(r) = 
        \begin{cases}
        a_{l0}^{(B,D),\rm out} \, r^{-(l+1)}, & m = 0, \\
        a_{lm}^{(B,D),\rm out} \, h_l^{(1)}(k\, m\, r), & m \neq 0,
\end{cases}
\label{eqn:flmAndalm}
\end{equation}
where $h_l^{(1)}$ is the spherical Hankel function (SHF; \citealt{arfken_mathematical_2005}) of the first kind, and $k = \Omega / c$. The coefficients $a_{lm}^{\rm in}$ and $a_{lm}^{\rm out}$ are related through the boundary conditions at the stellar surface, which ensure continuity of the radial component of the magnetic field, and of the tangential component of the electric field. In particular, for the magnetic coefficients, we have
\begin{equation}
    a_{l0}^{B,{\rm out}} = a_{l0}^{B,{\rm in}},
    \label{eqn:almOutAndIn_m0}
\end{equation}
and, for $m\neq 0$,
\begin{equation}
    a_{lm}^{B,{\rm out}} =
    \frac{a_{lm}^{B,{\rm in}}}
    {R_{\star}^{\,l+1}\,h_l^{(1)}(k\, m\, R_{\star})}\;,
    \label{eqn:almOutAndIn}
\end{equation}
where $R_{\star}$ is the stellar radius. The tangential electric-field boundary conditions yield, for the axisymmetric modes ($m=0$),
\begin{align}
a_{l+1,0}^{D,\rm{out}}
&=
-\frac{
\epsilon_0 R_\star^{\,l+3}\Omega
\sqrt{l(l+2)}\, J_{l+1,0}\,
f_{l0}^{B,\mathrm{in}}(R_\star)}{l+1},
\label{eq:bcD_lplus1_0}
\\
a_{l-1,0}^{D,\rm{out}}
&=
\frac{
\epsilon_0 R_\star^{\,l+1}\Omega
\sqrt{(l-1)(l+1)}\, J_{l,0}\,
f_{l0}^{B,\mathrm{in}}(R_\star)}{l-1},
\label{eq:bcD_lminus1_0}
\end{align}
and, for the non-axisymmetric modes ($m>0$),
\begin{align}
a_{l+1,m}^{D,\mathrm{out}}
&=
\frac{
\epsilon_0 R_\star \Omega
\sqrt{l(l+2)}\, J_{l+1,m}\,
f_{lm}^{B,\mathrm{in}}(R_\star)}
{\left.
\partial_r \!\left[ r\, h_{l+1}^{(1)}(k\, m \,r) \right]
\right|_{r=R_\star}},
\label{eq:bcD_lplus1_m}
\\
a_{l-1,m}^{D,\mathrm{out}}
&=
-\frac{
\epsilon_0 R_\star \Omega
\sqrt{(l-1)(l+1)}\, J_{l,m}\,
f_{lm}^{B,\mathrm{in}}(R_\star)}
{\left.
\partial_r \!\left[ r\, h_{l-1}^{(1)}(k\, m\, r) \right]
\right|_{r=R_\star}},
\label{eq:bcD_lminus1_m}
\end{align}
where
\begin{equation}
J_{lm}=\sqrt{\frac{l^2-m^2}{4l^2-1}}
\label{eq:Jlm_For_almD_out}
\end{equation}
\citepalias{petri_multipolar_2015} and $\epsilon_{0}$ is the permittivity of free space. These relations show that a magnetic multipole of order $l$ couples to electric multipoles of orders $l-1$ and $l+1$, except in the dipolar case ($l=1$), where only the $l+1$ contribution is present.
The coefficients $g_{lm}^{B,{\rm out}}$ and $g_{lm}^{D,{\rm out}}$\footnote{
The coefficients $g_{lm}$ appear only in the exterior solution. We retain
the superscript ``out'' for notational consistency with the corresponding
$f_{lm}^{\rm out}$ and $a_{lm}^{\rm out}$ coefficients.
}
follow the linear relations \citepalias{petri_multipolar_2015}
\begin{equation}
    g_{lm}^{B,{\rm out}} = -i\mu_{0}m\Omega\,f_{lm}^{D,{\rm out}},
\end{equation}
\begin{equation}
    g_{lm}^{D,{\rm out}} = i\epsilon_{0}m\Omega\,f_{lm}^{B,{\rm out}},
\end{equation}
where $\mu_{0}$ is the magnetic permeability in vacuum. Thus, once the internal or surface
field is specified through Equation~\ref{eqn:magfieldBinside}, the
corresponding exterior coefficients can be determined from the boundary
conditions and the above linear relations, from which the complete
exterior electromagnetic field (Equations~\ref{eqn:magfieldB} and \ref{eqn:elecfieldD}) follows.

The physical magnetic field corresponding to the SVM2F solution given by Equation~\ref{eqn:magfieldB} is obtained by taking the real part of the expression. The standard terminology reads as $l = 1$ for dipole (which is essentially the rotating solution first given by \citealt{deutsch_1955_deutsch_the_1955}), $l =2$ for quadrupole, $l = 3$ for octupole\footnote{In contrast to \citetalias{petri_multipolar_2015}, we adopt the standard convention, referring to the  $l = 3$ component as an octupole rather than a hexapole.}, and so on.

\subsection{Quantifying Field Complexity} \label{subsec:completeness}
Because the VSH basis is orthonormal and complete for field expansions, any sufficiently regular vector field on the sphere can be represented in this basis. In this section, we compare different magnetic field descriptions and quantify their complexity within this common centered-expansion framework. For a given accuracy, we define the complexity of a field by the minimal multipolar content required to reproduce it, as characterized by the highest multipole order needed, $l_{\max}$, and the strength distribution of the corresponding expansion coefficients.

The coefficients, $a_{lm}^{B,{\rm in}}$, can be calculated for any input surface magnetic field using the following method. Equation~\ref{eqn:magfieldBinside}, whose explicit form is derived in Appendix~\ref{app:multipole_expansion} (Equation~\ref{eqnApp:MagFieldVecBin}), is here rewritten explicitly as,
\begin{align}
\label{eqn:MagFieldVecB_in}
    \bsym{B}_{\rm in}(r, \theta, \phi, t) & =  \sum_{l=1}^{\infty}  \sum_{m=-l}^{l} e^{- i m \Omega t} \nonumber \\
    & \left( - \frac{\sqrt{l(l+1)}}{r} f_{lm}^{B,{\rm in}} \bsym{Y}_{lm} - \frac{1}{r} \partial_{r}(r f_{lm}^{B,{\rm in}}) \bsym{\Psi}_{lm}  \right).
\end{align}
For the centered interior-field expansion, evaluated at the stellar surface, the radial functions are given by Equation~\ref{eqn:flmInside}.
Therefore, at the stellar surface, $r=R_{\star}$, the two terms on the right-hand side of Equation~\ref{eqn:MagFieldVecB_in} become
\begin{align}
\label{eqn:MagFieldVecB_in_m0}
    - \frac{\sqrt{l(l+1)}}{r} f_{lm}^{B,{\rm in}} \bsym{Y}_{lm} \implies & - \frac{\sqrt{l(l+1)}}{R_{\star}^{(l+2)}} a_{lm}^{B,{\rm in}} \bsym{Y}_{lm} \nonumber\\
    - \frac{1}{r} \partial_{r}(r f_{lm}^{B,{\rm in}}) \bsym{\Psi}_{lm} \implies & - \frac{1}{r} a_{lm}^{B,{\rm in}} \partial_{r}(r^{-l})  \bsym{\Psi}_{lm} \nonumber\\
    & = \frac{l}{R_{\star}^{(l+2)}} a_{lm}^{B,{\rm in}} \bsym{\Psi}_{lm}.
\end{align}
Multiplying each side by the complex conjugate of $\bsym{Y}_{lm}$ (i.e., $\bsym{Y}_{l'm'}^{*}$) and integrating over the solid angle $d\Omega$, we utilize the orthogonality of $\bsym{Y}_{lm}$ (given in Appendix \ref{appsub:VSHProperties}) to obtain
\begin{align}
    \int \bsym{B}_{\rm in}(R_{\star}, \theta, \phi, t) \cdot \bsym{Y}_{l'm'}^{*} d\Omega = & - \frac{\sqrt{l'(l'+1)}}{R_{\star}^{l'+2}} a_{l'm'}^{B,{\rm in}}, \\
    a_{l'm'}^{B,{\rm in}} =  - \frac{ S_1  R_{\star}^{l'+2}}{\sqrt{l'(l'+1)}},
\end{align}
where, $S_1 = \int \bsym{B}_{\rm in}(R_{\star}, \theta, \phi, t) \cdot \bsym{Y}_{l'm'}^{*} d\Omega$. Likewise, using the orthogonality property of $\bsym{\Psi}_{lm}$, we obtain
\begin{equation}
    a_{l'm'}^{B,{\rm in}} =  \frac{ S_2  R_{\star}^{l'+2}}{l'},
\end{equation}
where $S_2 = \int \bsym{B}_{\rm in}(R_{\star}, \theta, \phi, t) \cdot \bsym{\Psi}_{l'm'}^{*} d\Omega$. These expressions imply that
\begin{equation}
    S_1=-S_2\sqrt{\frac{l+1}{l}}.
\end{equation}
Thus, the input field $\bsym{B}_{\rm in}$ can be expressed in terms of its centered VSH expansion by using the coefficients $a_{lm}^{B,{\rm in}}$ in Equation \ref{eqn:MagFieldVecB_in}.

This opens a pathway to quantitatively measure the mathematical complexity of any field description in this expansion framework. In practice, an input (``truth") field on the stellar surface, $\bsym{B_{\star}}$, is projected onto the VSH basis, and a truncated expansion through degree $l$, $\bsym{B}_{\rm ex}^{(l)}$, is constructed. To quantify how well the truncated expansion approximates the input field, we define a convergence metric ($\delta {\cal C}$) as the mean value of the relative field deviation over the unit sphere,
\begin{equation}
\delta {\cal C}(l) = \frac{1}{4 \pi} \int_0^\pi \int_0^{2\pi}  \frac{\delta B}{B} [\theta, \phi; l] \sin(\theta) \, d\phi \, d\theta,
\end{equation}
where 
\begin{equation}
\frac{\delta B}{B} = \sqrt{ \frac{(\bsym{B}_{\star} - \bsym{B}_{\rm ex}^{(l)}) \cdot (\bsym{B}_{\star} - \bsym{B}_{\rm ex}^{(l)})}{\bsym{B}_{\star} \cdot \bsym{B}_{\star}} }.
\end{equation}
For a prescribed accuracy, the required truncation order $l_{\max}$, together with the distribution of the included coefficients, serves as an applicable measure to estimate complexity. This is demonstrated using an example field configuration below.

\subsubsection{Coefficients for static offset dipole-plus-quadrupole fields} \label{subsubsec:coeff_SVF}
\begin{figure*}[tb]
    \centering
    \includegraphics[width=0.98\linewidth]{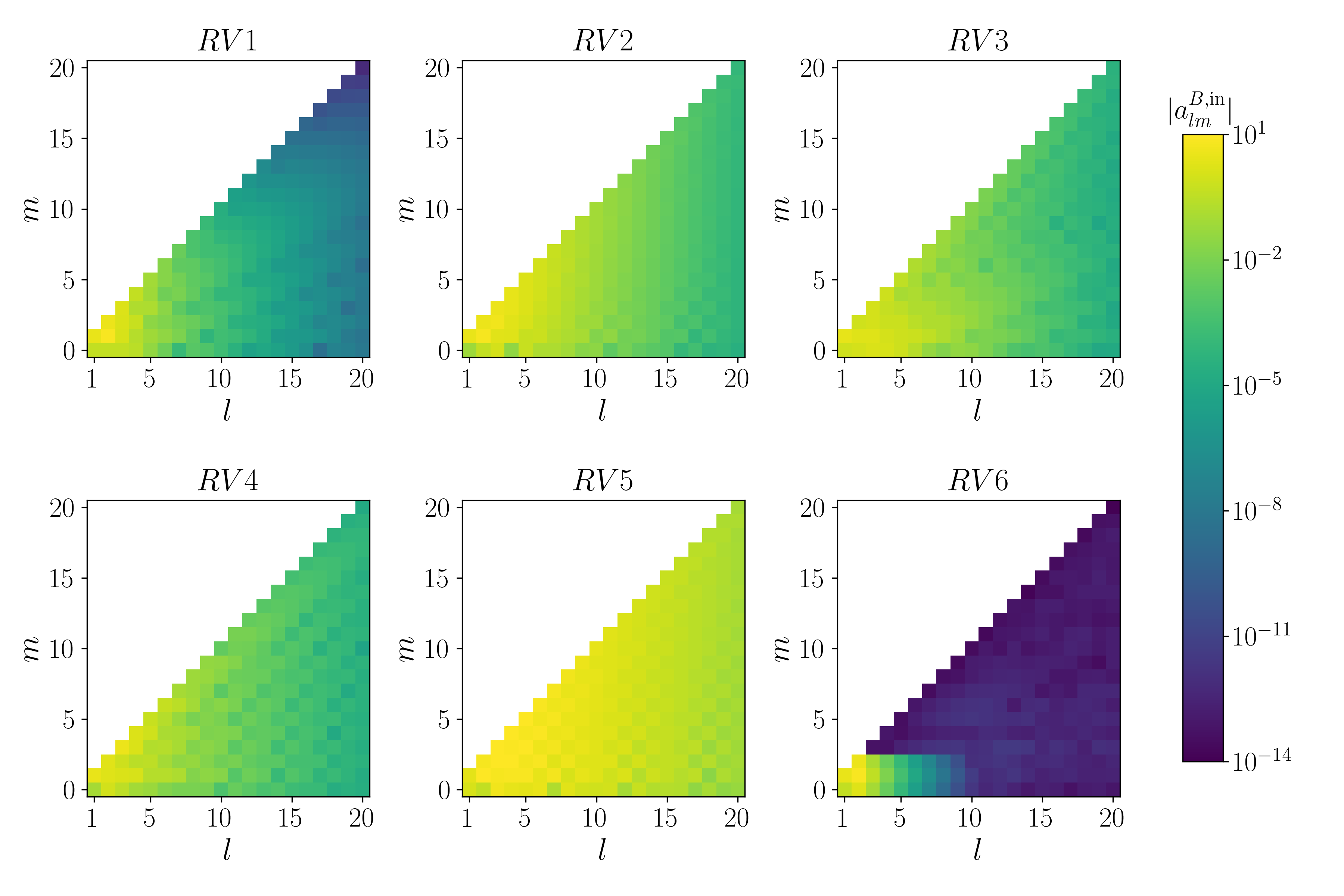}
    \caption{The $x$- and $y$-axes indicate the indices $l$ and $m$, respectively, while the colors represent the absolute magnitude of the multipolar coefficients, $|a_{lm}^{B,{\rm in}}|$, on the stellar surface. The log-linear color bar shows the true values. Each panel corresponds to one of the vacuum solutions of \citetalias{kalapotharakos_multipolar_2021}. All $m=\{0,\ldots,l\}$ cases are shown up to $l=20$.}
    \label{fig:acoeff}
\end{figure*}
\begin{figure*}[tbp]
    \centering
    \includegraphics[width=0.80\linewidth]{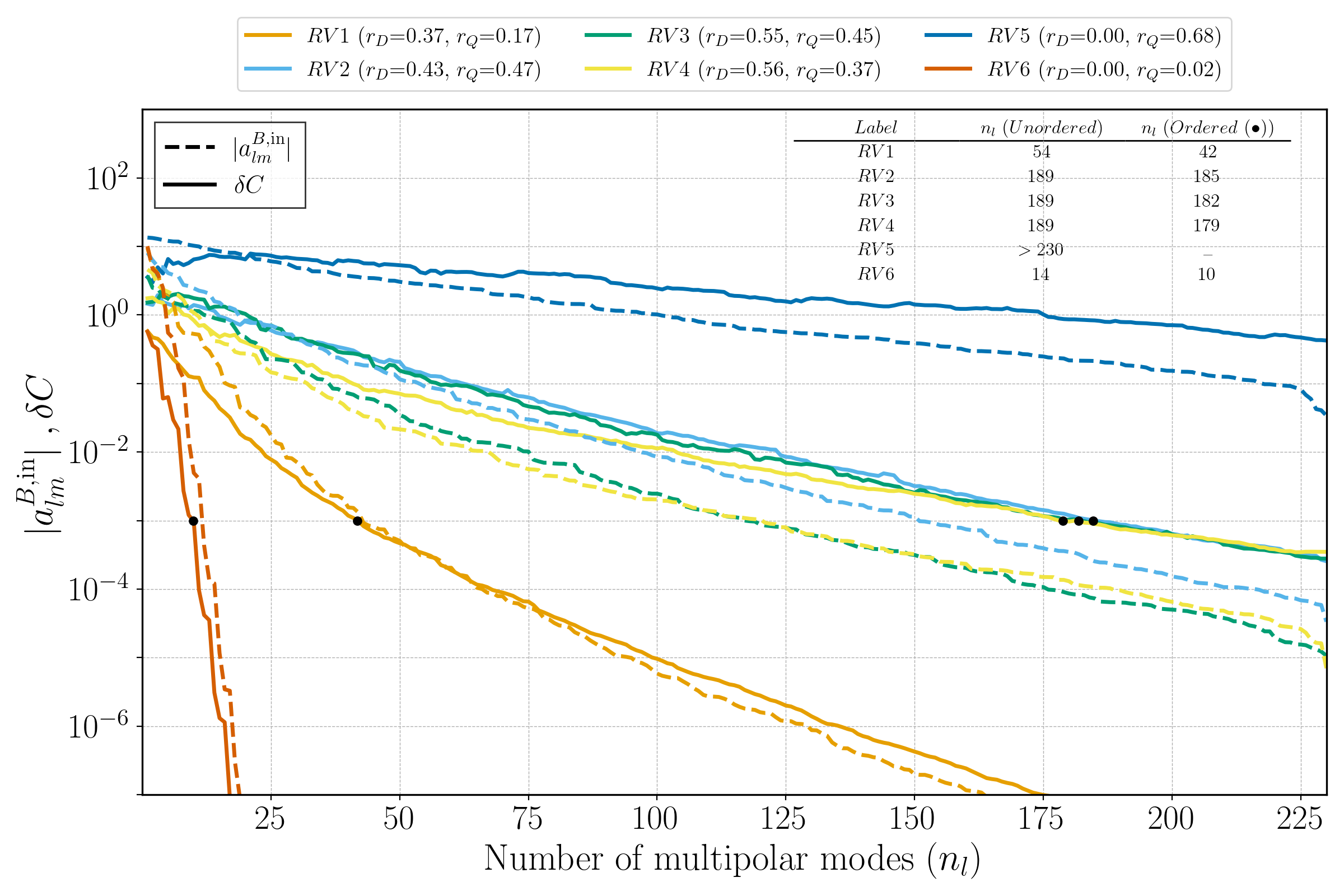}
    \caption{The absolute magnitudes of the multipolar components, $|a_{lm}^{B,{\rm in}}|$, arranged in descending order (dashed lines), together with the corresponding $\delta {\cal C}$ values (solid lines) for the same cases as in Figure~\ref{fig:acoeff}. The total number of multipolar modes, $n_l$, is shown on the $x$-axis. The legend at the top indicates the radial offsets, in units of $R_{\star}$, of the dipole ($r_D$) and quadrupole ($r_Q$) components of the input field solution. The inset table lists the $n_l$ values for $\delta {\cal C} = 10^{-3}$ computed from the unordered and ordered $|a_{lm}^{B,{\rm in}}|$; only the ordered case is shown as black dots.}
    \label{fig:CM_CK21_RV}
\end{figure*}
We use the analytical static offset dipole-plus-quadrupole field from \citetalias{kalapotharakos_multipolar_2021} as the input field $\bsym{B_{\star}}$ and compute numerical values\footnote{Analytic expressions for the coefficients are possible, but the derivation are unwieldily for high-order terms.} of the coefficients $a_{lm}^{B,{\rm in}}$ for several highest-likelihood field configurations from \citetalias{kalapotharakos_multipolar_2021}. Using their six vacuum field solutions based on the \MR values from \cite{riley_nicer_2019}, we estimate the maximum multipole order $l_{\max}$ required for the truncated VSH expansion to approximate these fields. For a convergence threshold of $\delta {\cal C}=10^{-3}$, we find the corresponding $l_{\max}$ values for the solutions $\{RV1, RV2, RV3, RV4, RV5, RV6\}$ (following their labels) to be  $\{9, 18, 18, 18, >20, 4\}$. However, $l_{\max}$ alone is only a proxy for ``complexity," because power is distributed non-monotonically across $(l,m)$ and can extend to higher $l$ without following a simple decay law.

To reveal the actual spectral content,  we plot the numerical values of $|a_{lm}^{B,{\rm in}}|$ up to $l=20$ with $m = \{0,\ldots,l\}$ in Figure~\ref{fig:acoeff}. Since $a_{l,-m}^{B,{\rm in}} = (-1)^{m} (a_{l,m}^{B,{\rm in}})^{*}$, it
follows that $|a_{l,-m}^{B,{\rm in}}|=|a_{l,m}^{B,{\rm in}}|$. Therefore, we show only the $m\geq 0$ terms and account for the $m<0$ ones by symmetry, practically doubling the $m>0$ contributions when reconstructing the field. The spectra exhibit resonant, non-monotonic behavior, consistent with the harmonic basis of the expansion. On average the amplitudes decrease with $l$, but not in a way that makes $l_{\max}$ alone a reliable metric for complexity. While the input surface field has the magnetic field orientations, linear offsets, and strength of the field components as free parameters, we find that the offsets contribute more strongly than the orientations to the resulting complexity.

We therefore evaluate $\delta {\cal C}$ as a function of the number of included modes sorted by coefficient magnitude, independent of $(l,m)$. Figure~\ref{fig:CM_CK21_RV} plots the ordered $|a_{lm}^{B,{\rm in}}|$ (dashed lines) together with $\delta {\cal C}$ (solid lines) as modes are added in decreasing order of $|a_{lm}^{B,{\rm in}}|$. A single $l$ contributes $l+1$ modes and the cumulative count up to $l$, $n_l$, is $l(l+3)/2$. Thus, the $x-$axis shows $230$ multipolar modes corresponding to $l = 20$. The legend shows the radial dipole and quadrupole offset, $r_D$ and $r_Q$ for all six cases, normalized to $R_{\star}$. The inset table lists the $n_l$ values required to achieve a target accuracy of $\delta {\cal C}\leq 10^{-3}$ for both unordered (not shown) and ordered coefficients, whose corresponding values are shown as black points. While the values are broadly consistent between the two cases, the coefficient-ordered calculation provides a more robust measure of complexity than $l_{\max}$ alone (i.e., the unordered case) and yields a more reliable estimate of the effective number of modes required.

Only two cases require fewer than $50$ terms to achieve convergence at the level of $10^{-3}$, whereas three of the solutions reach this level near $200$ multipolar modes due to the significant dipole and quadrupole offsets. The $RV5$ solution, despite having a zero dipole offset, is the slowest converging one because its quadrupole offset exceeds half of $R_{\star}$. The $RV6$ solution with no dipole offset and only a small quadrupole offset ($\sim 0.02 R_{\star}$) is effectively a centered multipolar field, and hence, converges rapidly, requiring only a small number of multipolar modes. This interpretation is also supported by Figure~\ref{fig:acoeff}, where $RV6$ exhibits a considerably smaller number of terms with meaningfully high coefficients than the other cases. Although some small residual non-zero coefficients remain visible at higher $(l,m)$, these are close to the precision limits and are therefore not physically meaningful.

The aim of this part of the study is to establish a way of comparing different magnetic field prescriptions, specifically by using the centered VSH expansion and $\delta {\cal C}$. In this way, we assess how an offset static-vacuum field and a centered SVM2F relate to one another in terms of how much expansion content is needed to represent them to a given accuracy. This does not assign an absolute physical simplicity to either description; rather, it quantifies their relative mathematical compactness in the adopted centered-expansion framework.

\section{Compact parameterization of the SVM2F expansion}\label{sec:application_SVM2F}
For the analysis of the NICER X-ray light curve of J0030, we now recast the SVM2F magnetic field expansion into a form suitable for efficient numerical implementation. Using the standard properties of the scalar and vector spherical harmonics, the corresponding field expressions are derived in Appendix~\ref{app:multipole_expansion}. Once the centered interior coefficients $a_{lm}^{B,\mathrm{in}}$ are specified, the remaining coefficients of the full analytic solution follow from the boundary conditions discussed in Section~\ref{sec:mf_expansion}. The main task, therefore, is to introduce a compact and physically meaningful parameterization of these independent coefficients.

Following \citetalias{petri_multipolar_2015}, we parameterize the
centered interior magnetic coefficients $a_{lm}^{B,\mathrm{in}}$ of each multipolar family by an overall amplitude $B_l$, a set of generalized (hyperspherical) angles $\{\chi_{l1},\ldots,\chi_{ll}\}$ that determine the relative weights of the different $m$-modes, and phase angles $\{\phi_{lm}\}$ that fix their azimuthal orientations. This choice provides a compact and physically transparent description of the field geometry while ensuring that, for fixed $B_l$, different choices of the $\chi$-angles correspond only to redistributions among the $m$-modes at fixed total magnetic energy outside the star. Specifically, we write
\begin{equation}
a_{lm}^{B,\mathrm{in}}
=
B_l R_\star^{\,l+2}\, q_{lm}\, e^{im\phi_{lm}},
\qquad m=0,\ldots,l,
\label{eq:abin_param}
\end{equation}
with $\phi_{l0}\equiv 0$. We note that, for the dipole axisymmetric coefficient ($l=1, m=0$), an additional minus sign must be included relative to Equation~\ref{eq:abin_param} in order to recover the standard dipole-field convention, so that $\chi_D=0$ corresponds to a dipole aligned with the $+\hat{z}$ axis. The real coefficients $q_{lm}$ are written as
\begin{equation}
q_{lm}=C_{lm}\,u_{lm}(\chi_{l1},\ldots,\chi_{ll}),
\label{eq:qlm_param}
\end{equation}
where the $u_{lm}$ are the standard hyperspherical coordinates on the
unit $l$-sphere,
\begin{align}
u_{l0} &= \cos\chi_{l1} \label{eq:ulm_def_m0},
\\
u_{lm} &=
\left(\prod_{j=1}^{m}\sin\chi_{lj}\right)\cos\chi_{l,m+1},
\qquad 1\le m\le l-1 \label{eq:ulm_def_m1lm1},
\\
u_{ll} &= \prod_{j=1}^{l}\sin\chi_{lj} \label{eq:ulm_def_ml},
\end{align}
with $\chi_{l1},\ldots,\chi_{l,l-1}\in[0,\pi]$ and
$\chi_{ll}\in[0,2\pi]$. The normalization constants are
\begin{equation}
C_{lm} = \sqrt{\frac{16\pi}{3l(1+\delta_{m0})}},
\label{eq:Clm_general}
\end{equation}
where $\delta_{m0}$ is the Kronecker delta, so that for fixed $B_l$, the magnetic energy outside the star is independent of the particular distribution among the $m$-modes:
\begin{equation}
W_l
=
\frac{l B_l^2 R_\star^3}{4\mu_0}
\left(2q_{l0}^2+\sum_{m=1}^{l}q_{lm}^2\right)
=
\frac{4\pi}{3\mu_0} B_l^2 R_\star^3.
\label{eq:Wl_general}
\end{equation}
Thus $B_l$ sets the overall strength of the $l$th multipolar family, the $\chi$-angles determine how that strength is distributed among the allowed $m$-modes, and the $\phi_{lm}\in[0,2\pi]$ specify the azimuthal orientations of the non-axisymmetric components. With the expressions for $a_{lm}^{B,\mathrm{in}}$ known, the $a_{lm}^{B,\mathrm{out}}$ and $a_{lm}^{D,\mathrm{out}}$ expressions are described using Equations~\ref{eqn:almOutAndIn}-\ref{eq:Jlm_For_almD_out}. We list these expressions\footnote{Once the coefficients $a_{lm}^{B,\mathrm{in}}$ for a given input field are determined, as described in Section~\ref{subsubsec:coeff_SVF}, the corresponding SVM2F parameter set ($\chi_{lm}, \phi_{lm}, B_l$) can be reconstructed by inverting the numerical $a_{lm}^{B,\mathrm{in}}$ using the relations above, up to any desired multipole order. This procedure allows any input field to be interpreted in terms of the SVM2F parameters, enabling a direct comparison.} up to an octupole in Table~\ref{tab:coeffs} (Appendix~\ref{app:multipole_expansion}).

Focusing on the low-order components retained in our J0030 analysis, we include multipoles up to $l=3$, namely the dipole, quadrupole, and octupole families. We denote their overall amplitudes by $B_D$, $B_Q$, and $B_O$, respectively, and express their relative strengths through the ratios $B_Q/B_D$ and $B_O/B_D$, with the dipole setting the reference normalization. In this notation, the dipolar component is described by one hyperspherical angle and one phase parameter, denoted by $\chi_D$ and $\phi_D$. The quadrupolar component requires two hyperspherical angles and two phase parameters, denoted by $\chi_{Q1}$, $\chi_{Q2}$ and $\phi_{Q1}$, $\phi_{Q2}$, respectively. Likewise, the octupolar component is described by three hyperspherical angles and three phase parameters, denoted by
$\chi_{O1}$, $\chi_{O2}$, $\chi_{O3}$ and $\phi_{O1}$, $\phi_{O2}$, $\phi_{O3}$. 

With this notation, the full parameter set up to $l=2$ contains seven free parameters, namely $\{\chi_D,\allowbreak \phi_D,\allowbreak \chi_{Q1},\allowbreak \chi_{Q2},\allowbreak \phi_{Q1},\allowbreak \phi_{Q2},\allowbreak B_Q/B_D\}$, while including the octupole adds seven more, $\{\chi_{O1},\allowbreak \chi_{O2},\allowbreak \chi_{O3},\allowbreak \phi_{O1},\allowbreak \phi_{O2},\allowbreak \phi_{O3},\allowbreak B_O/B_D\}$, bringing the total to 14. To keep the dimensionality of the parameter space comparable to that used in \citetalias{kalapotharakos_multipolar_2021}, we impose the additional constraints $\phi_{Q1}=\phi_{Q2}$ and $\phi_{O1}=\phi_{O2}=\phi_{O3}$. As a result, the effective parameter space is reduced to 6 free parameters for dipole-plus-quadrupole models and 11 for dipole-plus-quadrupole-plus-octupole models, instead of 7 and 14, respectively. Therefore, the full set of parameters adopted in this work becomes $\{\chi_D,\allowbreak \phi_D,\allowbreak \chi_{Q1},\allowbreak \chi_{Q2},\allowbreak \phi_{Q},\allowbreak B_Q/B_D,\allowbreak \chi_{O1},\allowbreak \chi_{O2},\allowbreak \chi_{O3},\allowbreak \phi_{O},\allowbreak B_O/B_D\}$.

Replacing the static field prescription of \citetalias{kalapotharakos_multipolar_2021} with SVM2F offers the advantage of retaining the near-surface multipolar structure while also incorporating the rotation-induced sweepback of the magnetic field. The former controls the local field geometry close to the stellar surface, while the latter becomes increasingly important toward the light-cylinder radius $R_{\rm LC}$.

\section{Methodology} \label{sec:methodology}
Our approach follows \citetalias{kalapotharakos_multipolar_2021} in modeling the NICER bolometric X-ray ($0.25 - 1.45$ keV) light curve of J$0030$ \citep{bogdanov_constraining_2019}, but replaces their static vacuum field with the SVM2F and its associated parameterization. The implementation of the complex SVM2F expressions (Section~\ref{sec:mf_expansion}, ~\ref{sec:application_SVM2F}) up to the quadrupole components in an independent Fortran module was calibrated by reproducing magnetic field lines and polar cap footprints. The polar caps are calculated by determining the field lines that cross the light cylinder (open field lines), which are integrated using an adaptive fourth-fifth order Runge-Kutta method. The field lines are initialized uniformly over the stellar surface with a fixed grid resolution\footnote{Unless stated otherwise, grid resolution refers to a uniform grid in $\phi$ and cos $\theta$.}, and the full surface is described by employing bilinear interpolation. We reproduced several orientations from \citetalias{petri_multipolar_2015} to benchmark the general field geometry produced by our implementation. However, the polar-cap figures presented there are based on a point-multipole approximation and therefore are not suitable for validating the absolute size of the finite-multipole polar caps. For that reason, we calibrated the finite dipole polar caps by reproducing the exact Deutsch solutions \citep{deutsch_1955_deutsch_the_1955}. In addition, we present several octupole-only cases in Appendix~\ref{app:PolarCaps_l3}.

The photon trajectories are identical to those used in \citetalias{kalapotharakos_multipolar_2021}, generated using their ray-tracing code, \texttt{GIKS} (Geodesic Integration in Kerr Spacetime)\footnote{We note that while \texttt{GIKS} uses the Kerr metric, frame-dragging effects are negligible in this regime, making the results effectively equivalent to Schwarzschild spacetime.}, which integrates null geodesics, i.e., photon trajectories, from a distant image plane, effectively at infinity, to the NS surface in a Kerr spacetime (an approach similar to \citealt{johannsen_testing_2010, baubock_ray-tracing_2012, psaltis_ray-tracing_2011}). For each geodesic, we tabulate the impact position and the arrival time at the stellar surface, the corresponding position on the image plane, and the zenith angle, $\theta_{z}$. These quantities are stored in a library implemented in an MPI-Fortran code, which is used to identify the photons intersecting the polar caps, i.e., hotspots, and to map their arrivals to rotational phases. Photon weights follow \citet{psaltis_pulse_2014} and include Doppler boosting and gravitational redshift. Following \cite{riley_nicer_2019}, we adopt their median values $M_{\star} = 1.34~M_{\odot}$, $R_{\star} = 12.71$~km, and the observer viewing angle, $\zeta = 53.85^\circ$. Furthermore, following \citetalias{kalapotharakos_multipolar_2021}, the temperature is assumed to be the same for all hotspots, and a simplified anisotropic atmosphere model with specific intensity proportional to $\cos^b{\theta_{z}}$ with $b=1$ is adopted.
For all calculations, we use $P = 4.87$~ms, and therefore, $R_{LC}$ is approximately $233 {\, \rm km}$, or equivalently $\approx 18 R_{\star}$.

For each set of field configuration parameters, we determine the instantaneous polar-cap footprints using the SVM2F module, and synthesize the corresponding model thermal X-ray light curve over one stellar rotation by sampling the precomputed geodesic/photon library across one rotational phase, using the same phase binning as the NICER data (64 bins). Throughout this paper, we refer to this as the \emph{SVM2F physical model}.

The MCMC methodology is the same as that employed by \citetalias{kalapotharakos_multipolar_2021} using the stretch-move algorithm \citep{goodman_ensemble_2010}. The MCMC sampler explores the parameter space of magnetic field configurations by evaluating, at each time step, the likelihood of the model-generated light curves against NICER’s bolometric X-ray light-curve data and accepting or rejecting  proposed moves according to the corresponding MCMC acceptance criterion, thereby drawing samples from the posterior distribution rather than merely maximizing a log-likelihood.

We investigate two truncations of the SVM2F expansion: (i) a model including dipolar and quadrupolar components ($l\leq 2$), \UQ, and (ii) a model that additionally incorporates the octupole component ($l\leq 3$), \UO. After applying the additional phase constraints described in Section~\ref{sec:application_SVM2F}, these models have 6 and 11 free parameters, respectively.

\subsection{NN Surrogate Models}\label{subsec:NNsurrograteModels}
Robust posterior inference requires on the order of $10^{9}$ log-likelihood evaluations during MCMC. Even with a highly parallel affine-invariant sampler running on $\sim 4000$ CPU cores, where each high-fidelity forward evaluation of a light curve takes minutes, this would correspond to wall times of $\gtrsim$ year-scale to obtain a converged posterior, making robust inference computationally impractical. As mentioned in the introduction, to overcome this barrier, \citetalias{olmschenk_pioneering_2025} developed NN surrogates trained on model light curves generated from the static offset dipole–plus–quadrupole prescription of \citetalias{kalapotharakos_multipolar_2021}. These NNs reproduced the physical model with high fidelity and enabled $>400\times$ faster likelihood evaluations. On the same $\sim 4000$ cores, their NN-enabled MCMC reached equilibrium for J0030 in $\sim$1 day. Additionally, for a physical model, the calculation of each light curve requires the identification of hotspots, and hence, field-line integration to determine open line regions. As a result, the runtime varies with the structure of the magnetic field, unlike a trained NN surrogate where evaluation time is fixed across cases.

The reliability of their NNs was validated through several complementary tests: (i) direct comparisons of posterior distributions obtained from full MCMC runs using the physical model versus the NN; (ii) continuation runs in which a physical-model MCMC was resumed from the NN-converged state, showing no significant divergence; and (iii) stability checks across NNs trained on data sets of different sizes. Agreement was quantified using multiple distributional metrics, Wasserstein distance, Jensen–Shannon divergence, and Kolmogorov–Smirnov statistic \citep{panaretos_statistical_2019, menendez_jensen-shannon_1997, massey_jr_kolmogorov-smirnov_1951}, as well as the credible interval required to encompass the median of the distribution from the other method. Overall, the NN- and physical-model posteriors were in good agreement typically well within the $1\sigma$ credible region, and showed only small differences across the adopted comparison metrics.

Building on the demonstrated reliability of the trained NN in
\citetalias{olmschenk_pioneering_2025}, we also employ the NN surrogate\footnote{\url{https://github.com/golmschenk/haplo} \citep{haplo2025}}, integrated into our MCMC framework. The convolutional NN model is trained on a dataset generated using the \UO model, covering $11$ magnetic field parameters and the corresponding model light curves. We followed the same training procedure described in \citetalias{olmschenk_pioneering_2025}; briefly, the network minimizes a loss function normalized by the median flux of the input light curves to ensure scale invariance.

Using a dataset consisting of $5 \times 10^7$ parameter--light-curve pairs (with $10^5$ samples reserved for validation and $10^5$ for testing), the NN was trained for $24$ hours across 128 Nvidia A100 GPUs. For the \UQ truncation, with $6$ free parameters, we initialized the training from the pre-trained \UO NN and continued training for an additional $12$ hours utilizing same GPU resources and a dataset of the same size but for an extended parameter range that also includes cases with zero octupole strength. Training the NN on a dataset an order of magnitude larger did not result in a significant performance difference \citepalias{olmschenk_pioneering_2025}. 

We verified that the SVM2F-trained NN exhibits comparable performance to the static vacuum model. In particular, the distribution of loss-function values on the test sets closely match in the two cases. This provides strong evidence that a training dataset of size $5 \times 10^7$ offers sufficient accuracy for SVM2F as well. Additionally, several reliability tests, analogous to those performed for the NN trained on the static offset field by \citetalias{olmschenk_pioneering_2025}, were conducted for the NN trained on the SVM2F model.

To maximize training efficiency, we exploit the rotational symmetry of the light curves. From each physical calculation, we generate 64 phase-shifted light curves (matching the 64 phase bins of the NICER J0030 data) by rotating the phase parameters of all $(l,m)$ modes over $2\pi$ in 64 steps. Model light curves inevitably include very small numerical discrepancies arising from the finite surface grid and the finite photon budget in the ray-tracing library. These effects vary smoothly across the parameter space and do not bias our results. The synthetic phase rotations introduce similarly minor differences, since a rotated light curve is not perfectly identical to one recalculated from the correspondingly rotated parameters at finite resolution. However, these discrepancies originate from the same numerical limitations, remain small and smoothly varying, and are typically much smaller than the uncertainties of the observed signal, as well as smaller than the residual errors of the NN surrogate in reproducing the physical-model light curves.

\subsection{MCMC execution}
We explore the parameter space\footnote{We set an upper limit of $15$ on the relative magnetic strengths. The allowed ranges for the rest of the parameters are discussed in Section~\ref{sec:application_SVM2F}.} for the \UQ and \UO truncations using the NN surrogate model integrated with the MCMC algorithm for $\sim 10^{10}$ samples with randomly initialized parallel chains (\emph{Run 1}). To improve the performance of the stretch move, we studied the variation of an adjustable scale parameter, $a_{scale}$, in the MCMC algorithm, which controls the step size for walkers, and hence, influences the acceptance rate (the fraction of proposed steps that are accepted, \citealp{foreman-mackey_emcee_2013}). For \UO, it was fixed to $12$ (as opposed to the default value of $2$; \citealt{goodman_ensemble_2010}). For \UQ, however, an additional $\sim 10^9$ evaluations were performed with varying $a_{scale}$ values, to eliminate samples that were clustered in a mode with sub-optimal log-likelihood values.

To ensure physically meaningful solutions, chains that had reached equilibrium using the NN surrogate were subsequently continued with the physical model (\emph{Run 2}). The resulting peak log-likelihood samples were then used as initial points for $40$ independent serial chains, allowing refined exploration of each solution island and identification of the ultimate best-fit parameters (\emph{Run 3}). A summary of the MCMC execution flow for each run is provided in Table~\ref{tab:MCMCexecutionDetails} in Appendix~\ref{app:MethodsTables}, along with an index of the corresponding output figures which are presented in the following sections.

\section{Results} \label{sec:results}
\begin{figure*}
    \centering
    \begin{subfigure}{0.98\linewidth}
        \centering
        \includegraphics[width=\linewidth]{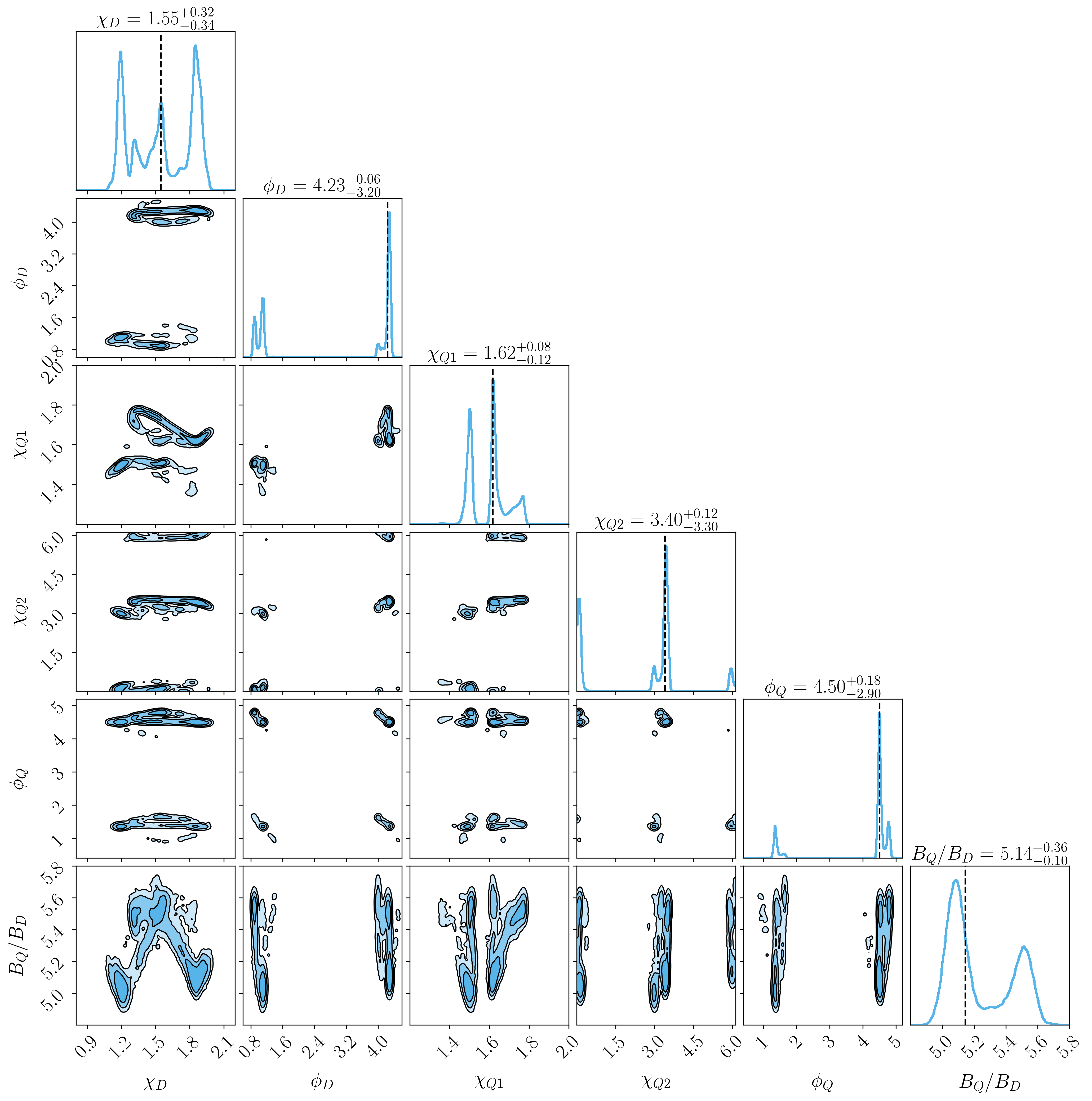}
    \end{subfigure}
    \caption{Posterior probability distributions of the six parameters for the \UQ truncation are displayed, with black contour boundaries and shaded regions corresponding to the $1\sigma$, $2\sigma$, and $3\sigma$ confidence levels, from darkest to lightest sky blue. The diagonal panels show the marginal $1D$ distributions, with dashed vertical lines indicating the median values, and the corresponding $\pm \sigma$ uncertainties are given at the top of each panel.}
    \label{fig:corner_l12}
\end{figure*}
\begin{figure*}
    \centering
    \begin{subfigure}{0.98\linewidth}
        \centering
        \includegraphics[width=\linewidth]{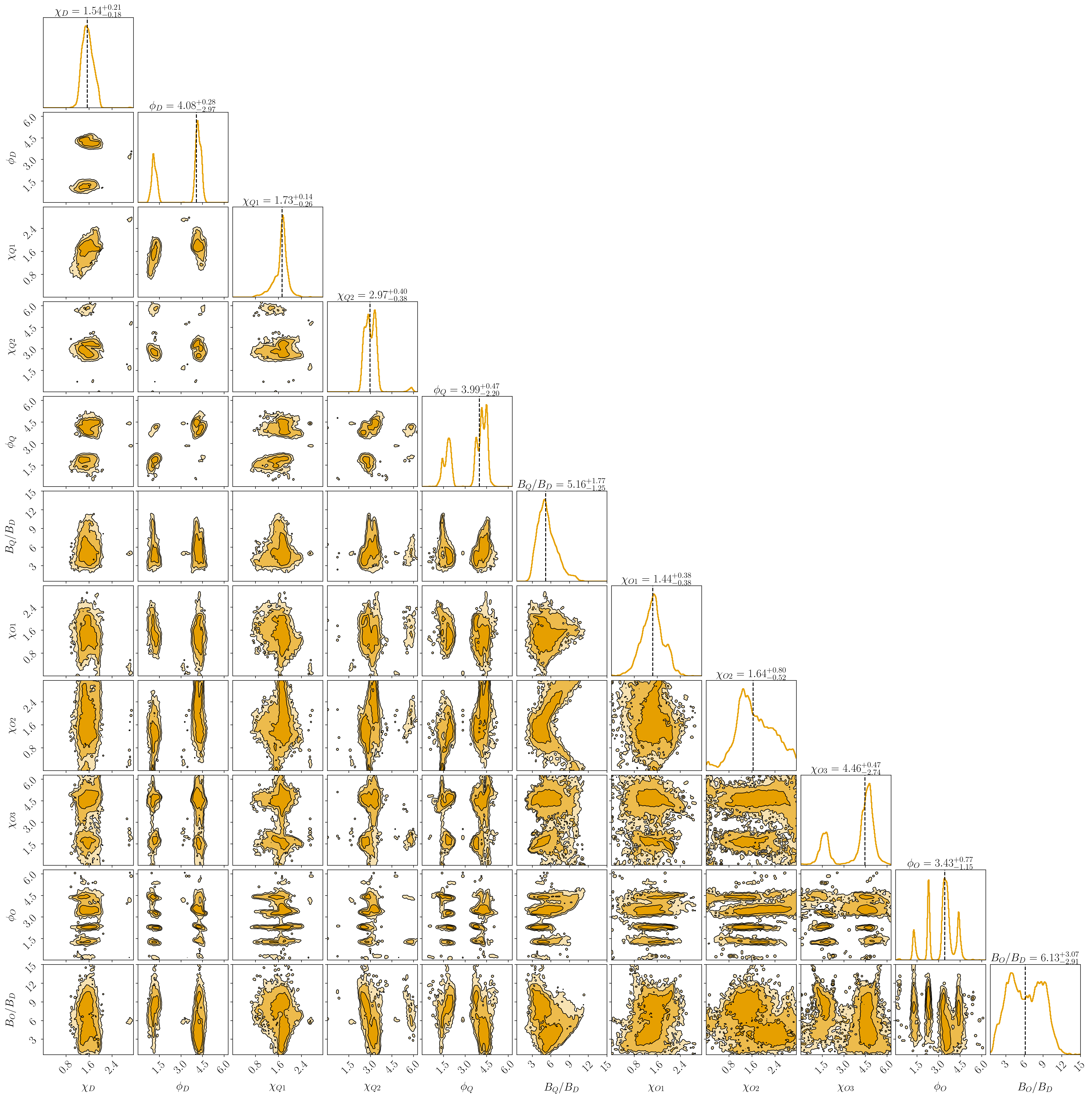}
    \end{subfigure}
    \caption{Posterior probability distributions of the $11$ parameters for the \UO truncation are displayed, with black contour boundaries and shaded regions corresponding to the $1\sigma$, $2\sigma$, and $3\sigma$ confidence levels, from darkest to lightest orange. The diagonal panels show the marginal $1D$ distributions, with dashed vertical lines indicating the median values, and the corresponding $\pm \sigma$ uncertainties are given at the top of each panel.}
    \label{fig:corner_l123}
\end{figure*}
The posterior distribution for the \UQ truncation is presented in Figure \ref{fig:corner_l12}, with contours extending to the $3\sigma$ level. The posterior is highly structured. Several marginalized distributions are strongly asymmetric, and a number of projections show two to three local maxima, indicating apparent multi-modality rather than a single compact mode. To assess whether these features correspond to distinct solution populations, we applied several clustering algorithms, including K-Means (see \citealp{pedregosa_scikit-learn_2011}). However, no sharply distinct clusters were identified. Figure~\ref{fig:corner_l123} shows the posterior distribution for \UO, with contours also extending to the $3\sigma$ level. This distribution is similarly structured with several high-probability regions that are more isolated in some projections, again consistent with multi-modality.

The orientation of the dipolar component, $\chi_{D}$, is strongly constrained to be near orthogonal in both cases, with a median value of $\approx 88^{\circ}$. This is consistent with the observed X-ray light curve morphology of J$0030$ \citep{bogdanov_constraining_2019}, which exhibits an asymmetric double-peaked structure indicative of a near-orthogonal magnetic field geometry. Our inferred value agrees with previous estimates provided by \citet{chen_numerical_2020, kalapotharakos_multipolar_2021, petri_constraining_2023}
and lies within the robust range suggested by \cite{CaoYang_J0030_2026}.

The phase parameters are multi-modal, with two distinct modes in both $\phi_{D}$ and $\phi_{Q}$ for each truncation. In most cases, the two peaks are separated by approximately $\pi$, except for $\phi_{Q}$ in the \UO case, where the modes exhibit pronounced substructure. The distribution of $\phi_{O}$ shows four modes, each separated by roughly $\pi/2$. Although most modes display regular spacing, the distributions remain asymmetric, as reflected in the unequal peak amplitudes. This indicates that the posterior weight is not distributed uniformly across the corresponding modes. 

The quadrupolar component is inferred to be approximately five to six times stronger than the dipolar component in all cases. This further supports the broader conclusion that a purely dipolar magnetic field geometry is insufficient to reproduce the bolometric thermal X-ray light curve of J$0030$, and that a significant non-dipolar contribution is required.
\begin{figure}[htbp]
    \centering
    \includegraphics[width=0.98\linewidth]{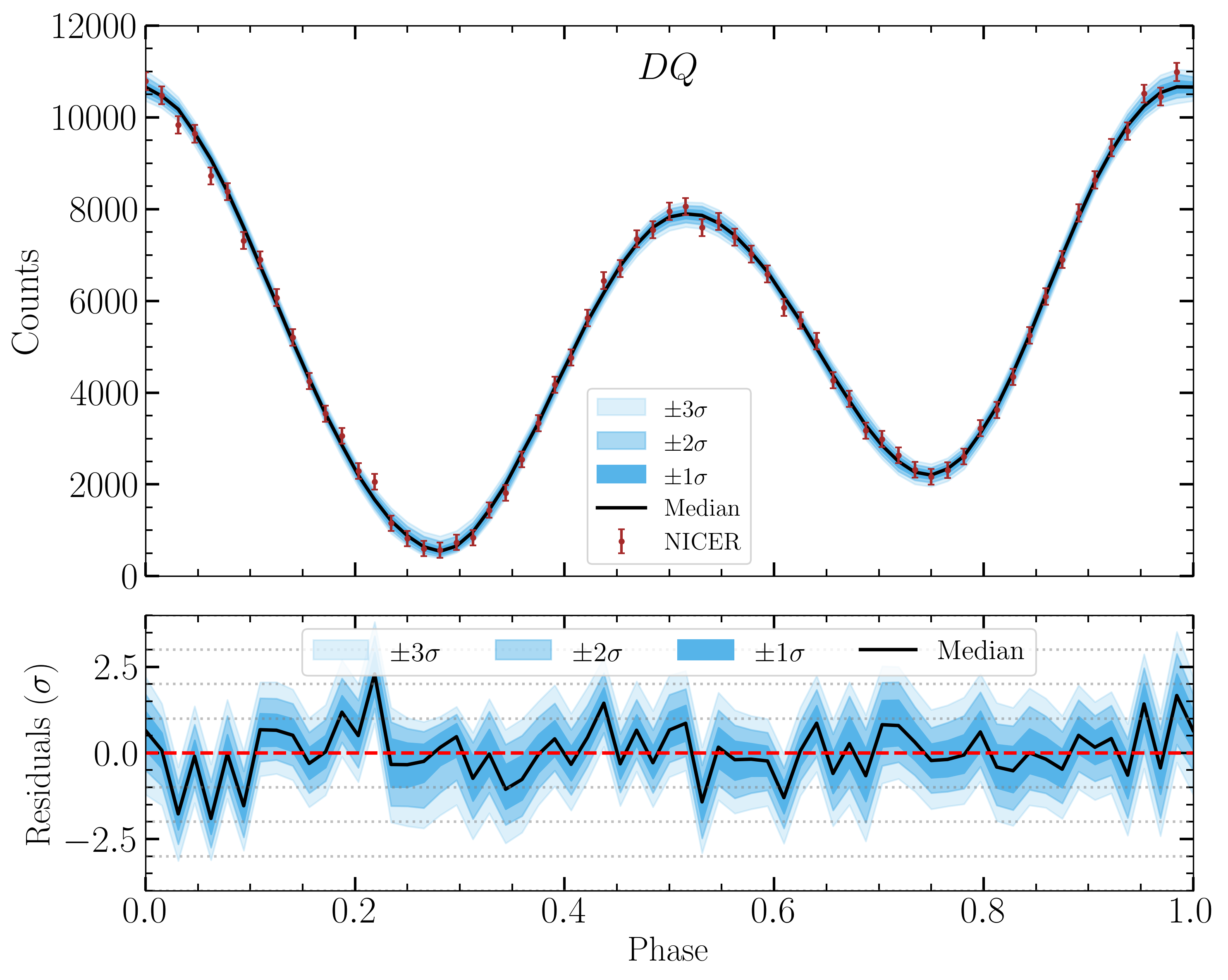}
    \includegraphics[width=0.98\linewidth]{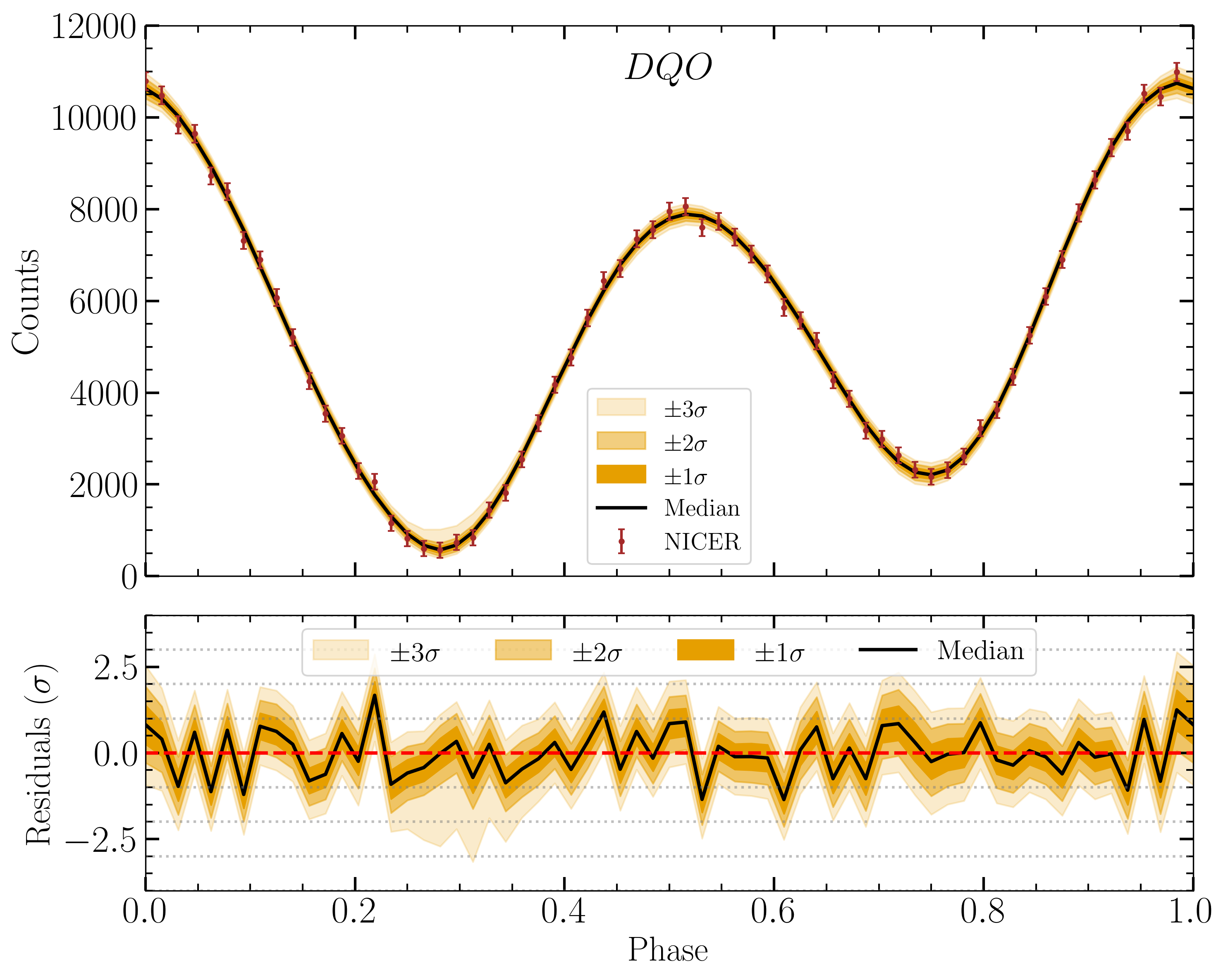}
    \caption{Light-curve ensembles for \UQ (top, sky blue) and \UO (bottom, orange) configurations are shown with the median in black and $1\sigma$, $2\sigma$, and $3\sigma$ regions shaded from darkest to lightest. The ensembles are generated using the same samples as shown in Figures~\ref{fig:corner_l12} and ~\ref{fig:corner_l123}. NICER data are shown in brown. In the lower panels, residuals are displayed with a red dashed line at 0 for reference.}
    \label{fig:lightCurvesEnsemble}
\end{figure}

In Figure~\ref{fig:lightCurvesEnsemble}, we present the light-curve ensembles corresponding to the posterior samples shown in Figures~\ref{fig:corner_l12} and \ref{fig:corner_l123}. For each parameter set, a light curve is computed using the NN model, and the resulting set of light curves defines, at each phase bin, a distribution of predicted counts across the ensemble. The \UQ and \UO truncations are shown in the top and bottom panels, respectively, with shaded regions indicating confidence intervals up to $3\sigma$ around the median ensemble profile. The residuals of the median model together with the residual ranges corresponding to the three confidence levels are displayed in the lower panels. The light-curve ensembles closely reproduce the NICER X-ray light-curve observation,\footnote{Note that our calculations are referenced to the phase zero of the X-ray light curve, in contrast to \citetalias{kalapotharakos_multipolar_2021}, who adopted the radio phase zero to synchronize the X-ray and $\gamma$-ray light curves.} which is shown in brown, adopting the same background treatment and uncertainty estimation as \citetalias{kalapotharakos_multipolar_2021}. The residuals remain within the corresponding $1\sigma$ to $3\sigma$ bands, indicating that the models adequately capture the variations in NICER data.
\begin{figure*}[htbp]
    \centering
    \begin{subfigure}[b]{0.49\linewidth}
        \centering
        \includegraphics[width=\linewidth]{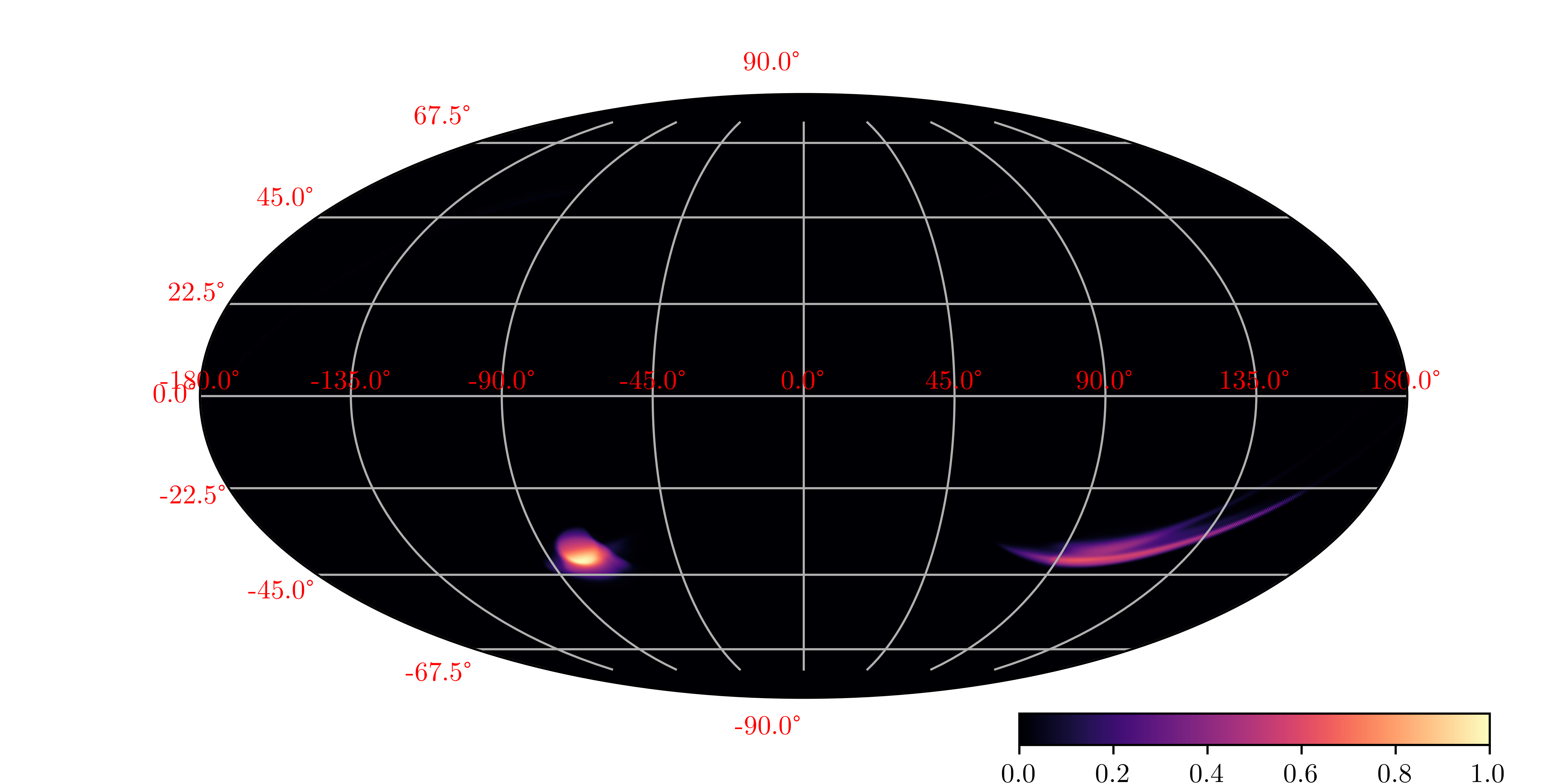}
        \caption{}
        \label{fig:l12HotspotsEnsemble}
    \end{subfigure}
    \begin{subfigure}[b]{0.49\linewidth}
        \centering
    \includegraphics[width=\linewidth]{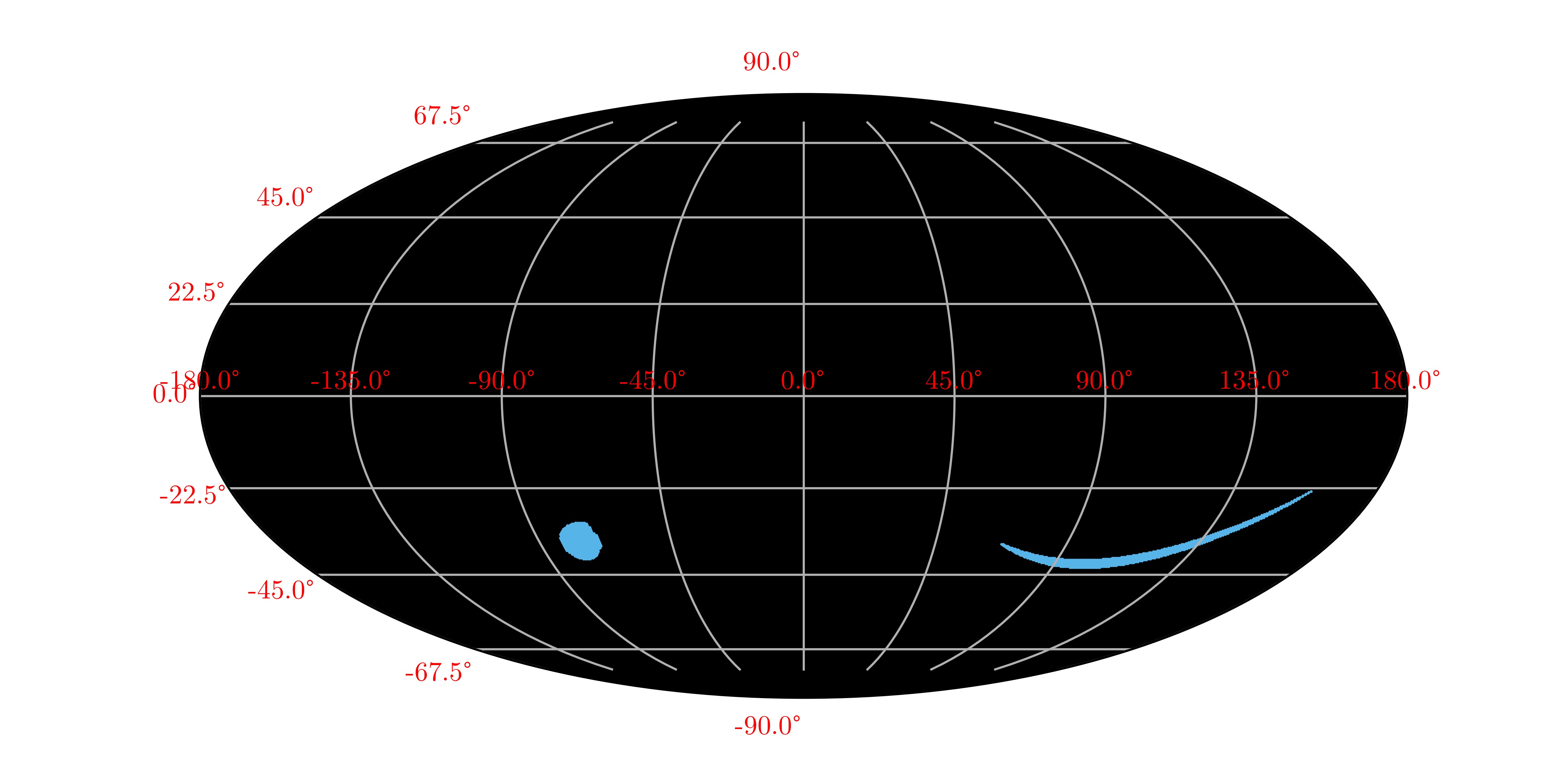}
        \caption{}
    \end{subfigure}
    \begin{subfigure}[b]{0.49\linewidth}
        \centering
    \includegraphics[width=\linewidth]{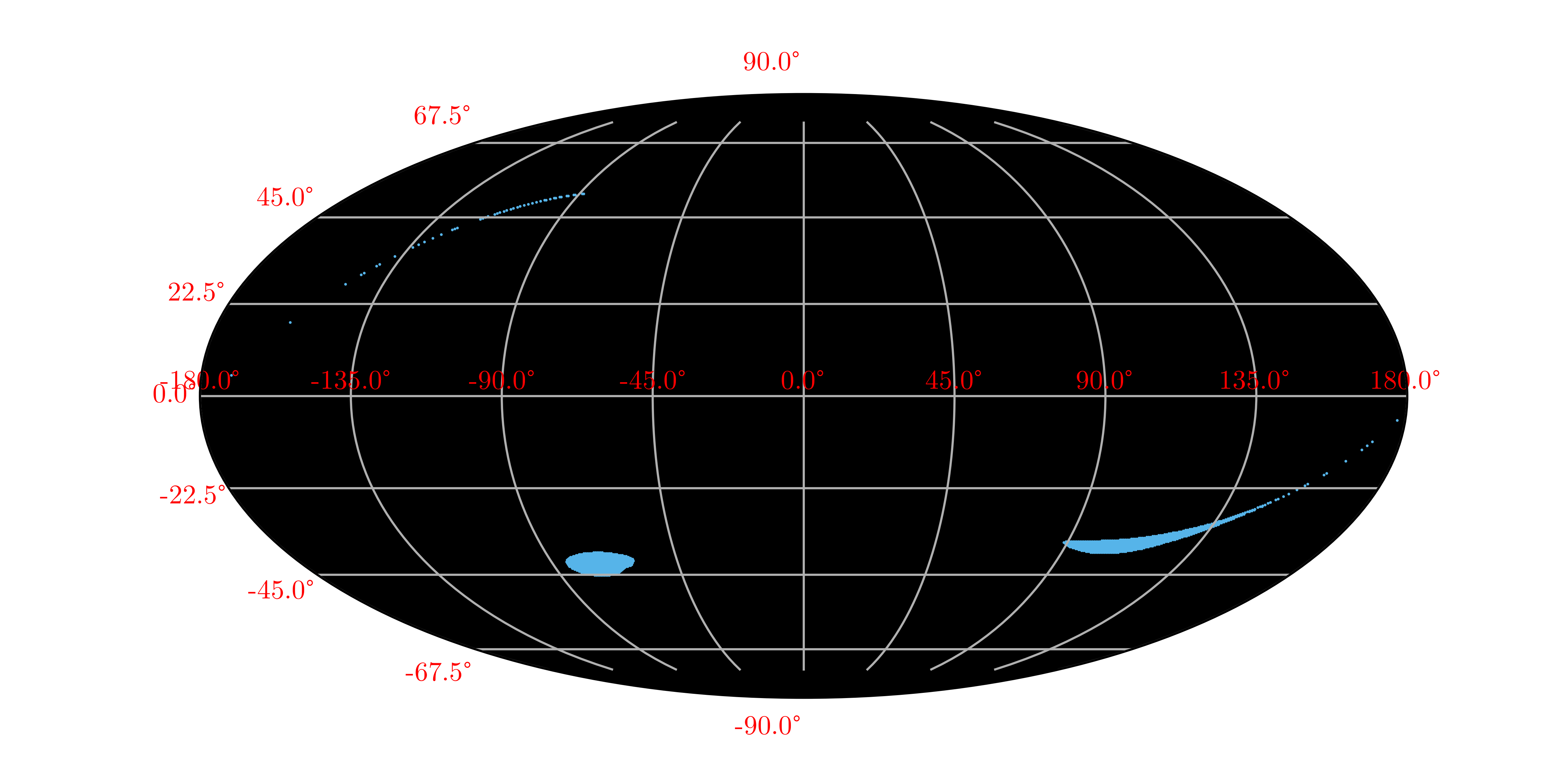}
        \caption{}
    \end{subfigure}
    \begin{subfigure}[b]{0.49\linewidth}
        \centering
    \includegraphics[width=\linewidth]{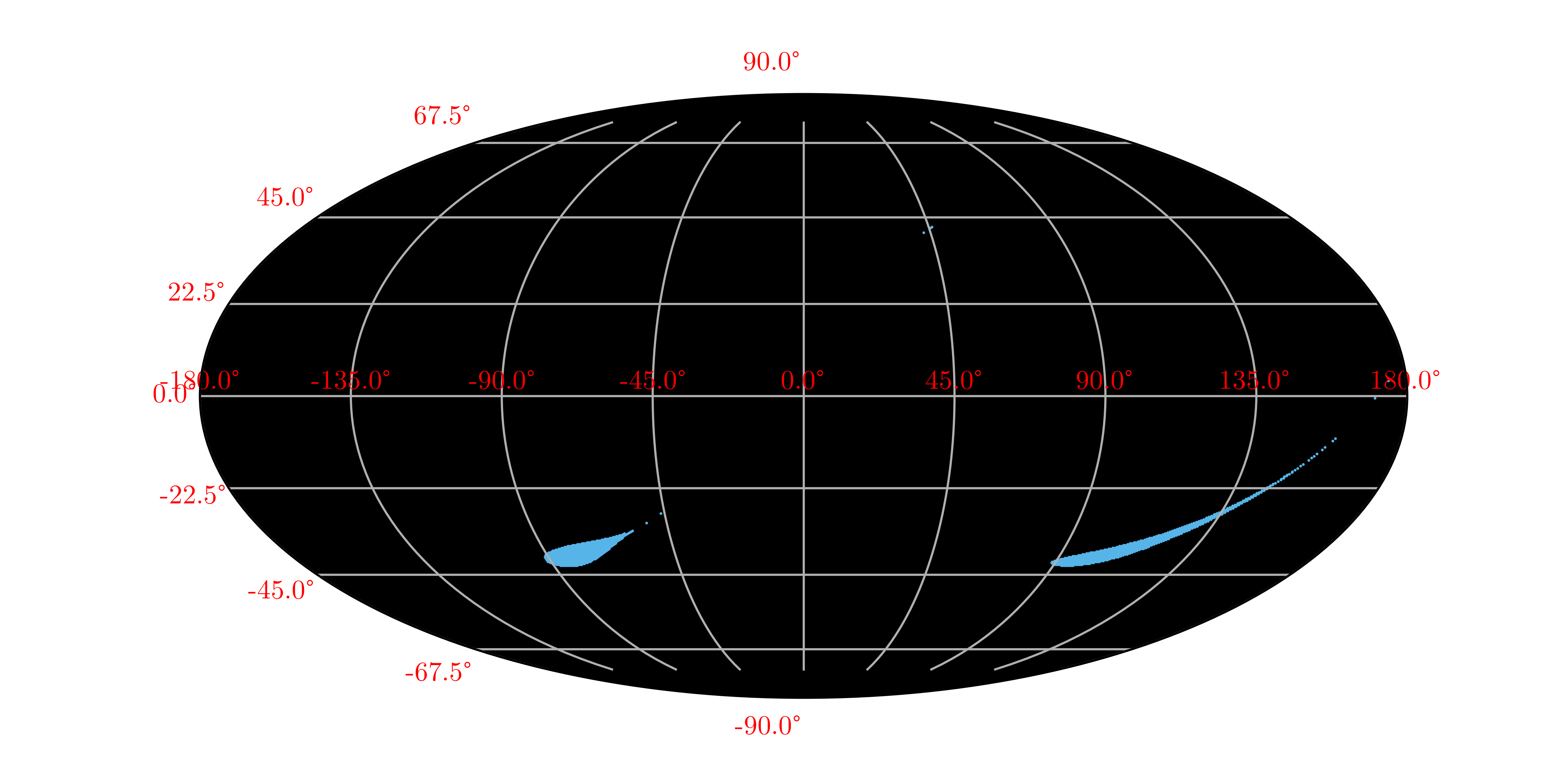}
        \caption{}
    \end{subfigure}
    \caption{Hotspots for the \UQ truncation on a Mollweide projection. (a) Ensemble of $10^5$ weighted hotspots, generated using the same samples as shown in Figure~\ref{fig:corner_l12}, with the colorbar indicating the transition from low density (black) to high density (yellow-white) regions. (b), (c), and (d) show individual hotspots (sky blue) for three randomly selected cases, illustrating the variety of hotspot configurations.}
    \label{fig:l12HotspotsAll}
\end{figure*}
\begin{figure*}[htbp]
    \centering
    \begin{subfigure}[b]{0.49\linewidth}
        \centering
        \includegraphics[width=\linewidth]{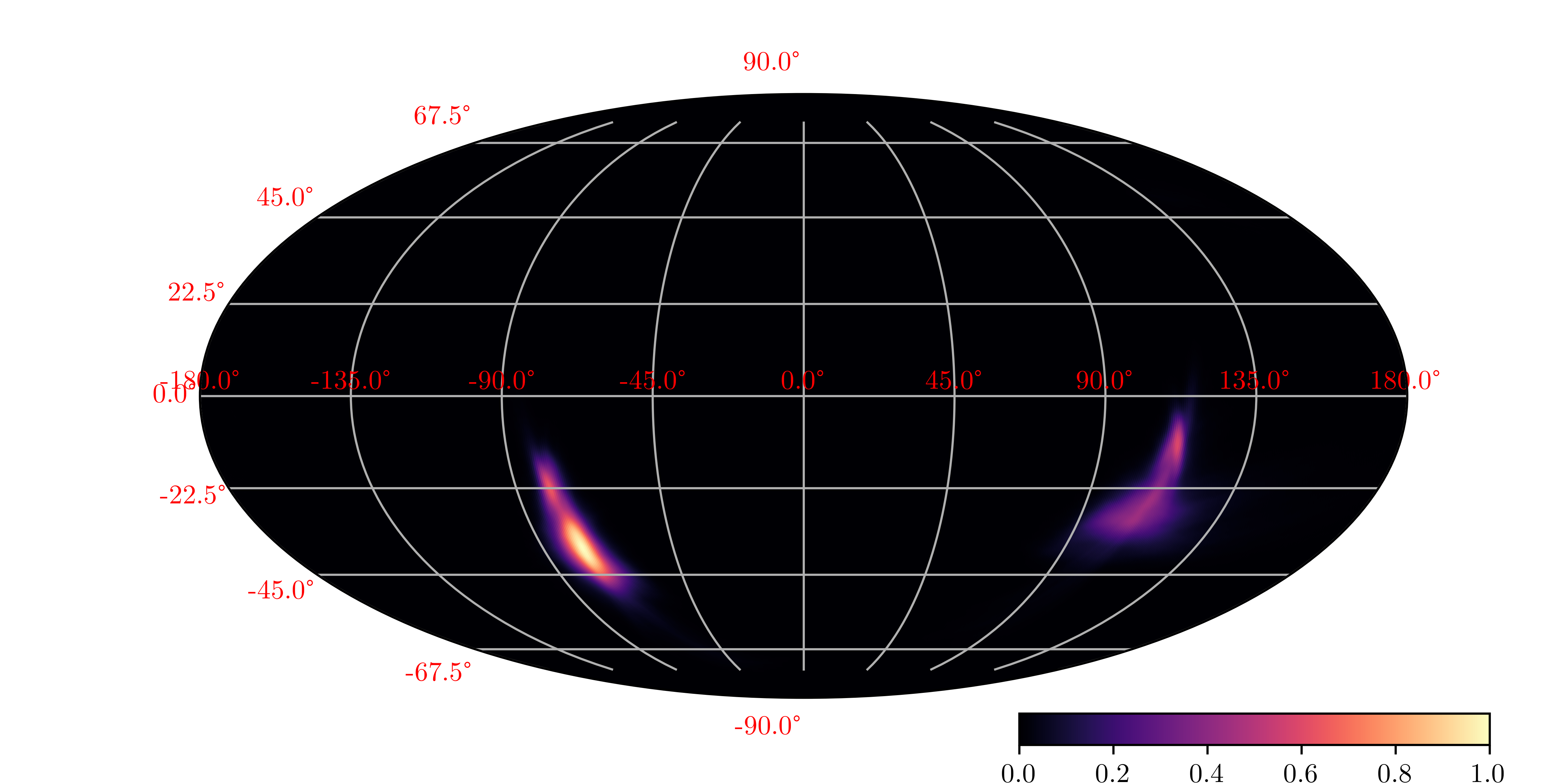}
        \caption{}
        \label{fig:l123HotspotsEnsemble}
    \end{subfigure}
    \begin{subfigure}[b]{0.49\linewidth}
        \centering
        \includegraphics[width=\linewidth]{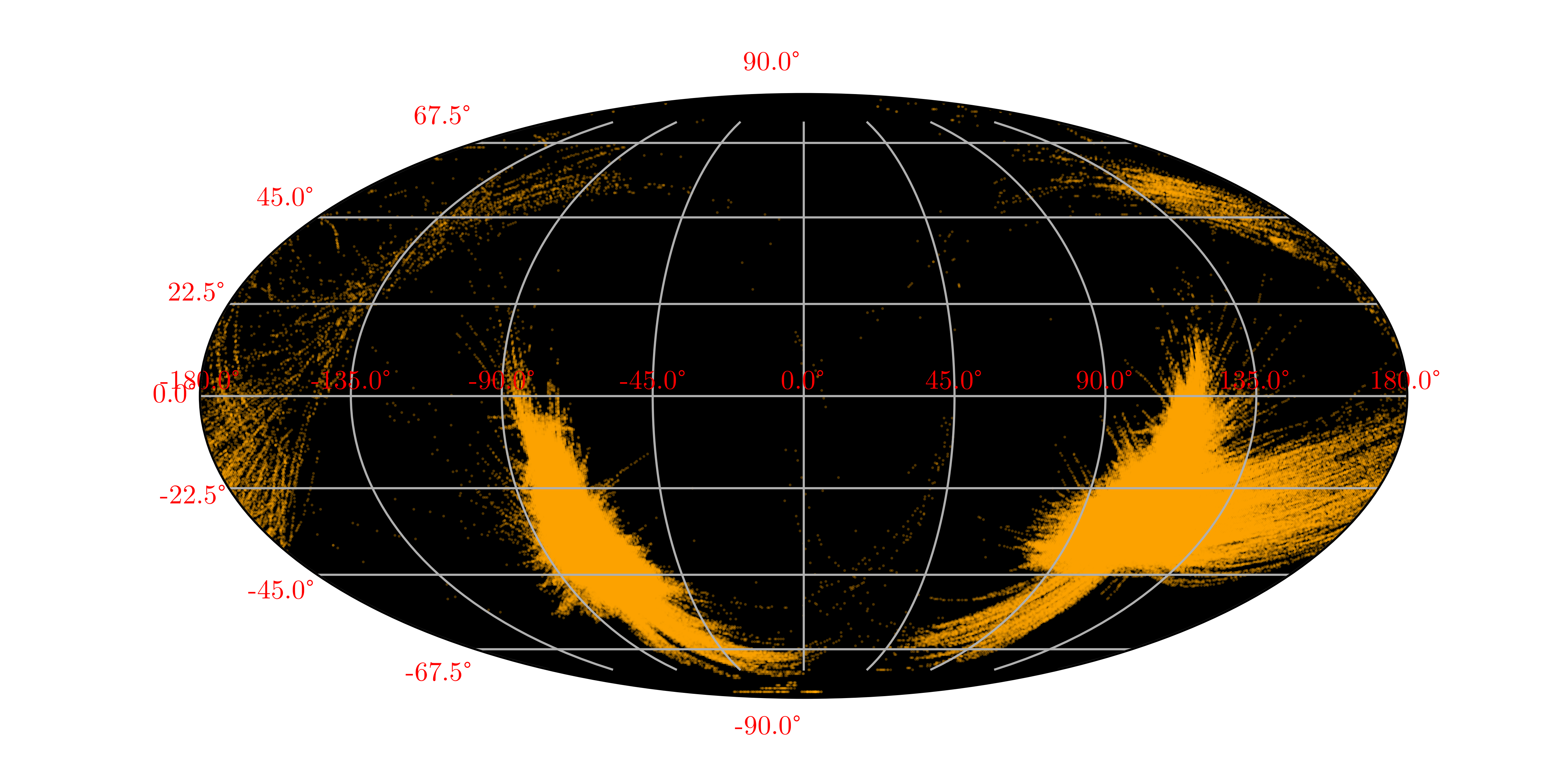}
        \caption{}
        \label{fig:l123HotspotsScatter}
    \end{subfigure}
    \begin{subfigure}[b]{0.49\linewidth}
        \centering
        \includegraphics[width=\linewidth]{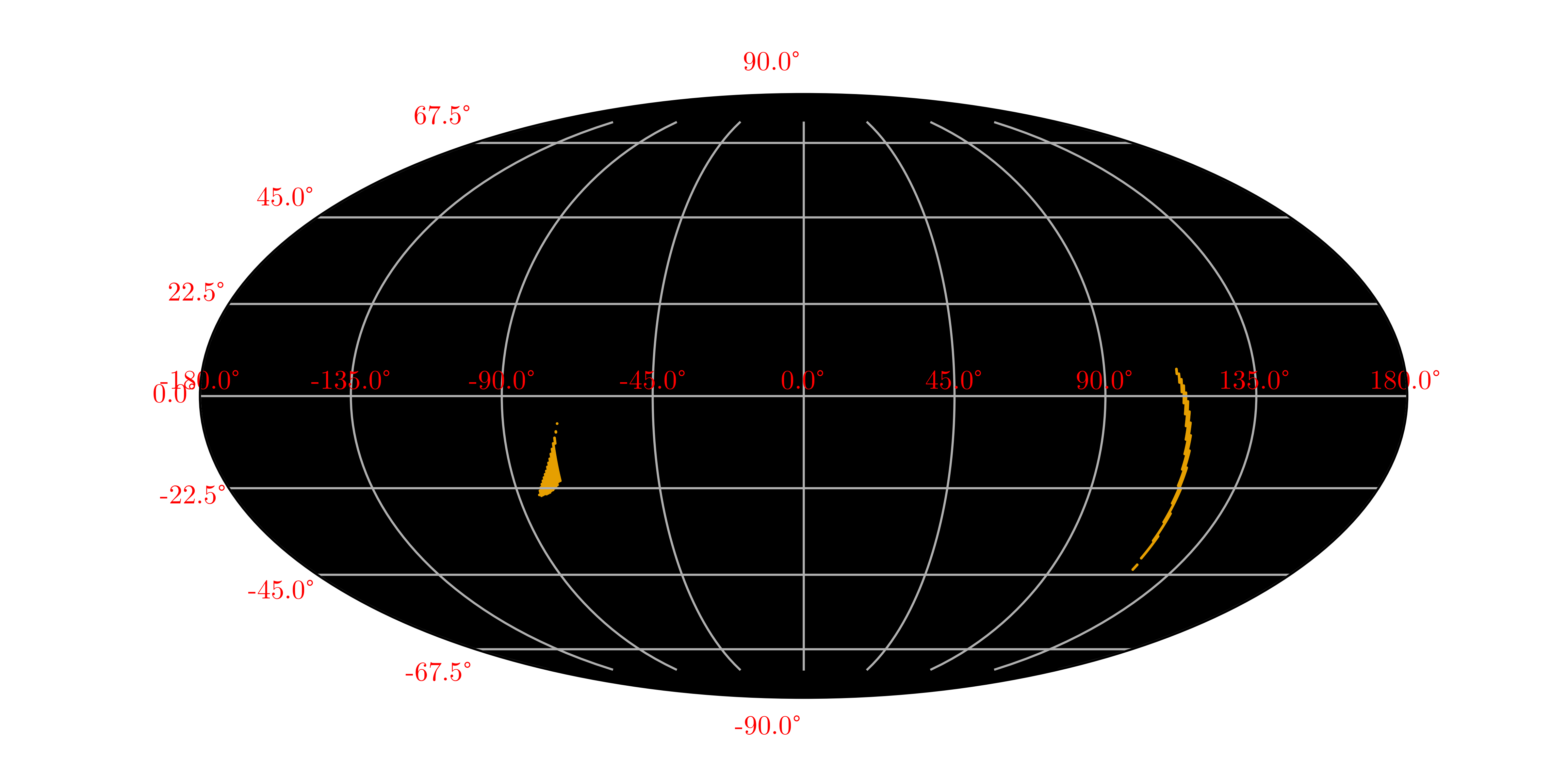}
        \caption{ }
    \end{subfigure}
    \begin{subfigure}[b]{0.49\linewidth}
        \centering
        \includegraphics[width=\linewidth]{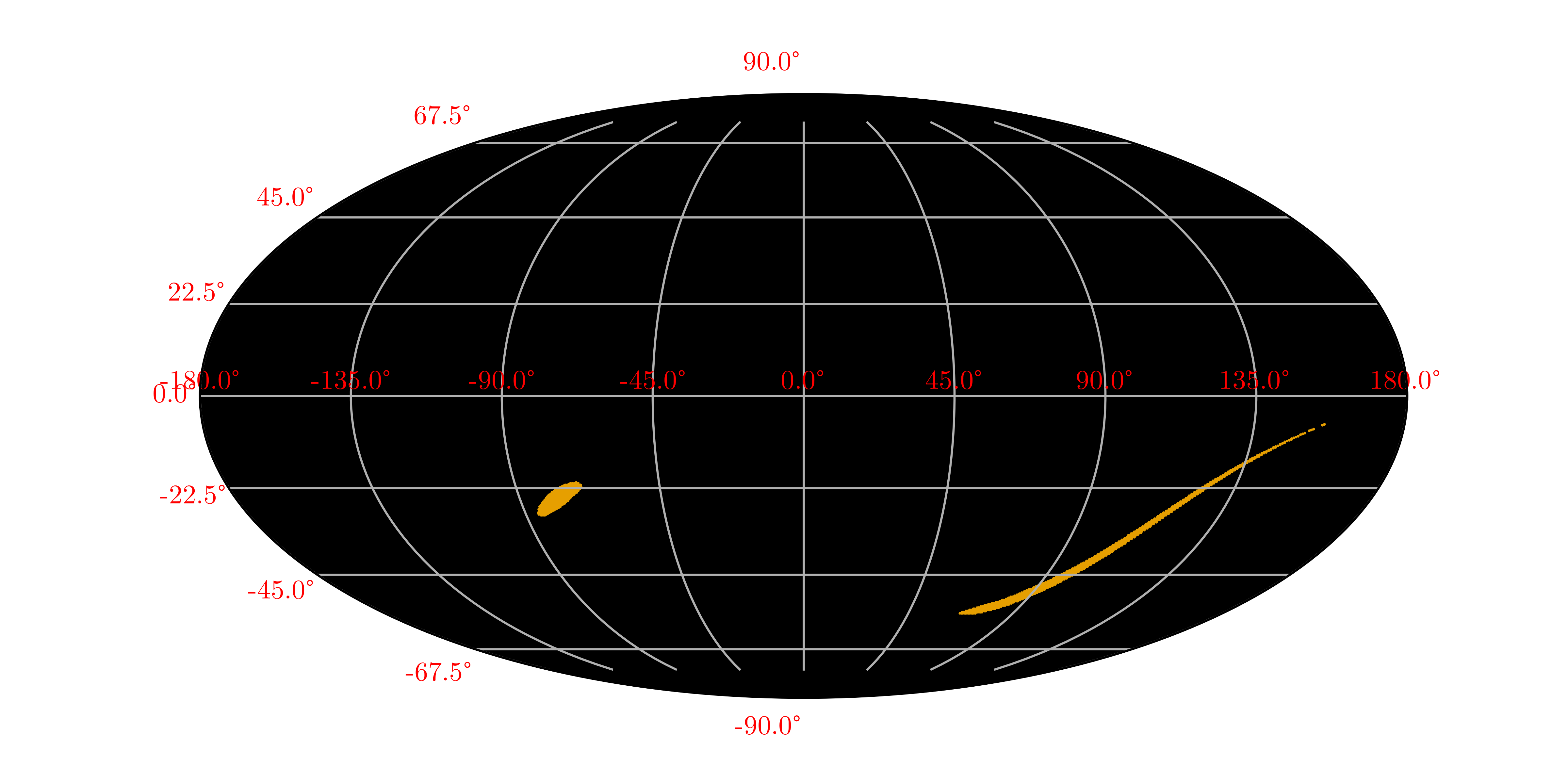}
        \caption{ }
    \end{subfigure}
    \begin{subfigure}[b]{0.49\linewidth}
        \centering
        \includegraphics[width=\linewidth]{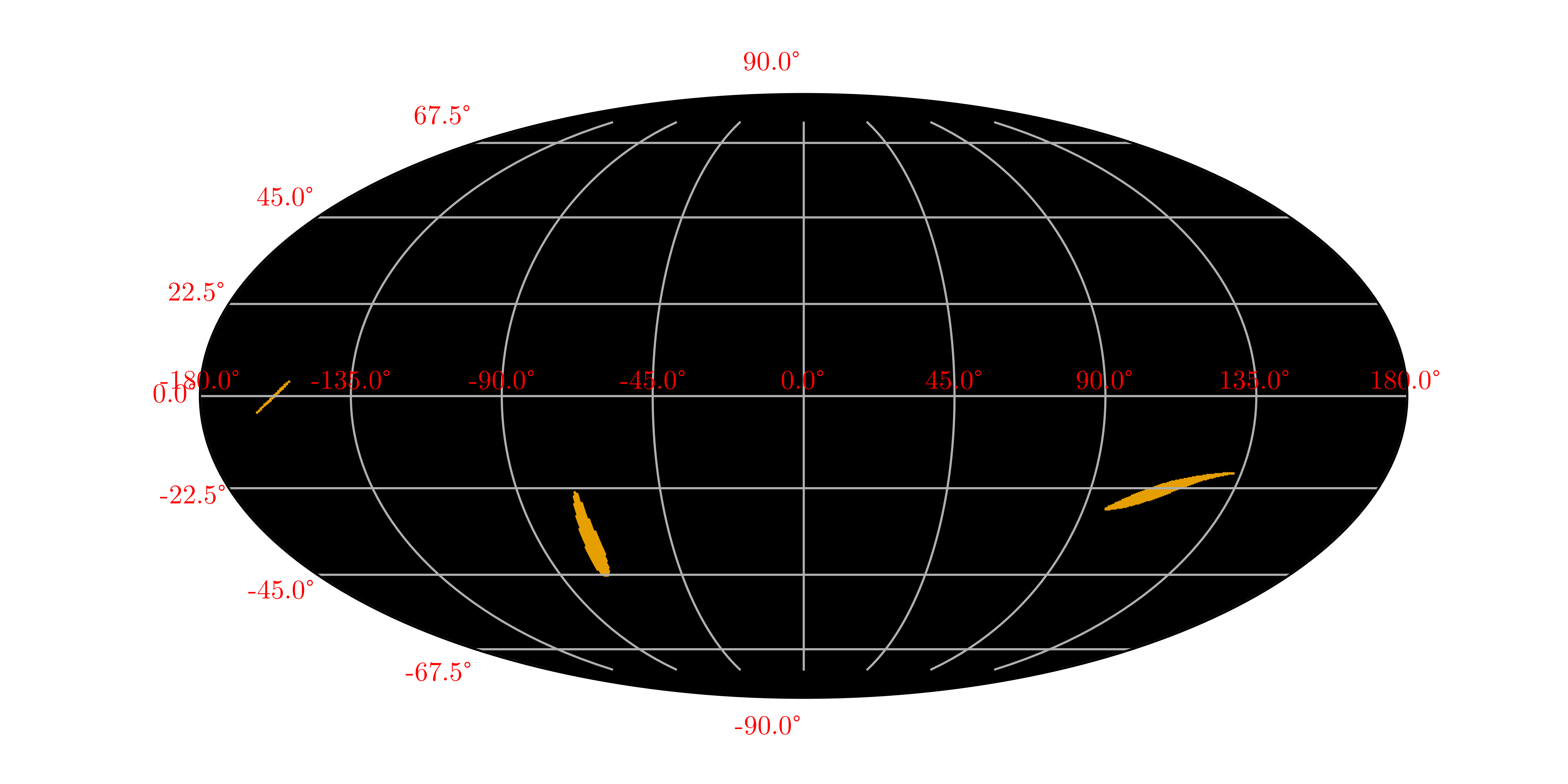}
        \caption{ }
    \end{subfigure}
    \begin{subfigure}[b]{0.49\linewidth}
        \centering
        \includegraphics[width=\linewidth]{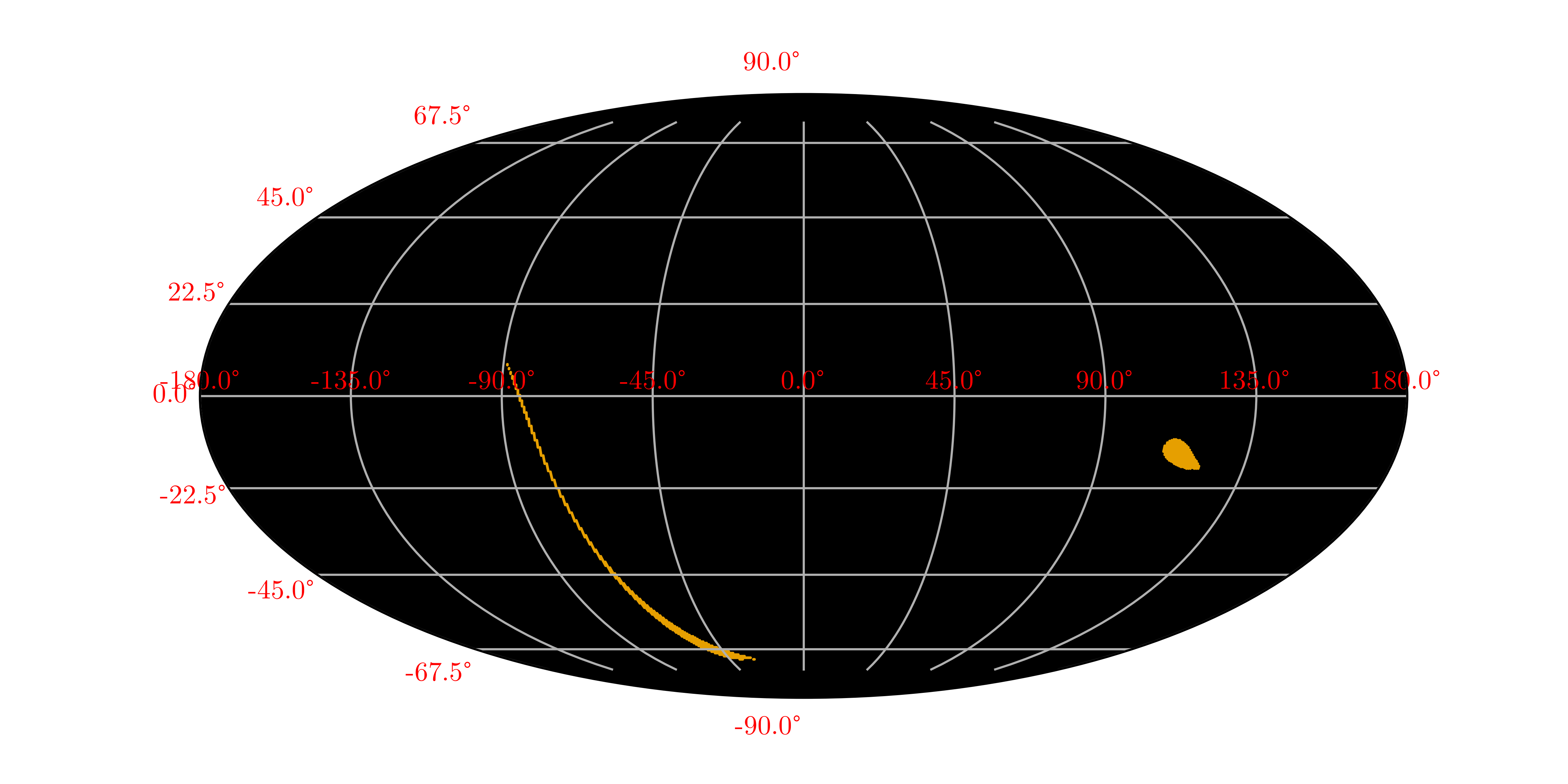}
        \caption{ }
    \end{subfigure}
    \begin{subfigure}[b]{0.49\linewidth}
        \centering
        \includegraphics[width=\linewidth]{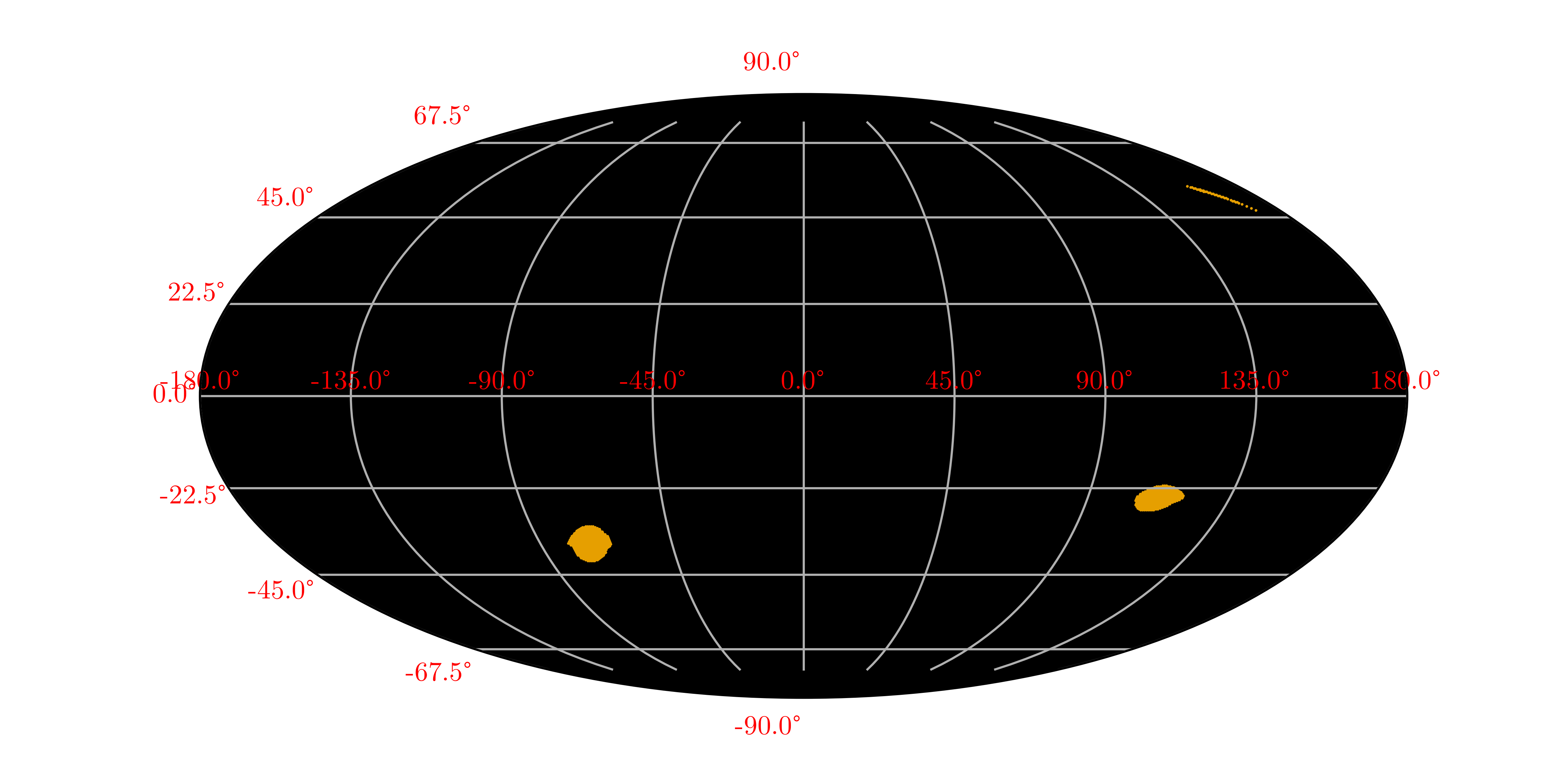}
        \caption{ }
    \end{subfigure}
    \begin{subfigure}[b]{0.49\linewidth}
        \centering
        \includegraphics[width=\linewidth]{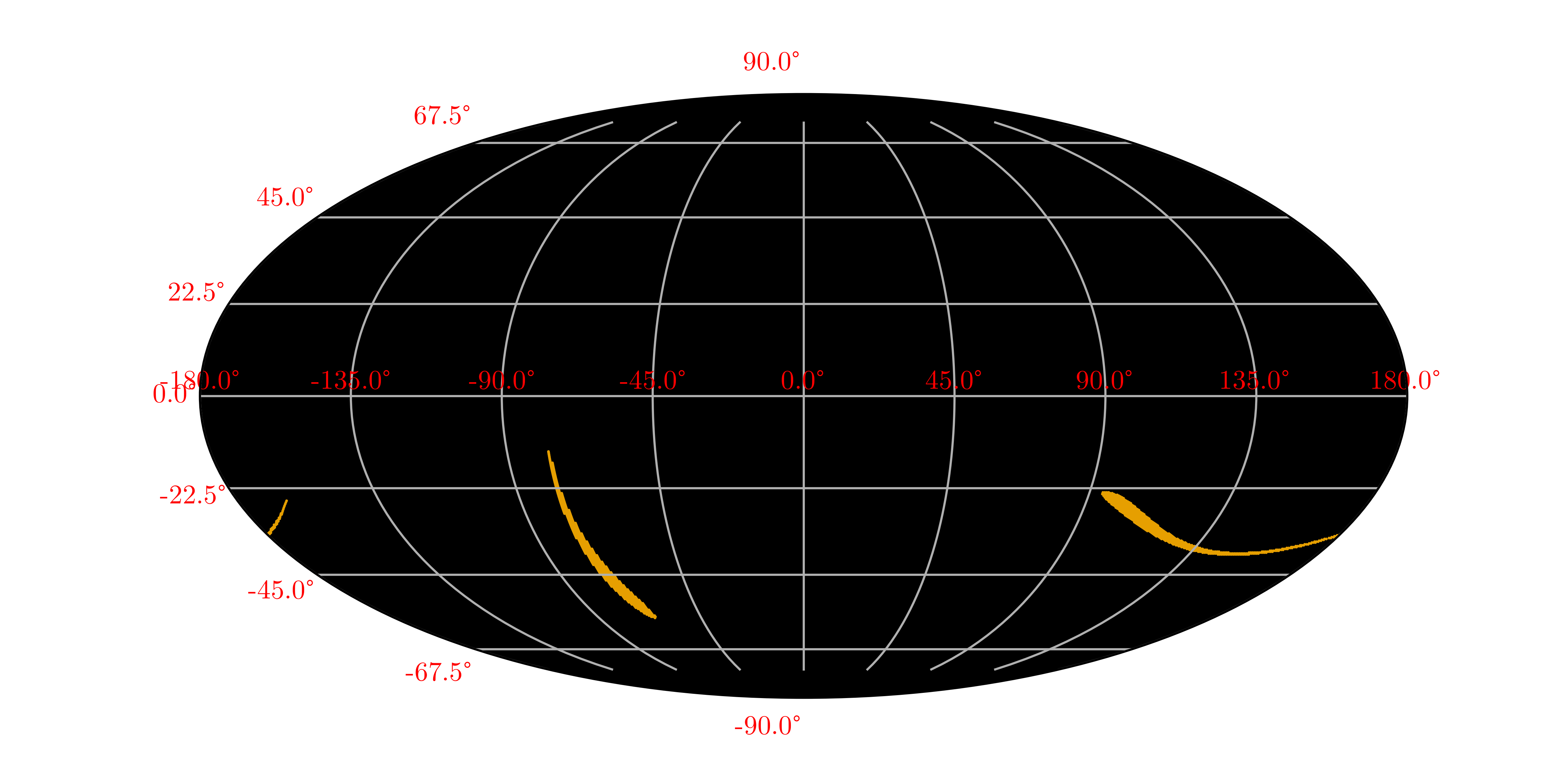}
        \caption{}
    \end{subfigure}
    \caption{Hotspots for the \UO truncation on a Mollweide projection. (a) Ensemble of $10^5$ weighted hotspots, generated using the same samples as shown in Figure~\ref{fig:corner_l123}, with the colorbar indicating the transition from low density (black) to high density (yellow-white) regions. (b) to (h) illustrate the wide variety of hotspot configurations (orange) where (b) shows overlap of non-weighted hotspots for $500$ randomly selected samples, and the rest show individual hotspots for six random cases.}
    \label{fig:l123HotspotsAll}
\end{figure*}

Figure~\ref{fig:l12HotspotsAll} presents the hotspot configurations for the \UQ truncation. The ensemble constructed from the same posterior samples shown in Figure~\ref{fig:corner_l12} is displayed in Figure~\ref{fig:l12HotspotsEnsemble}, while the remaining panels show three representative individual realizations, selected to illustrate the diversity of hotspot morphologies. The hotspots are generated using the physical model for each parameter set on a surface grid with a resolution of $600 \times 600$. For the ensemble map, each surface element is weighted by the frequency with which it appears as part of a hotspot across the posterior samples. 

Similarly, for the \UO truncation, the hotspot ensemble is shown in Figure~\ref{fig:l123HotspotsEnsemble}. In addition, Figure~\ref{fig:l123HotspotsScatter} displays the unweighted overlap of 500 randomly selected samples, along with several individual realizations in the remaining panels of Figure~\ref{fig:l123HotspotsAll}. These figures illustrate the greater geometrical and spatial diversity of hotspot configurations in the \UO case compared to \UQ. Overall, the hotspots are preferentially confined to one rotational hemisphere, consistent with previous results reported for J$0030$ in the literature.

\subsection{Best-fit solutions}
\newcommand{\bestfitparamsColumns}{$\chi_D$ $[0, \pi]$      & 1.90 & 1.56 \\
$\phi_D$ $[0, 2\pi]$      & 4.27 & 4.10 \\
$\chi_{Q1}$ $[0, \pi]$   & 1.65 & 1.87 \\
$\chi_{Q2}$ $[0, 2\pi]$   & 3.37 & 3.40 \\
$\phi_Q$ $[0, 2\pi]$      & 4.45 & 4.48 \\
$B_Q/B_D$ $[0.5, 15]$     & 4.96 & 6.23 \\
$\chi_{O1}$ $[0, \pi]$   & --   & 1.38 \\
$\chi_{O2}$ $[0, \pi]$   & --   & 2.31 \\
$\chi_{O3}$ $[0, 2\pi]$   & --   & 4.79 \\
$\phi_O$ $[0, 2\pi]$      & --   & 3.45 \\
$B_O/B_D$ $[0.5, 15]$    & --   & 5.18 \\
\hline
$\chi_r^2$    & 0.61 & 0.59 \\
}
\setlength{\tabcolsep}{0.04\textwidth}
\begin{deluxetable}{lcc}
\tablecaption{Best-fit parameters for the \UQ and \UO truncation models, and the corresponding $\chi_r^2$ values are listed. All the angular parameters are expressed in radians. \label{tab:Prms_SolutionsTable}}
\tablehead{
\colhead{Parameter [Range]} & \colhead{\UQ} & \colhead{\UO}
}
\startdata
\bestfitparamsColumns
\enddata
\end{deluxetable}
\begin{figure}[htbp]
    \centering
    \includegraphics[width=0.98\linewidth]{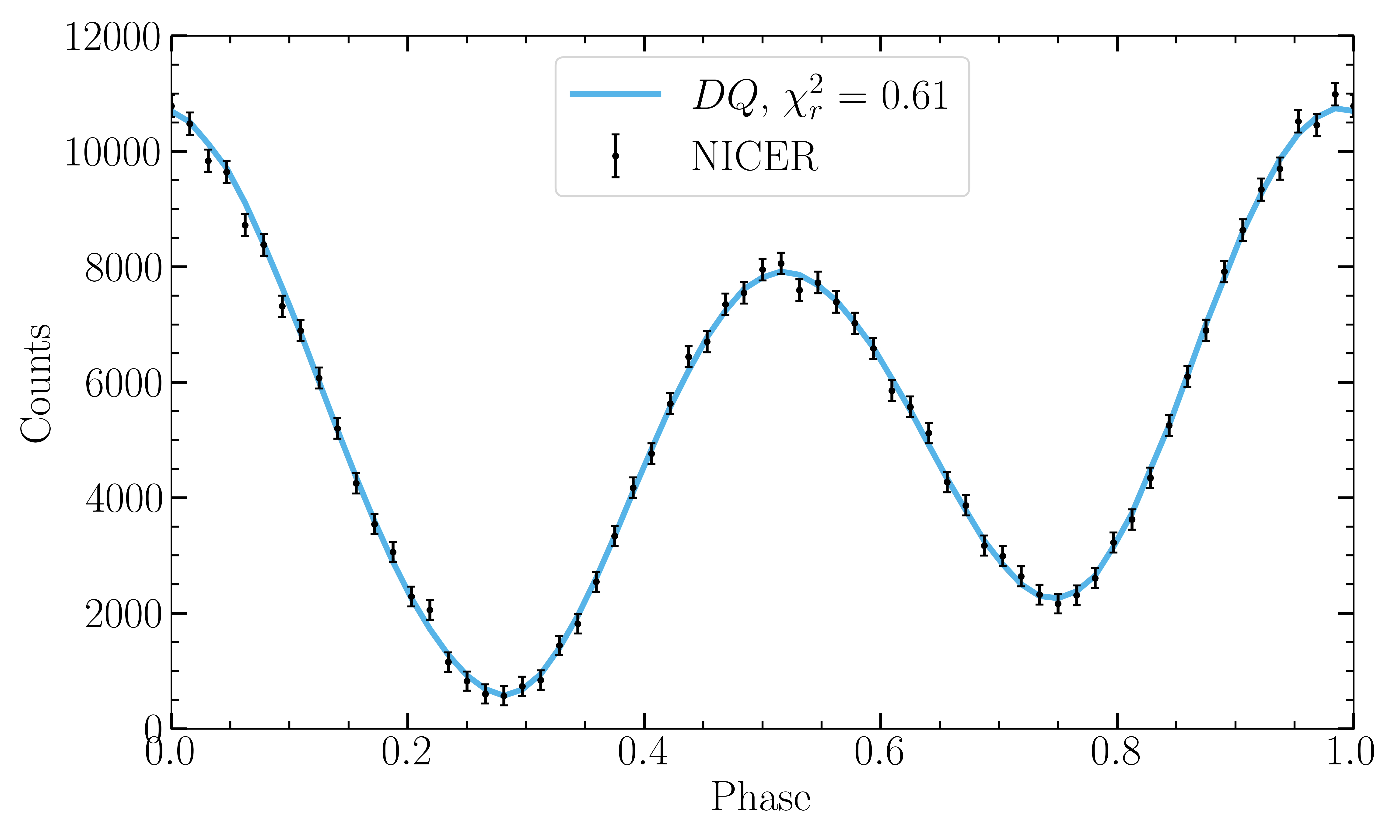}
    \includegraphics[width=0.98\linewidth]{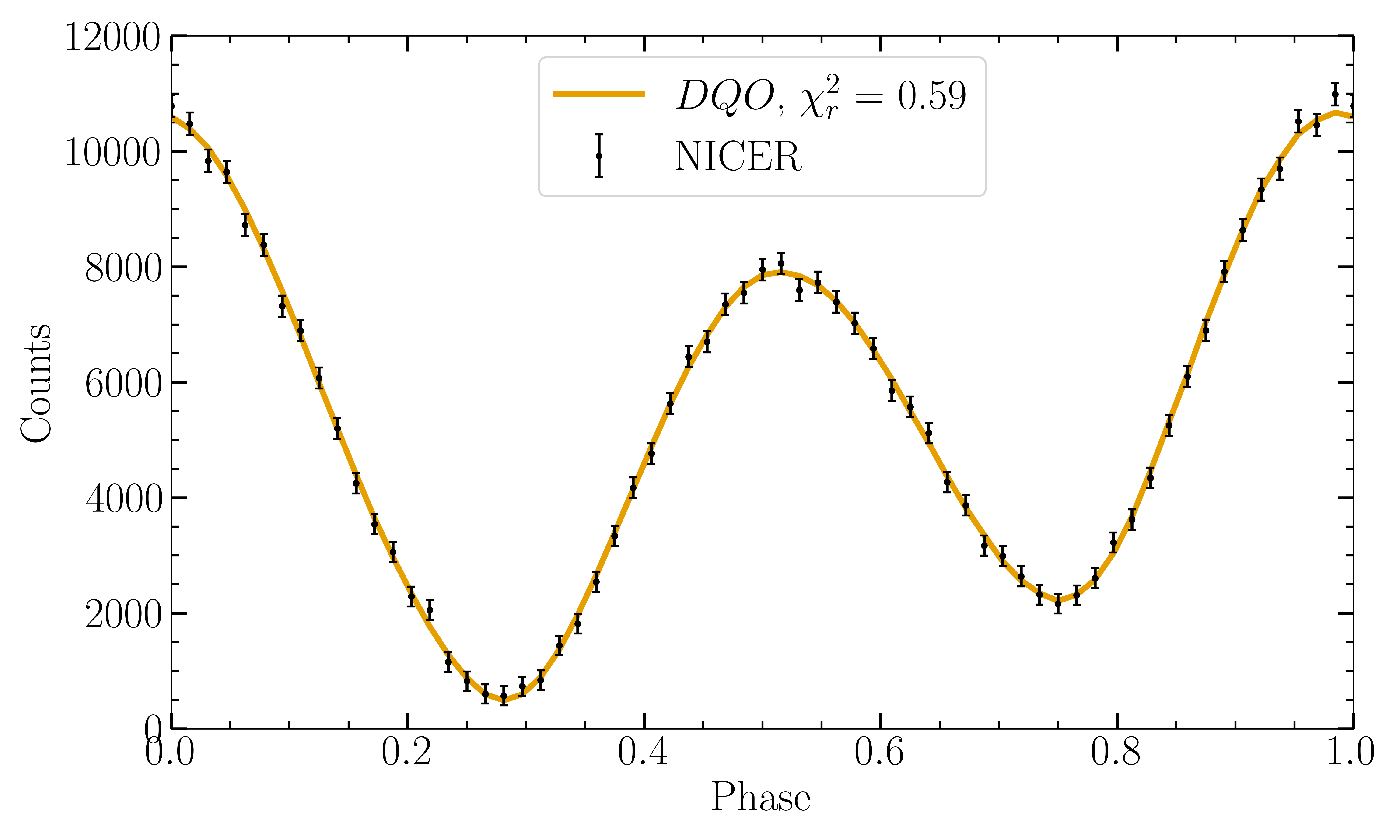}
    \caption{Best-fit light curve solutions for the \UQ (top, sky blue) and \UO (bottom, orange) configurations (see Table~\ref{tab:Prms_SolutionsTable}). NICER data are shown in black and the legends indicate $\chi^2_{r}$ for each case.}
    \label{fig:BestFitLightCurves}
\end{figure}

The best-fit solutions for both truncations are listed in Table~\ref{tab:Prms_SolutionsTable}, and the corresponding light curves are shown in Figure~\ref{fig:BestFitLightCurves}. The top and bottom panels display the best-fit light curves for \UQ and \UO, respectively. 
The NICER X-ray data are shown in black. The legend reports the reduced chi-squared statistic, defined as $\chi_r^2 = \chi^2/\mathrm{dof}$, where $\mathrm{dof} = n - k$ denotes the degrees of freedom. Here, $n = 64$ is the number of phase bins, and $k$ is the dimensionality of the parameter space (6 for \UQ and 11 for \UO). The resulting values are $0.61$ for \UQ and $0.59$ for \UO. Although reduced chi-squared values near unity are expected for typical good fits, the maximum-likelihood solution corresponds to the minimum $\chi^2$ over the sampled parameter space and can therefore naturally yield a value below unity. In this case, the best-fit $\chi_r^2$ values are sufficiently below unity that they should not be interpreted too literally as a definitive goodness-of-fit measure.

\begin{figure*}[htbp]
    \centering
    \begin{interactive}{js}{field_lines_best_l12_INTERACTIVE.zip}
    \begin{subfigure}[b]{0.45\linewidth}
        \includegraphics[width=0.90\linewidth]{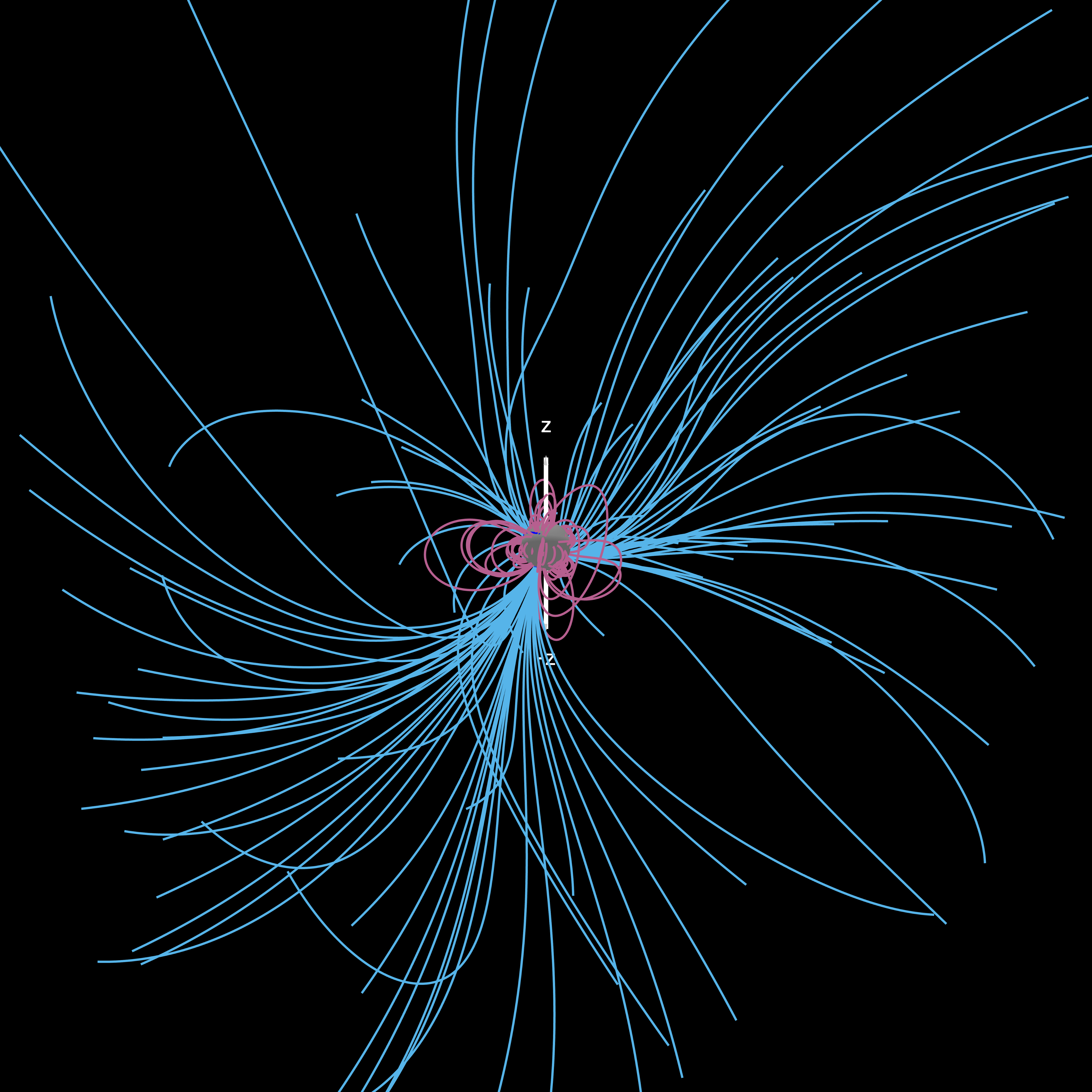}
        \caption{}
    \end{subfigure}
    \begin{subfigure}[b]{0.45\linewidth}       
        \includegraphics[width=0.90\linewidth]{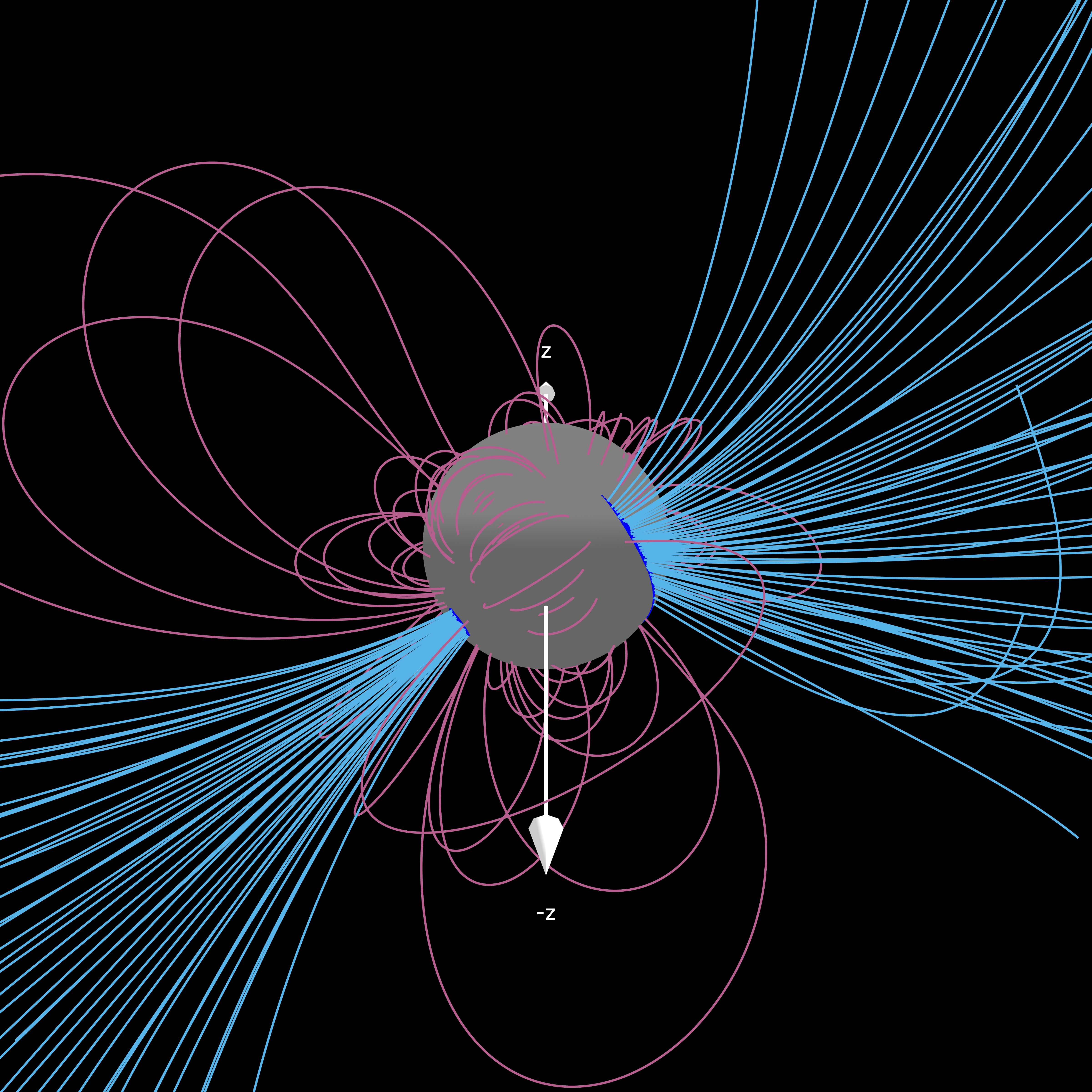}
        \caption{}
    \end{subfigure}
    \end{interactive}
    \caption{Magnetic field structure for the best-fit \UQ configuration (Table~\ref{tab:Prms_SolutionsTable}). (a) shows the full magnetic field structure, while (b) presents a zoomed-in view of the inner magnetospheric region. Open field lines are shown in sky blue, and closed field lines in magenta. The $z$-axis (rotation axis) is indicated in white. The star is shown in gray at the center, and the hotspots are presented in dark blue on the surface. For clarity, the hotspots are also shown as white contours in Figure~\ref{fig:BestFitSurfaceMagneticField}. This figure is available online as an interactive figure. The interactive version allows the reader to rotate, zoom, and inspect the 3D magnetic-field topology from arbitrary viewing angles. Open and closed field lines are colored as described above; hovering over a field line displays its Cartesian coordinates.}
    \label{fig:BestFitFieldLines1}
\end{figure*}

\begin{figure*}[htbp]
    \centering
    \begin{interactive}{js}{field_lines_best_l123_INTERACTIVE.zip}
    \begin{subfigure}[b]{0.45\linewidth}
        \includegraphics[width=0.90\linewidth]{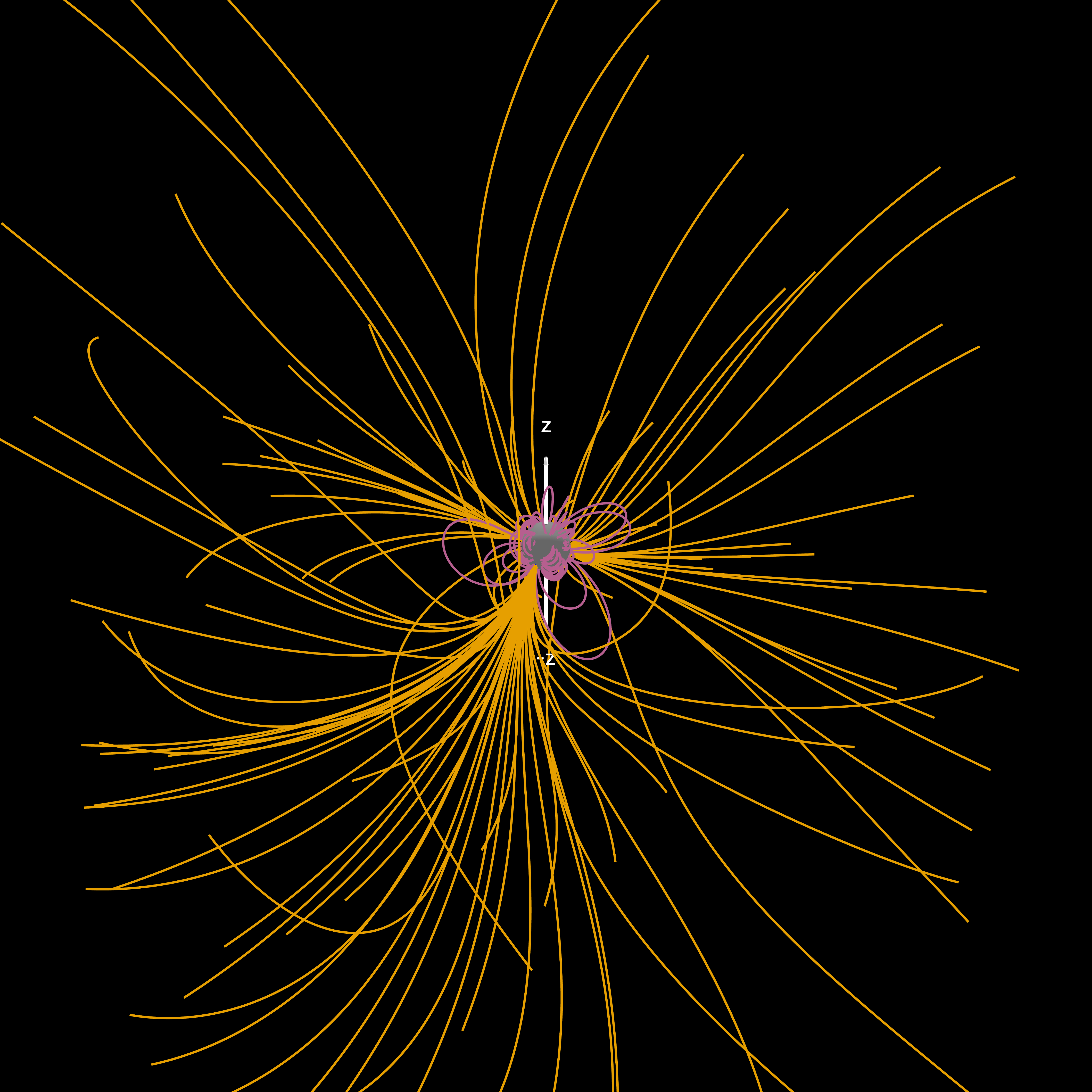}
        \caption{}
    \end{subfigure}
    \begin{subfigure}[b]{0.45\linewidth}
        \includegraphics[width=0.90\linewidth]{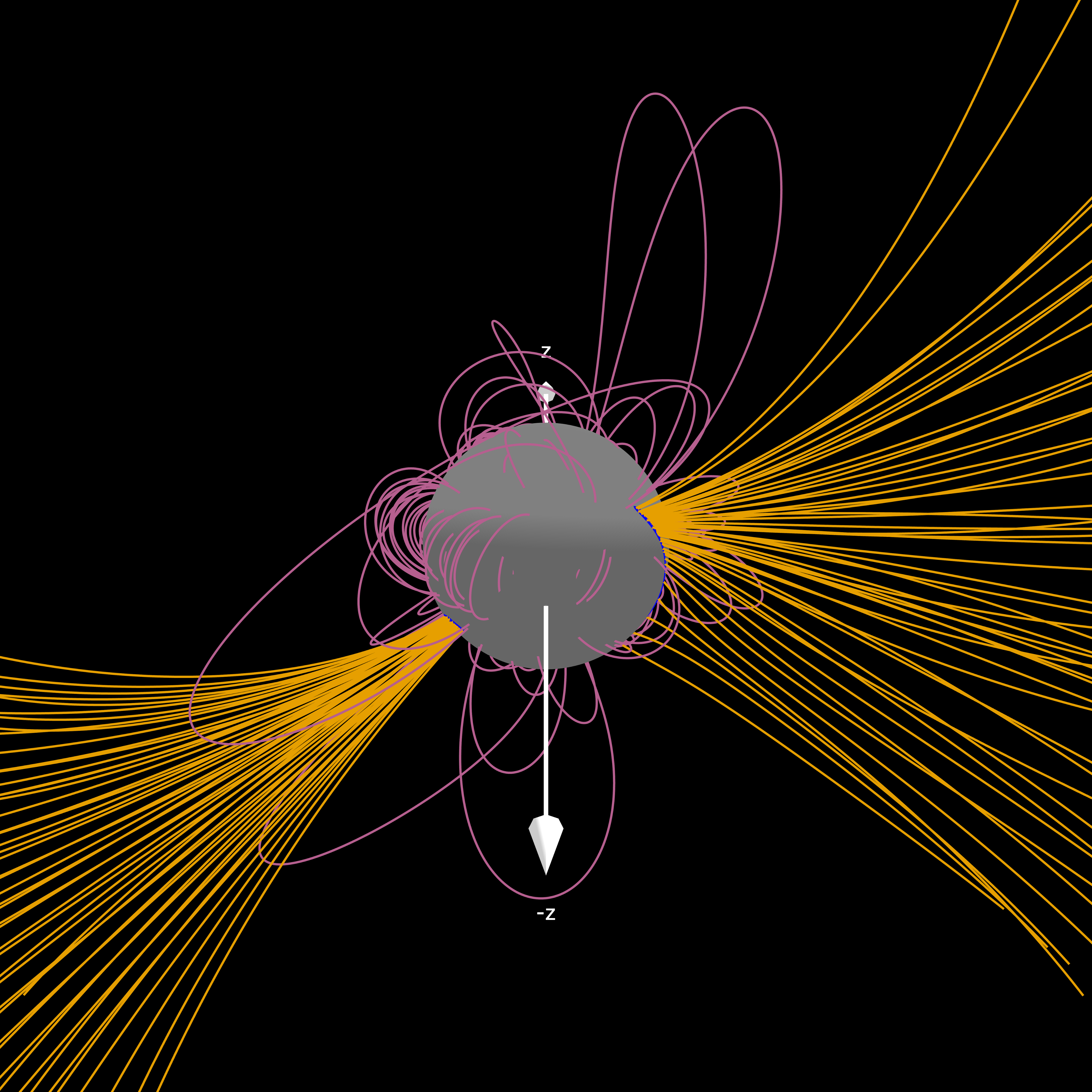}
        \caption{}
    \end{subfigure}
    \end{interactive}
    \caption{Magnetic field structure for the best-fit \UO configuration (Table~\ref{tab:Prms_SolutionsTable}). (a) shows the full magnetic field structure, while (b) presents a zoomed-in view of the inner magnetospheric region. Open field lines are shown in orange, and closed field lines in magenta. The $z$-axis (rotation axis) is indicated in white. The star is shown in gray at the center, and the hotspots are presented in dark blue on the surface. For clarity, the hotspots are also shown as white contours in Figure~\ref{fig:BestFitSurfaceMagneticField}. This figure is available online as an interactive figure. The interactive version allows the reader to rotate, zoom, and inspect the 3D magnetic-field topology from arbitrary viewing angles. Open and closed field lines are colored as described above; hovering over a field line displays its Cartesian coordinates.}
    \label{fig:BestFitFieldLines2}
\end{figure*}

The magnetic field structures corresponding to the best-fit solutions are shown in Figures~\ref{fig:BestFitFieldLines1} and~\ref{fig:BestFitFieldLines2} for \UQ and \UO truncations, respectively. In each figure, the right panel provides a zoomed-in view of the left panel. Open field lines are shown in sky blue or orange, and their footpoints coincide with the surface hotspots (indicated in dark blue). Both configurations exhibit an oval-shaped hotspot together with a more extended emission region. Closed field lines are shown in magenta. At larger distances from the star, several field lines display a twisted and swept-back morphology, whereas the near-surface structure is dominated by multipolar loops.
\begin{figure}[htbp]
    \centering
    \includegraphics[width=0.9\linewidth]{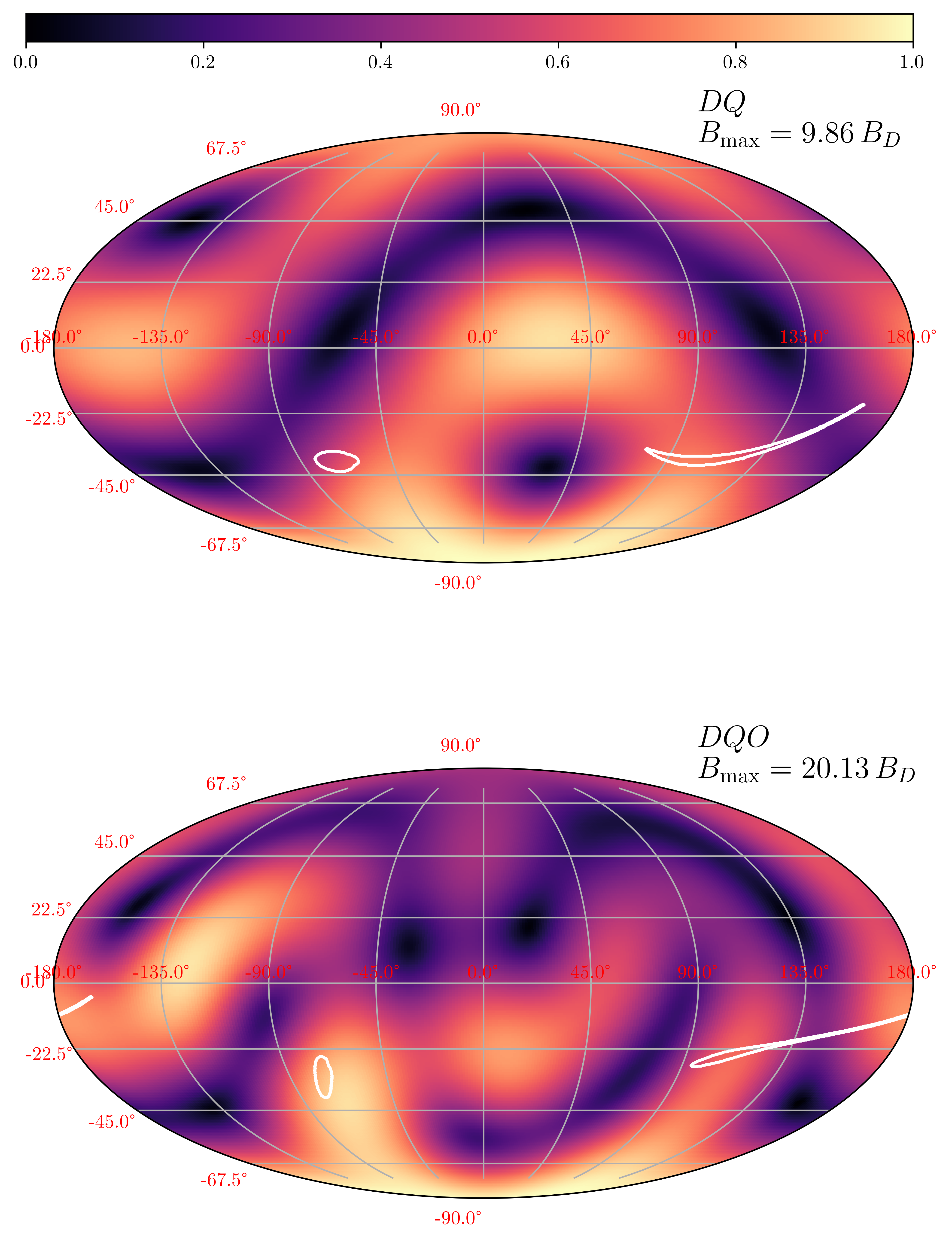}
    \caption{Total surface magnetic field for the best-fit solutions, \UQ on top and \UO at the bottom, on a Mollweide projection. The color bar represents normalized value, with $1$ being the maximum value on the surface which is given on top of every sub plot as $B_{\max}$ (in terms of the dipolar field $B_{D} = 1$). The contours of the corresponding hotspots are overlaid on top in white.}    \label{fig:BestFitSurfaceMagneticField}
\end{figure}

We also compute the total surface magnetic field strength as $B^{S}_{\rm tot} = \sqrt{\bsym{B}_{r}^2 + \bsym{B}_{\theta}^2 + \bsym{B}_{\phi}^2}$. The maximum magnitude of the individual dipolar, quadrupolar, and octupolar surface fields vary from $1$ to $2$, $\approx3.16$, and $\approx4.3$, respectively. The surface magnetic-field distributions for the best-fit solutions are shown in Figure~\ref{fig:BestFitSurfaceMagneticField}, with \UQ and \UO displayed in the top and bottom panels, respectively. The maps are computed on a $120 \times 60$ grid (uniform in $\phi$ and $\theta$) and rendered using cubic interpolation. Each panel is normalized to its corresponding maximum surface value, $B_{\max}$, indicated in the upper-right corner in units of the dipolar field strength $B_{D}$. For both truncations, $B_{\max}$ is $10 - 20$ times higher than the dipolar field-strength scale  alone. The contours of the associated hotspots (as shown in Figures~\ref{fig:BestFitFieldLines1} and~\ref{fig:BestFitFieldLines2}) are overlaid in white. In both cases, the surface-field distribution is strongly anisotropic, reflecting the asymmetry introduced by the higher-order multipolar components.

\section{Discussion and Conclusions} \label{sec:discussionsAndConclusions}
The broader aim of this line of research is to develop a physically grounded, self-consistent framework for constraining the fundamental parameters of NICER MSPs – and NSs in general – including their magnetic field configuration and \MR. In the longer term, such studies may contribute to tighter constraints on the nature of dense matter and to a better understanding of the magnetic field structures of MSPs. The implications of the present results, together with prospects for future work, are discussed below. 

\subsection{Complexity}
A key result is the establishment of a practical framework for comparing the mathematical complexity of different magnetic field configurations. Applying this framework to the static offset dipole-plus-quadrupole solutions of \citetalias{kalapotharakos_multipolar_2021}, we find that reproducing these fields with the centered SVM2F expansion generally requires multipolar content beyond the octupole order ($l>3$), and in some cases far beyond it. For nearly all of their best-fit vacuum solutions for J0030, the truncated expansion must extend to higher orders to achieve a comparable representation according to the $\delta {\cal C}$ metric; in the most extreme case (e.g., $RV5$, with a quadrupole-only offset), terms beyond $l=20$ are required.

\begin{figure}[htbp]
    \centering
    \includegraphics[width=0.99\linewidth]{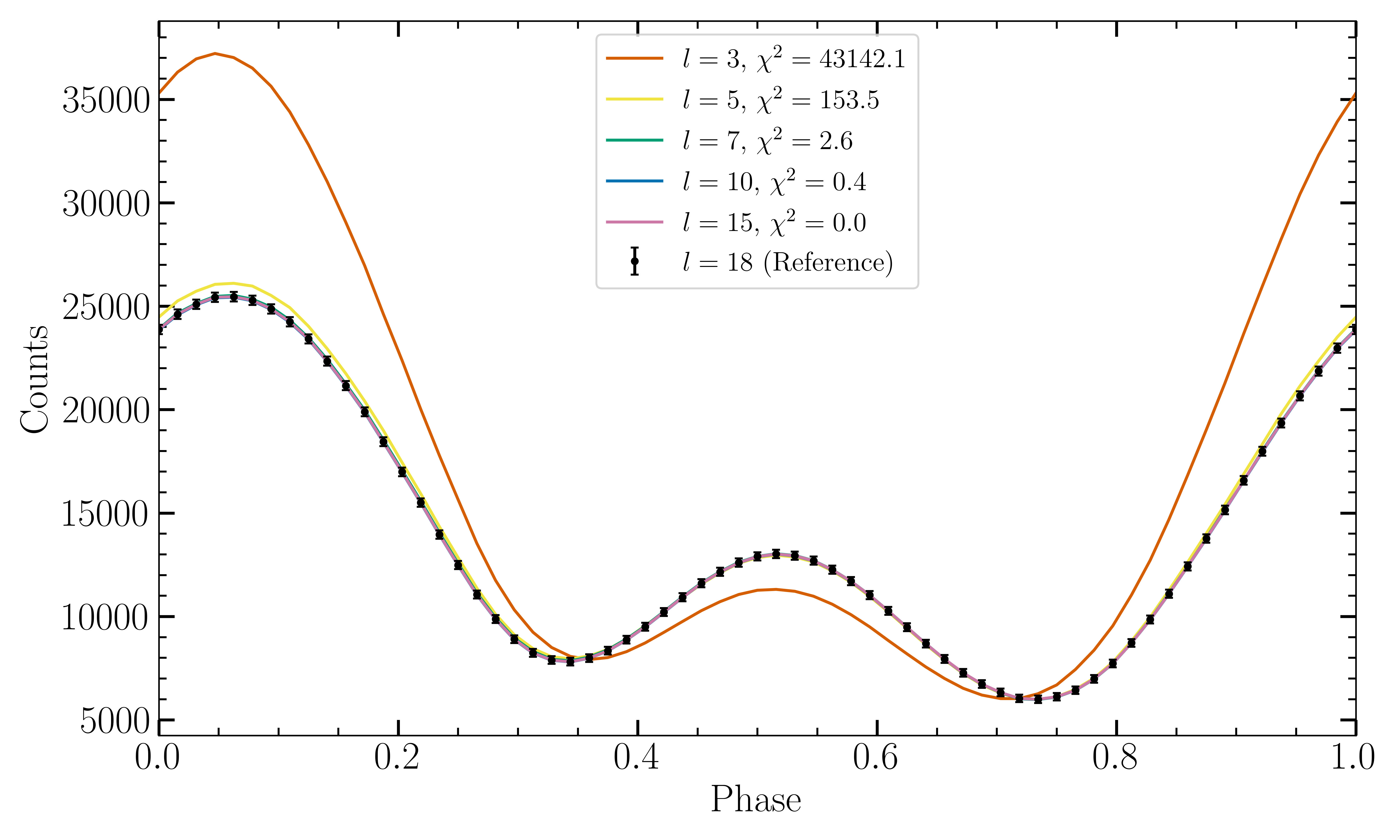}
    \includegraphics[width=0.99\linewidth]{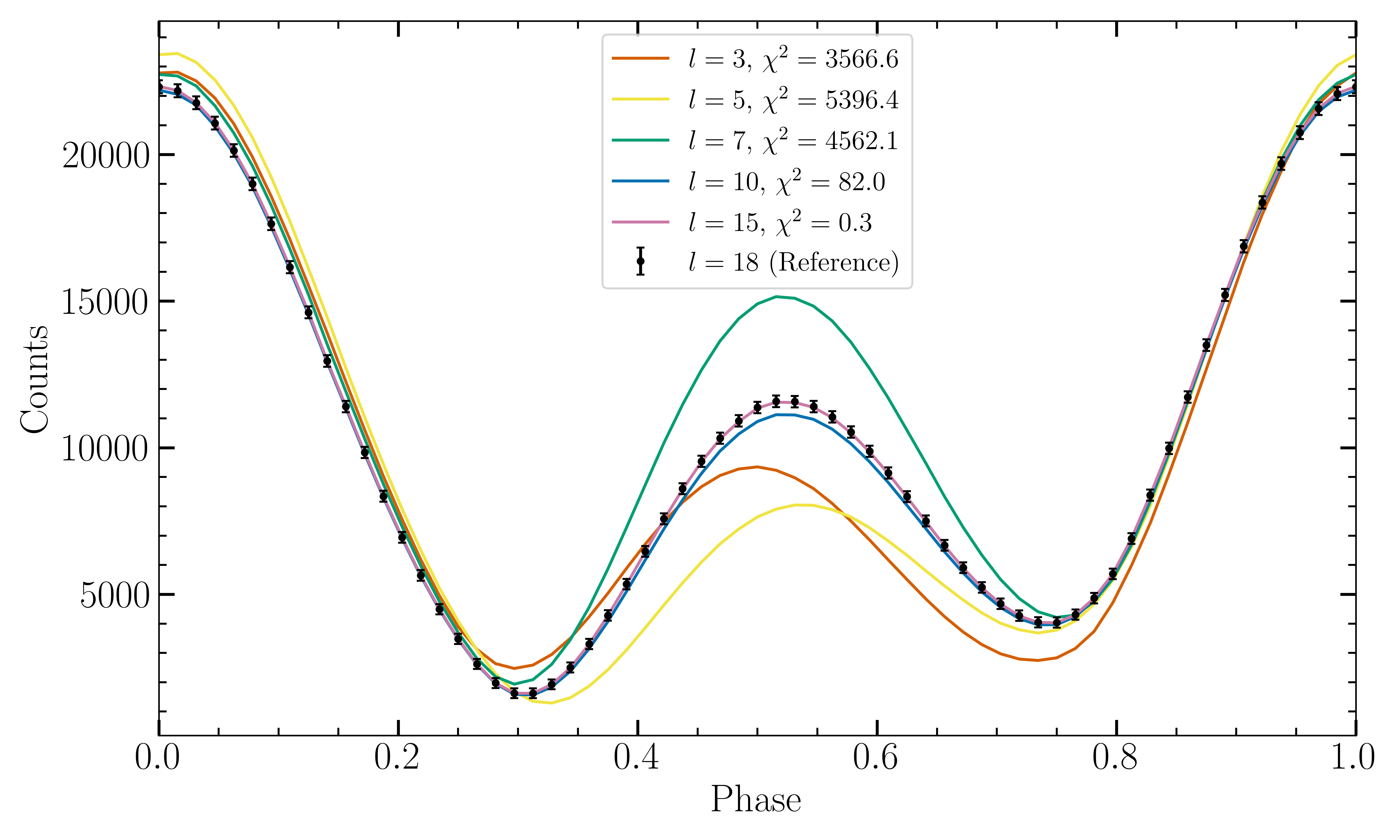}
    \includegraphics[width=0.99\linewidth]{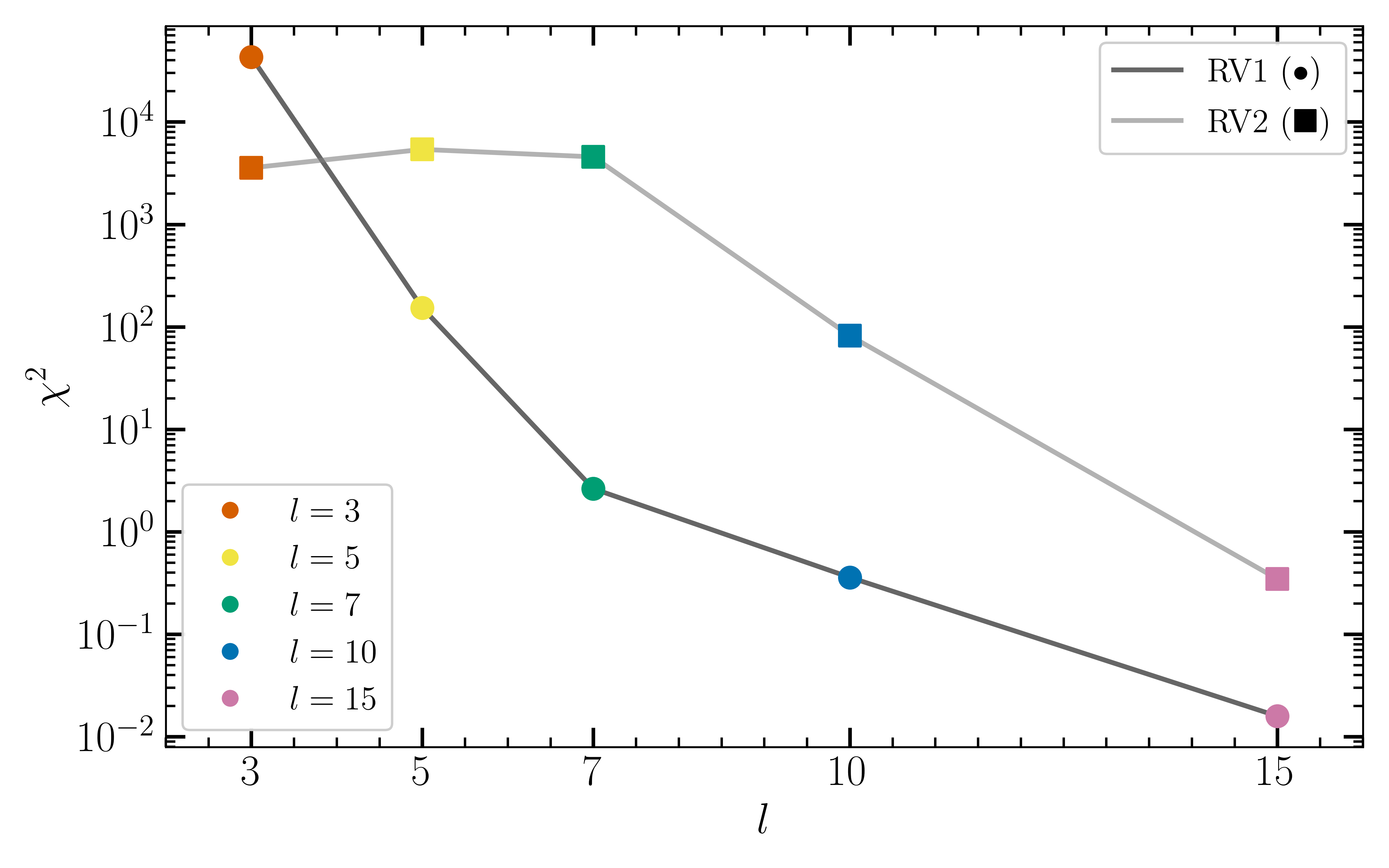}
    \caption{\textit{Top and middle panels:} Model light curves for the $RV1$ and $RV2$ input field configurations, respectively, at truncation orders $l \in \{3, 5, 7, 10, 15\}$ (colored lines), overlaid on the $l = 18$ reference (black points with NICER-like uncertainties). The $\chi^2$ values with respect to the reference light curve are indicated in the legend.\footnote{We note that these light curves would recover the exact $RV1$ and $RV2$ light curves of \citetalias{kalapotharakos_multipolar_2021} if the swept-back $\bsym{B}_{\rm out}$ were replaced by the corresponding static vacuum configuration.} \textit{Bottom panel:} $\chi^2$ as a function of multipolar truncation order $l$ for the above cases: $RV1$ (circles, black connecting line) and $RV2$ (squares, gray connecting line). Colors indicate the truncation order $l$, as shown in the legend.}
    \label{fig:chisq_vs_lmax_convergence}
\end{figure}

To connect the field-level complexity measured by $\delta{\cal C}$ with the actual observable used in pulse-profile modeling, we also examine the convergence of the corresponding light curves. This distinction is important because agreement between magnetic fields provides a useful and controlled diagnostic, but in this type of inference the practical truncation order is ultimately determined by whether additional multipolar terms produce observable changes in the modeled light curve. We therefore compute model light curves for the ($RV1$) and ($RV2$) input field configurations of \citetalias{kalapotharakos_multipolar_2021} at increasing multipolar truncation orders. As established in Section~\ref{subsubsec:coeff_SVF}, the field convergence threshold $\delta {\cal C} = 10^{-3}$ is reached at $l_{\max} = 9$ and $18$ for $RV1$ and $RV2$, respectively. We adopt $l = 18$ as the reference and assign NICER-like uncertainties (Section~\ref{sec:methodology}). The resulting light curves are shown in the top and middle panels  of Figure~\ref{fig:chisq_vs_lmax_convergence}, and the corresponding $\chi^2$ as a function of $l$ is shown in the bottom panel. For $RV1$, the light curve achieves steady convergence with increasing $l$, reaching $\chi^2 \lesssim 1$ by $l = 10$ and $\chi^2 \sim 0$ at the reference. In contrast, $RV2$ exhibits non-monotonic behavior at smaller truncation orders (discussed in Section~\ref{subsubsec:coeff_SVF}), with $\chi^2$ varying substantially across $l \in \{3, 5, 7\}$ before improving at $l = 10$ and reaching $\chi^2 \sim 0.3$ at $l = 15$. We note that the truncation orders required for convergence in the centered multipole representation are larger than those that would be needed in an offset description of the same field, since offset configurations are expected to concentrate power in fewer terms, suggesting that the centered expansion requires higher $l$ to achieve equivalent accuracy (as demonstrated at the field level in Section~\ref{subsubsec:coeff_SVF}). These results demonstrate that field and observable convergence are coupled: simple configurations like $RV1$ exhibit steady convergence with increasing truncation order, while complex field structures like $RV2$ exhibit non-monotonic patterns reflecting their underlying multipolar complexity. Both cases ultimately converge well within the observational uncertainties, indicating that once the model light curves agree with the reference to within the data uncertainties, the truncation order is sufficient for a physically meaningful description. The $\chi^2$ convergence with respect to observational uncertainties provides a physically motivated, observation-driven criterion for selecting the appropriate truncation order. Given NICER's exceptionally small uncertainties, this approach becomes a particularly powerful diagnostic, as the high data quality demands accurate reproduction of light curves and therefore naturally constrains the number of multipolar field components required.

In discussing ``complexity," however, we go beyond simply counting terms. $\delta {\cal C}$ and $\chi^2$ convergence of the light curve provide complementary operational measures of expansion compactness, but compactness alone is only one dimension of model evaluation. Additional criteria include: (i) parameter interpretability, where offset models have parameters that are geometrically intuitive (offset distance and direction), whereas centered multipoles follow a mathematically clean and controlled hierarchy; (ii) robustness and stability, since offset prescriptions may be sensitive to small parameter changes, while centered multipoles may vary more smoothly with data; (iii) predictive generalizability, as a model that is simpler under Occam’s razor should extrapolate better across different datasets or wavebands; and (iv) physical plausibility, since offsets might be justified by realistic stellar physics (e.g., crustal anomalies) or might simply be introduced ad hoc.

Our framework provides a consistent way to compare models across these facets. The key point is that expansion complexity does not equate to physical complexity. An offset model may appear mathematically complex in the VSH basis yet be physically motivated, while a centered expansion may appear compact but lack direct physical interpretability. In practice, for J0030, we find that the SVM2F framework can reproduce the required hotspot asymmetries with modest multipolar content, making them both implementable and stable for inference. However, the physical sustainability of any surface field configuration depends on the interior magnetic field structure, which remains unconstrained; it is unclear which field geometries, particularly those involving large offsets or higher multipolar components, may remain stable on $\gtrsim$Myr timescales. This uncertainty applies equally to both centered and offset approaches, and our results do not exclude the possibility that offset-based descriptions may still be warranted if future data, especially multiwavelength constraints, demand them.

\subsection{MCMC solutions}
We modeled the multipolar magnetic field structure of J0030 by fitting its bolometric thermal X-ray light curve within an MCMC framework using the SVM2F formalism. To reduce the computational cost, we employed an NN surrogate trained to generate light curves from magnetic field parameters, achieving a speed-up of $\sim1300$ times relative to the full physical model. 
For comparison, the NN trained on the static offset field of \citetalias{kalapotharakos_multipolar_2021} presented in \citetalias{olmschenk_pioneering_2025}, was $> 400$ times faster than its physical model. Because SVM2F includes higher-order multipolar components and the additional modes associated with them, it is itself slower than the static offset-field model by a factor of $\sim 3$.

Initially, we ran MCMC using a pure dipolar field and found that a centered physical dipolar model failed to reproduce the bolometric light curve, indicating that an asymmetric geometry is required. Using the NN surrogate framework, we extended the model to include quadrupolar (\UQ; 6 reduced free parameters) and octupolar (\UO; 11 reduced free parameters) components, which yielded converged posterior distributions in both cases. The light curves generated from these posterior distributions are in good agreement with the NICER observations (see Figure~\ref{fig:lightCurvesEnsemble}).

These results contrast with \cite{chen_numerical_2020}, who concluded that a centered vacuum dipole-plus-quadrupole configuration could not reproduce the NICER data without introducing a quadrupole offset. Their analysis, however, considered a more restricted centered dipole-plus-quadrupole vacuum family and did not involve a direct MCMC fit to the bolometric NICER light curve. The difference therefore likely reflects the broader and less restricted parameter-space exploration adopted here, together with the different fitting objective, since the present study directly samples the bolometric thermal X-ray posterior rather than targeting hotspot-like configurations within a more constrained vacuum setup.

Before comparing our results with \citetalias{kalapotharakos_multipolar_2021} in detail, several points about the comparison are worth noting. Among their reported solutions, $RV6$ and $RF6$ provide the closest counterparts to the present centered \UQ model in broad geometric terms, since they correspond to a centered dipole and a nearly centered quadrupole. Their static-vacuum solution $RV6$ achieves $\chi_r^2 = 0.85$ for the NICER X-ray light curve, whereas the corresponding FF solution $RF6$ yields a substantially poorer fit of $\chi_r^2 \approx 9$. Our centered \UQ model achieving $\chi_r^2 = 0.61$ is not directly comparable to either configuration, for two reasons. First, even though all three models start from the same general type of static-vacuum multipolar prescription at the stellar surface, they differ in how the field is continued into the exterior magnetosphere. The $RV6$ solution uses a static-vacuum exterior, $RF6$ uses an FF exterior that includes the effects of a plasma-filled magnetosphere, and the present \UQ model uses a swept-back vacuum exterior. Neither \citetalias{kalapotharakos_multipolar_2021} solution therefore provides a like-for-like comparison with the present model. The large difference between the $RV6$ and $RF6$ fits already indicates that changing the exterior-field prescription can significantly affect the resulting light curve. Second, the quadrupolar parameterizations used in the two studies are not equivalent. \citetalias{kalapotharakos_multipolar_2021} adopted an axisymmetric quadrupole corresponding to a pure $m=0$ in its principal-axis frame, whose orientation and offset were varied. In contrast, the present centered \UQ model includes all allowed $l=2$ azimuthal modes, $m=0,1,2$, subject to the phase constraint adopted in this work. A rotated axisymmetric quadrupole spans only a restricted subset of this broader quadrupolar family. The additional non-axisymmetric $m=1$ and $m=2$ contributions can therefore produce a wider range of hotspot geometries without requiring an offset. The good fit obtained here may therefore reflect the broader quadrupolar parameterization, the swept-back exterior field, or a combination of both. Isolating the effect of sweepback would require a controlled analysis using the same general centered \UQ parameterization and inference procedure while varying only the exterior-field prescription among static-vacuum, swept-back-vacuum, and FF configurations.

\begin{figure}[htbp]
    \centering
    \includegraphics[width=0.98\linewidth]{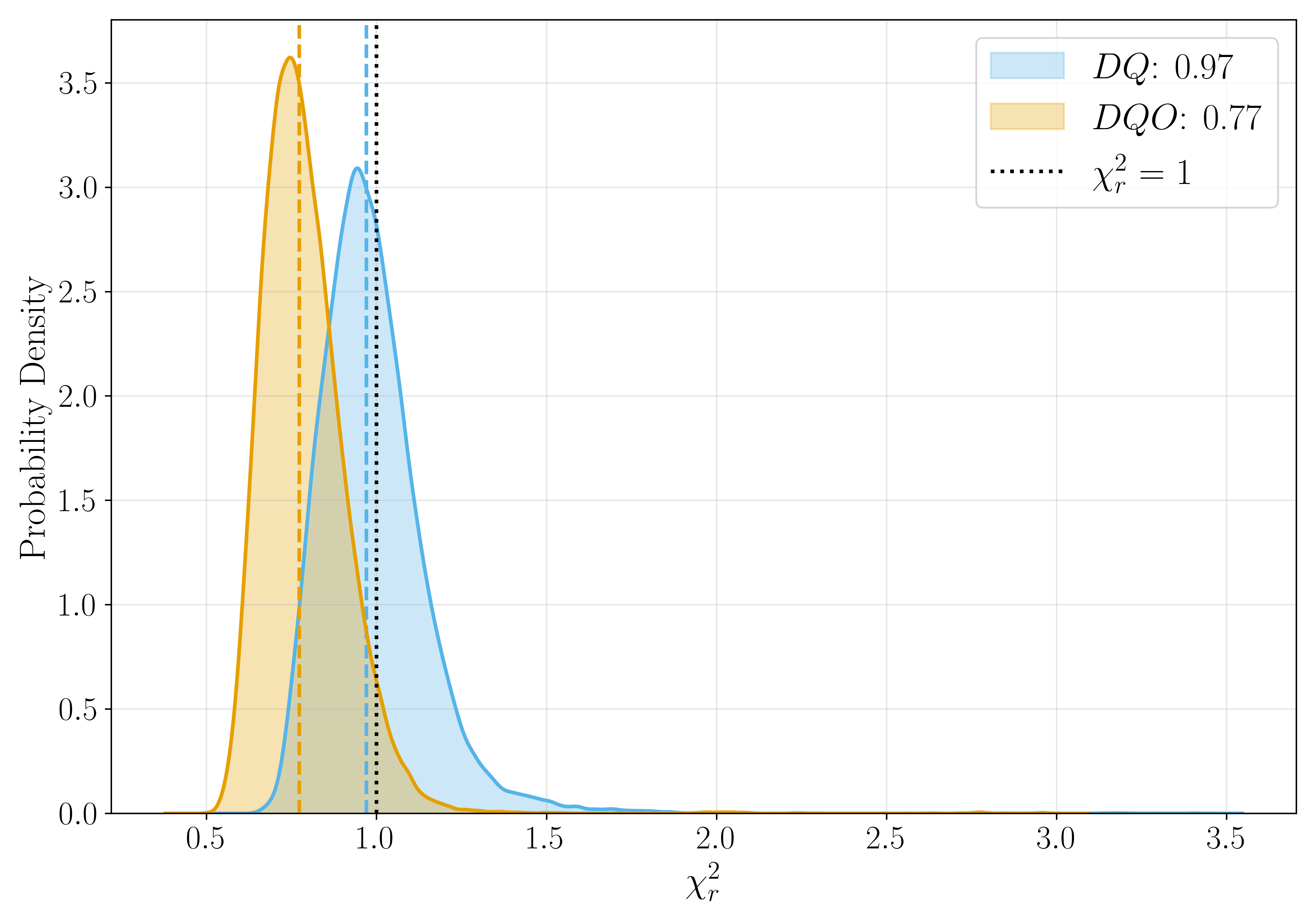}
    \caption{Distribution of the reduced chi-squared statistic, $\chi_r^2$, for the \UQ (sky blue) and \UO (orange) cases, using the same samples as in Figure~\ref{fig:corner_l12} and ~\ref{fig:corner_l123}. The median values are indicated by dashed lines in matching colors and are listed in the legend. The dotted black line denotes $\chi_r^2 = 1$ for reference.}
    \label{fig:ReducedChiSquared_DQ_DQO}
\end{figure}

The physical hotspots corresponding to the posterior distributions exhibit distinct morphologies. The \UQ model yields a limited set of three broadly similar hotspot families, whereas \UO produces a substantially broader diversity in both geometry and localization, indicating a stronger internal geometric degeneracy in the higher-order truncation; this is expected as the larger parameter space naturally permits a wider range of solutions. Field degeneracies were reported by \citetalias{kalapotharakos_multipolar_2021} and shown to be partially lifted through multiwavelength, particularly $\gamma$-ray, light-curve modeling, suggesting that a similar approach may help constrain the degeneracies seen here. In Figure~\ref{fig:ReducedChiSquared_DQ_DQO}, we show the distributions of the reduced chi-squared statistic for the same samples as in the posterior distribution. The distribution and the median values suggest that \UQ already provides an adequate description of the bolometric thermal X-ray light curve, with $\chi_r^2$ values clustered around unity, whereas \UO tends to populate systematically lower values. This suggests that the bolometric light curve alone may favor \UQ as the simpler sufficient description, while the additional flexibility of \UO may partly reflect an increased capacity to fit statistical fluctuations.

The hotspot morphologies found in the present work show partial qualitative overlap with those reported by \citetalias{kalapotharakos_multipolar_2021}. In particular, the \UQ truncation produces hotspot families that are broadly consistent with those associated with their $RV1$, $RV2$, $RV5$, and $RV6$ solutions, while the \UO truncation exhibits a wider range of configurations, including morphologies not apparent in \citetalias{kalapotharakos_multipolar_2021}, but also cases resembling their $RV3$ and $RV4$ solutions. At the parameter level, both \UQ and \UO yield $B_Q/B_D$ and $\chi_D$ distributions that are consistent with the detailed $RV1$ posterior presented by \citetalias{kalapotharakos_multipolar_2021}. The latter point is particularly noteworthy, because $\chi_D$ ($\alpha_D$ in \citetalias{kalapotharakos_multipolar_2021}) sets the inclination of the dipolar component, which dominates the field structure at large distances and is therefore expected to be especially relevant for the $\gamma$-ray emission produced near and beyond the light cylinder in the equatorial current sheet of close-to-force-free magnetospheres. Within our posterior distributions, the ratio ${B_O}/{B_D}$ spans a wider range than ${B_Q}/{B_D}$, indicating that the octupole contribution is more flexible and less tightly coupled to the dipole component.

The inferred surface-field configuration for the best-fit solutions indicate that the \UO solutions are generally more anisotropic than the \UQ ones, consistent with the stronger influence of higher-order multipolar structure on the stellar surface. The corresponding maximum surface field strengths, $B_{max}$, remain of order $10 - 20 B_D$, well below the most extreme values reported by \citetalias{kalapotharakos_multipolar_2021}, who found a broader range of $\sim 7 - 420 B_D$. This suggests that the centered SVM2F solutions explored here can reproduce the required hotspot asymmetries without invoking the largest surface-field contrasts found in the offset-vacuum solutions. 

\subsection{SVM2F in multiwavelength context}
\begin{figure}[htbp]
    \centering
    \includegraphics[width=0.99\linewidth]{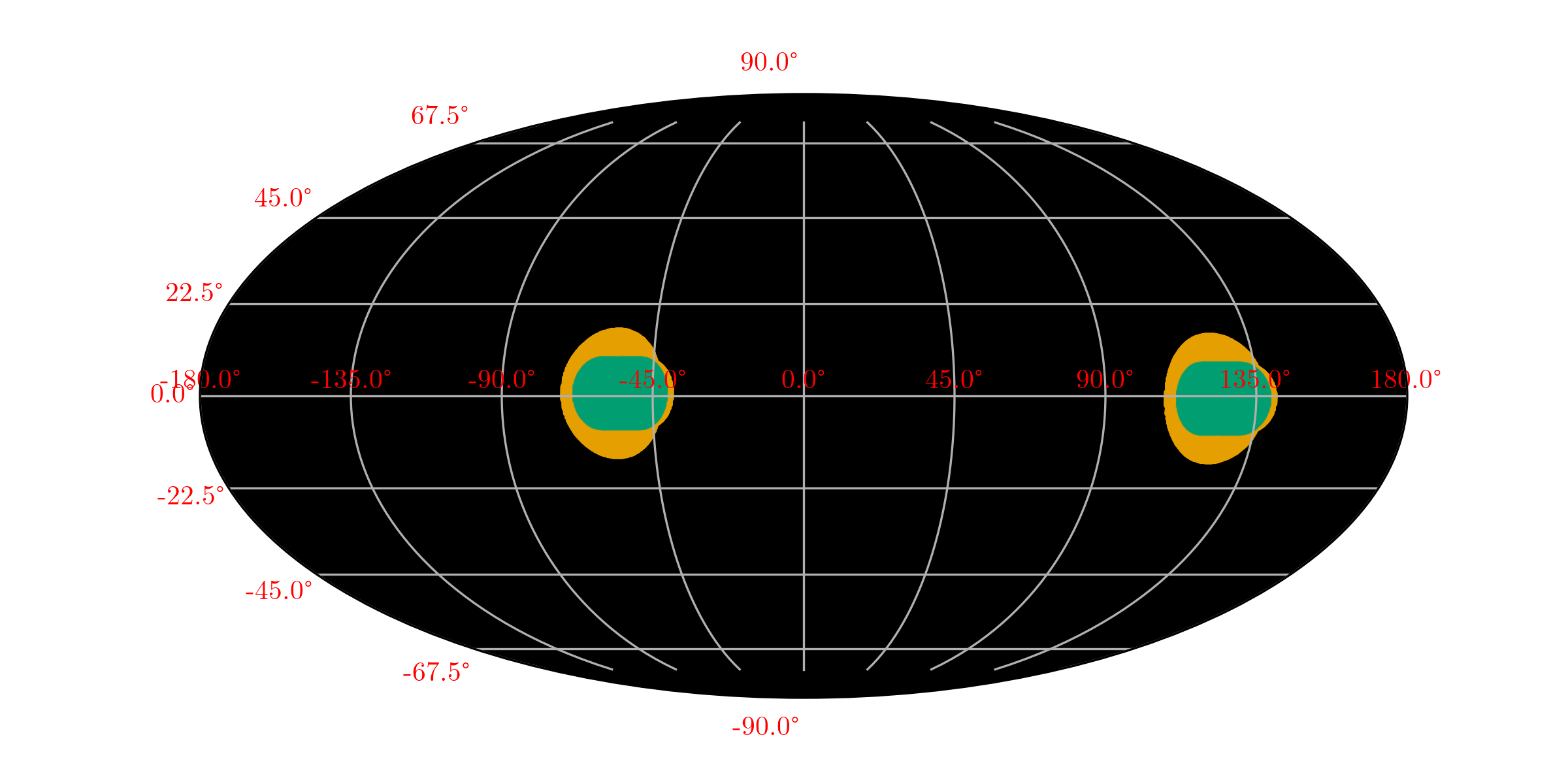}
    \includegraphics[width=0.99\linewidth]{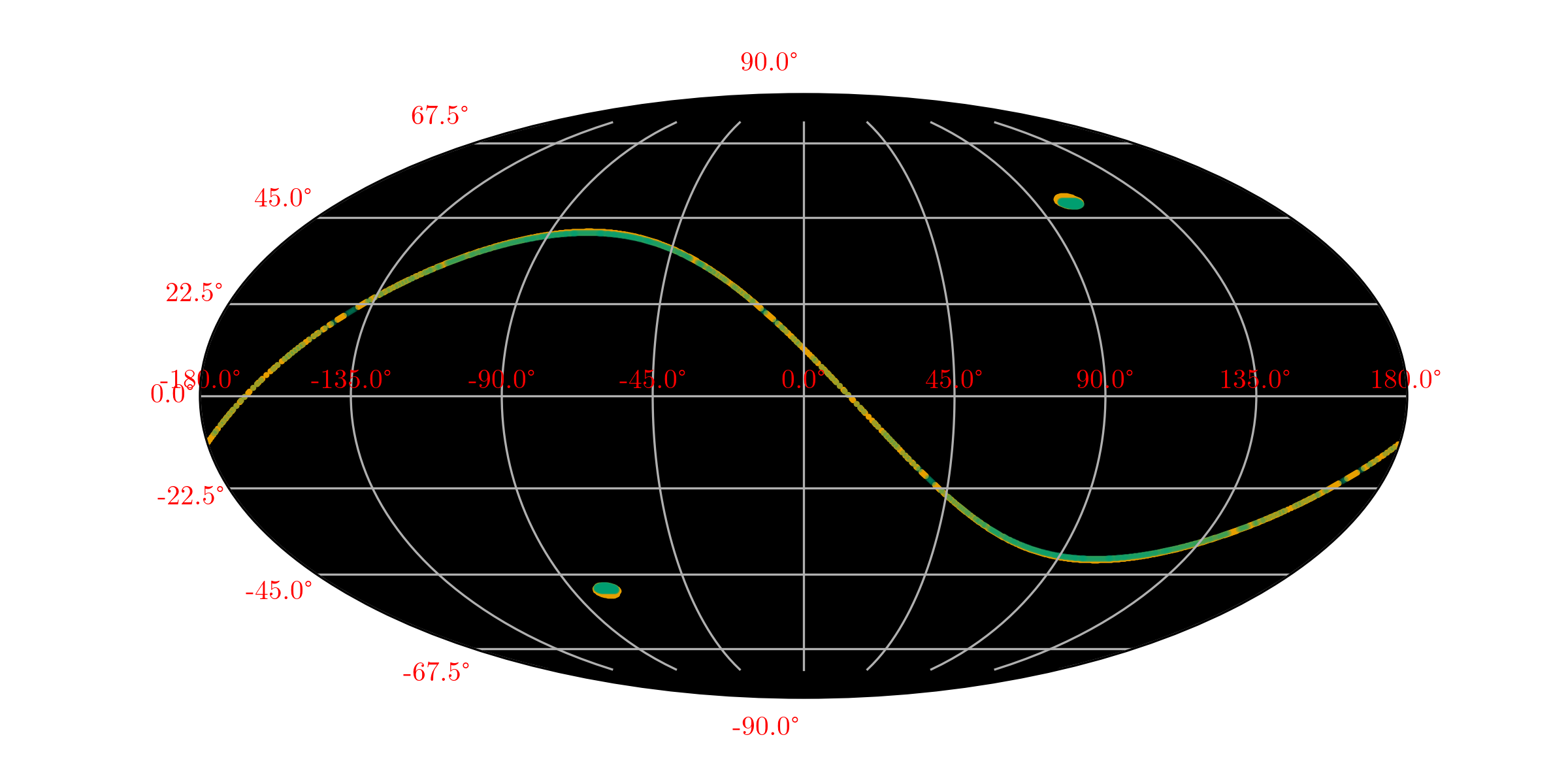}
    \includegraphics[width=0.99\linewidth]{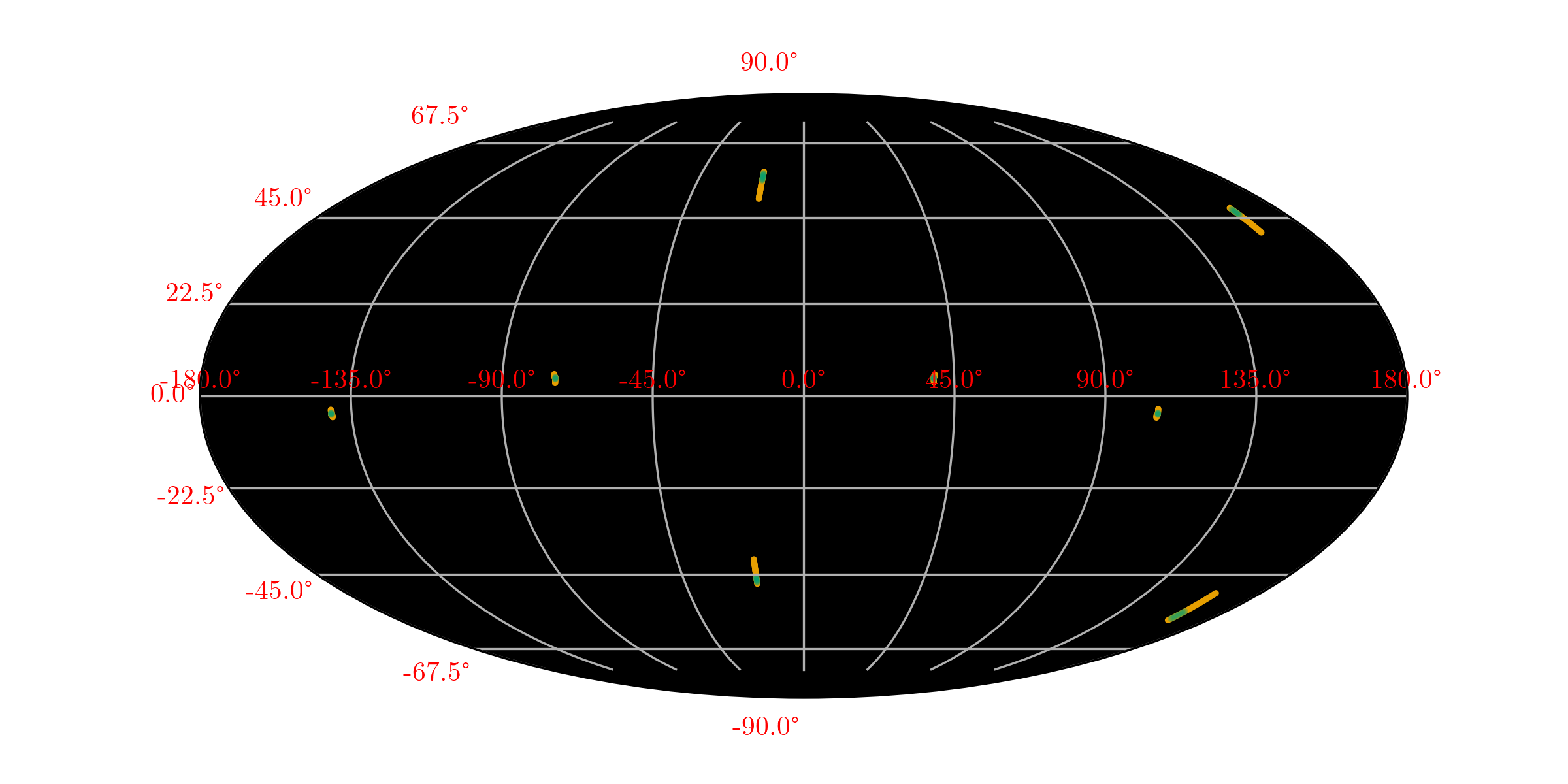}
    \includegraphics[width=0.99\linewidth]{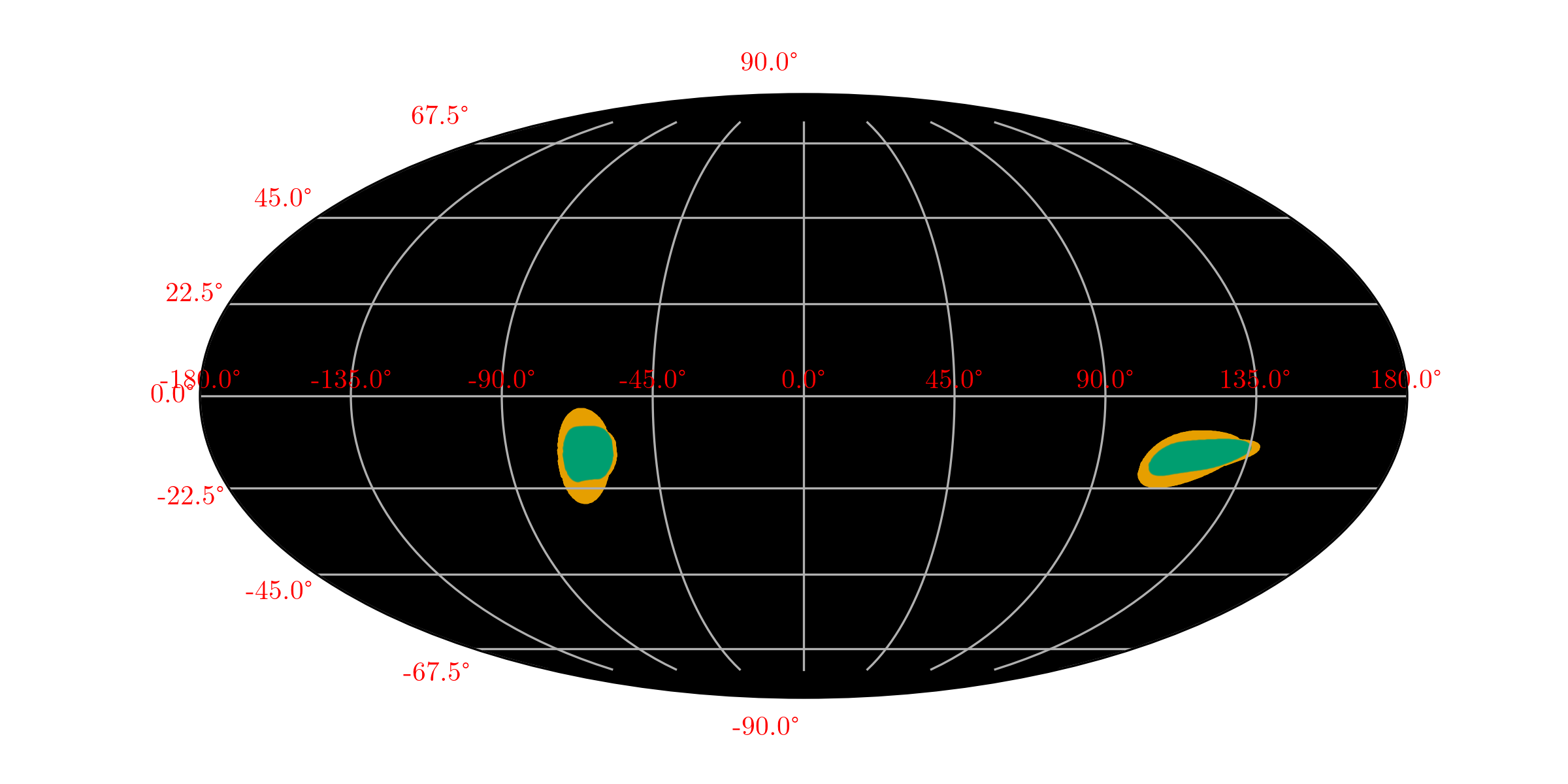}
    \includegraphics[width=0.99\linewidth]{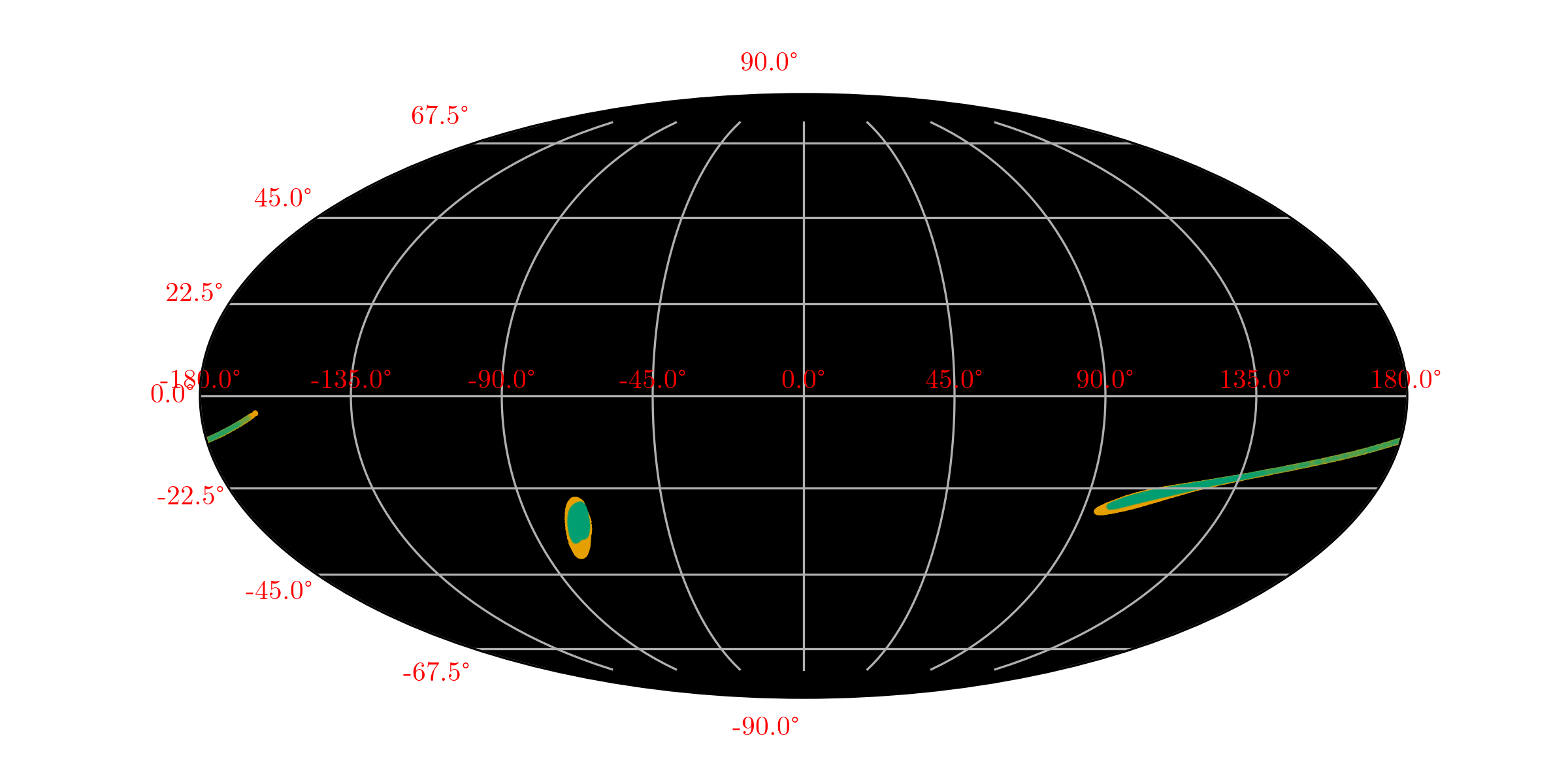}
    \caption{Hotspots for the \UO solution shown in a Mollweide projection for SVM2F (orange) and its corresponding static variant (bluish green). From top to bottom: dipole-only, quadrupole-only, octupole-only, \UO with relative field strengths set to unity, and the complete \UO solution.}   \label{fig:BestFitStaticVSSweepback_hotspots}
\end{figure}
Figure~\ref{fig:BestFitStaticVSSweepback_hotspots} compares the hotspot geometry for the static and swept-back configurations of the best-fit \UO solution. The SVM2F model and its corresponding static variant (calculated using the expressions for $a_{lm}^{B,\mathrm{in}}$ listed in Table~\ref{tab:coeffs}), are shown on a $1,000 \times 1,000$ grid. The five panels, from top to bottom, illustrate the following cases: dipole-only, quadrupole-only, octupole-only, \UO with relative field strengths set to unity, and the complete \UO solution. For all configurations, the static polar caps are clearly smaller and less distorted than those of the corresponding swept-back model \citep{dyks_rotational_2004}. When the individual multipolar components are assigned equal weights, the resulting geometry is closer to that of a dipole, because the higher-order components decay more rapidly with radius, as $1/r^{l+2}$ (see Figures~\ref{fig:BestFitFieldLines1} and~\ref{fig:BestFitFieldLines2}). When the relative strengths are fixed to the best-fit values for J0030, the higher-order components exert a much stronger influence on the open-field-line structure and on the resulting hotspot morphology.

This behavior is particularly relevant in the non-vacuum magnetospheres. The near-surface magnetic field structure determines the charge-density and current-density distributions over the polar caps, and hence the local $J/\rho c$ current regime. As shown by \citetalias{kalapotharakos_multipolar_2021}, multipolar field structures can produce charge-density and current-density patterns that differ substantially, and in some cases unexpectedly, from those of the pure dipole. Since the $J/\rho c$ regime directly affects the efficiency of pair production above the polar caps \citep{timokhin_arons_2013}, these differences can alter the plasma supply required to sustain currents in the outer magnetosphere. By incorporating both rotational sweepback and higher-order multipoles, the SVM2F configuration provides a more realistic description of the near-surface magnetic field structure and of the resulting polar cap structure.

Since SVM2F describes an exact rotating vacuum solution, it is expected to approximate realistic FF fields more closely than a static configuration, particularly in the outer magnetosphere \citep{kalapotharakos_gamma-ray_2014}. This proximity is valuable because FF configurations provide the most realistic large-scale magnetic field structures, for outer-magnetosphere $\gamma$-ray emission, while also improving the mapping between the global magnetosphere and the polar cap footprints that are important for NICER X-ray light-curve modeling. Direct FF exploration of a high-dimensional parameter space nevertheless remains computationally expensive. In that context, NN surrogates become even more valuable than in the present study, because the physical FF models are considerably slower, whereas the evaluation cost of the corresponding NN surrogates remains essentially unchanged. The main limitation is instead the generation of sufficiently large FF training sets. In T. Lechien et al. 2026 (in preparation), this difficulty is addressed by pre-training NNs on static vacuum solutions and then fine-tuning them using a much smaller amount of FF data, which appears to work well. The expectation is that NNs pre-trained on the more realistic SVM2F configurations will lie closer to the corresponding FF solutions than those pre-trained on static vacuum fields, thereby making the subsequent FF fine-tuning more efficient.

\subsection{Toward constraining the EoS}
Because all NICER MSPs are also Fermi $\gamma$-ray and radio pulsars, they are strong candidates for global, multiwavelength magnetospheric modeling. Other NICER MSPs that have been extensively analyzed include PSR J$0740+6620$ \citep{riley_nicer_2021, miller_radius_2021, dittmann_more_2024, salmi_radius_2024, petri_double_2025}, PSR J$0437-4715$ \citep{choudhury_nicer_2024, reardon_neutron_2024, MillerJ0437_2025, petri_2026_J0437}, PSR J$1231-1411$ \citep{salmi_nicer_2024}, and PSR J$0614-3329$ \citep{mauviard_nicer_2025}, making them compelling targets for future applications of this framework. The generality of SVM2F ensures its applicability to other NICER MSPs, and more broadly, to NSs, as normal pulsars also exhibit observational evidence for non-dipolar magnetic fields. 

In future work, we will allow \MR and $\zeta$ to vary in the parameter space to obtain independent constraints on both magnetic field parameters and size of the NSs. A further development will involve extending the MCMC exploration to include hotspot temperatures in order to model energy-dependent light curves (i.e., photon counts per phase per energy channel; \citealt{bogdanov_constraining_2019, bogdanov_constraining_2021}). This will require an accurate treatment of emission anisotropy across energy channels using a fully ionized hydrogen atmosphere model.  In parallel, a judicious multiwavelength analysis should incorporate FF field configurations whenever global magnetospheric structure, especially the non-thermal high-energy emission geometry, becomes important. Applying this framework to multiple MSPs and obtaining self-consistent constraints on magnetic structure and size of NSs will enable more stringent, physically motivated constraints on the EoS.

\begin{acknowledgments}
We thank the anonymous referee for their valuable feedback. AK would like to thank J\'er\^ome P\'etri, Louis du Plessis, and Dimitrios Skiathas for helpful discussions, and her colleagues at the Centre for Space Research (North-West University, South Africa) for their moral and caffeinated support. AK’s research was supported by an appointment to the NASA Postdoctoral Program at the NASA Goddard Space Flight Center, administered by Oak Ridge Associated Universities under contract with NASA (Grant number 80HQTR21CA005). Resources supporting this work were provided by the NASA High-End Computing (HEC) Program through the NASA Advanced Supercomputing (NAS) Division at Ames Research Center. This work is based on research supported wholly/in part by the National Research Foundation of South Africa (NRF; Grant Number 99072). The Grant holder acknowledges that opinions, findings and conclusions or recommendations expressed in any publication generated by the NRF-supported research is that of the author(s), and that the NRF accepts no liability whatsoever in this regard. We acknowledge the use of computational resources provided by the Centre for High Performance Computing (CHPC) in Cape Town under project ASTR1245. The material is based upon work supported by NASA under award numbers 80GSFC24M0006, 80GSFC21M0002, and 80NSSC21K1999, and under grants 21-ATP21-0116, 22-ADAP22-0142, 22-TCAN22-0027. We also acknowledge the use of the NASA Astrophysics Data Service (ADS).
\end{acknowledgments}
\software{Astropy \citep{astropy}, Mathematica \citep{mathematica}, Matplotlib \citep{matplotlib}, NumPy \citep{numpy}, Plotly \citep{plotly2015}, PyTorch \citep{paszke_pytorch_2019}, scikit-learn \citep{pedregosa_scikit-learn_2011}, and SciPy \citep{scipy}.}

\appendix

\section{Vector Spherical Harmonics} \label{app:VSH}
\subsection{Basis} \label{appsub:VSHBasis}
\cite{barrera_vector_1985} constructed vector functions analogous to the scalar spherical harmonics, with the aim of ensuring orthogonality (later demonstrated by \citealt{carrascal_vector_1991}) and completenesss properties. Using their notation for the three VSH, $\{ \bsym{Y}_{lm}, \bsym{\Psi}_{lm}, \bsym{\Phi}_{lm} \}$, and keeping the normalization the same as defined in Appendix A2 in \cite{petri_pulsar_2012}, we have,
\begin{subequations}
\label{VSH_definitions}
\begin{align}
    \label{VSH_Y_gen}
    \bsym{Y}_{lm} = & \,Y_{lm} \bsym{e}_{r},  \\ 
    \label{VSH_Psi_gen}
    \bsym{\Psi}_{lm} = & \,\frac{r}{\sqrt{l(l+1)}} \nabla Y_{lm}, \\
    \label{VSH_Phi_gen}
    \bsym{\Phi}_{lm} = & \,\frac{\bsym{r}}{\sqrt{l(l+1)}} \times \nabla Y_{lm},
\end{align}
\end{subequations}
where $Y_{lm}$ are the scalar spherical harmonics of order $(l,m)$. Note that this definition differs from the standard one, which does not include the denominator terms in the expressions for $\bsym{\Psi}_{lm}$ and $\bsym{\Phi}_{lm}$. This notation simplifies the orthogonality relations, which are provided in Appendix~\ref{appsub:VSHProperties}.

The gradient of $Y_{lm}$ is given by,
\begin{align*}
    \nabla Y_{lm} = & \, \frac{\partial Y_{lm}}{\partial r}  \bsym{e}_{r} + \frac{1}{r} \frac{\partial Y_{lm}}{\partial \theta} \bsym{e}_{\theta} + \frac{1}{r \sin{\theta}} \frac{\partial Y_{lm}}{\partial \phi} \bsym{e}_{\phi},
\end{align*}
where first term is zero since $Y_{lm} = Y_{lm}(\theta) e^{i m \phi}$ is only a function of $(\theta, \phi)$. Equation~\eqref{VSH_definitions} can be expanded for further usage (later in Appendix~\ref{app:multipole_expansion}) as,
\begin{subequations}
\label{VSH_definition_simplified}
\begin{align}
    \label{VSH_Y_gen_simplified}
    \bsym{Y}_{lm} = & \,Y_{lm} \bsym{e}_{r},  \\ 
    \label{VSH_Psi_gen_simplified}
    \bsym{\Psi}_{lm} = & \,\frac{1}{\sqrt{l(l+1)}} \left( \frac{\partial Y_{lm}}{\partial \theta} \bsym{e}_{\theta} + \frac{1}{\sin{\theta}} \frac{\partial Y_{lm}}{\partial \phi} \bsym{e}_{\phi} \right), \\
    \bsym{\Phi}_{lm} = & \,\frac{r \bsym{e}_{r}}{\sqrt{l(l+1)}} \times \left( \frac{1}{r} \frac{\partial Y_{lm}}{\partial \theta} \bsym{e}_{\theta} + \frac{1}{r \sin{\theta}} \frac{\partial Y_{lm}}{\partial \phi} \bsym{e}_{\phi} \right), \\
    \label{VSH_Phi_gen_simplified}
      = & \, \frac{1}{\sqrt{l(l+1)}} \left( \frac{\partial Y_{lm}}{\partial \theta} \bsym{e}_{\phi} - \frac{1}{\sin{\theta}} \frac{\partial Y_{lm}}{\partial \phi} \bsym{e}_{\theta} \right).
\end{align}
\end{subequations}

\subsection{Properties} \label{appsub:VSHProperties}
The conjugate property of VSH is described as,
\begin{align*}
    \bsym{Y}_{lm}^{*} = &  \,(-1)^{m} \bsym{Y}_{l, -m} , \\
    \bsym{\Psi}_{lm}^{*} = & \,(-1)^{m} \bsym{\Psi}_{l, -m}, \\    
    \bsym{\Phi}_{lm}^{*} = & \,(-1)^{m} \bsym{\Phi}_{l, -m},
\end{align*}

and the orthonormality property gives,
\begin{align*}
    \int \bsym{Y}_{lm}  \cdot \bsym{Y}_{l'm'}^{*} \, d\Omega = \, & \delta_{ll'} \delta_{mm'},  \\
    \int \bsym{\Psi}_{lm} \cdot \bsym{\Psi}_{l'm'}^{*} \, d\Omega =\, & \delta_{ll'} \delta_{mm'},  \\    
    \int \bsym{\Phi}_{lm} \cdot \bsym{\Phi}_{l'm'}^{*} \, d\Omega =\, & \delta_{ll'} \delta_{mm'}, \\
\int \bsym{Y}_{lm}  \cdot \bsym{\Psi}_{l'm'}^{*} \, d\Omega =\, & 0,  \\
    \int \bsym{Y}_{lm} \cdot \bsym{\Phi}_{l'm'}^{*} \, d\Omega = \,& 0, \\  
    \int \bsym{\Psi}_{lm} \cdot \bsym{\Phi}_{l'm'}^{*} \, d\Omega = \,& 0. 
\end{align*}

\section{Simplifying the multipole expansion} \label{app:multipole_expansion}
Using the relation \citep{barrera_vector_1985},
\begin{equation}
    \nabla \times \left[ f(r) \bsym{\Phi}_{lm} \right] = - \frac{\sqrt{l(l+1)}}{r} f \bsym{Y}_{lm}  - \frac{1}{r} \partial_{r}{(r f)}  \bsym{\Psi}_{lm},
\end{equation}
we rewrite the interior magnetic field expression, Equation~\ref{eqn:magfieldBinside}, as,
\begin{align}
    \label{eqnApp:MagFieldVecBin}
    \boldsymbol{B}_{\rm in}(r, \theta, \phi, t) =  & \sum_{l=1}^{\infty}  \sum_{m=-l}^{l} \left( - \frac{\sqrt{l(l+1)}}{r} f_{lm}^{B,{\rm in}} \bsym{Y}_{lm} - \frac{1}{r} \partial_{r}(r f_{lm}^{B,{\rm in}}) \bsym{\Psi}_{lm} \right) e^{- i m \Omega t},
\end{align}
where the temporal dependence is extracted as $e^{- i m \Omega t}$. Similarly, the exterior magnetic field expression, Equation~\ref{eqn:magfieldB}, is simplified as,
\begin{align}
    \label{eqnApp:MagFieldVecBout}
    \boldsymbol{B}_{\rm out}(r, \theta, \phi, t) =  & \sum_{l=1}^{\infty}  \sum_{m=-l}^{l} \left( - \frac{\sqrt{l(l+1)}}{r} f_{lm}^{B,{\rm out}} \bsym{Y}_{lm} - \frac{1}{r} \partial_{r}(r f_{lm}^{B,{\rm out}}) \bsym{\Psi}_{lm} + g_{lm}^{B,{\rm out}} \bsym{\Phi}_{lm} \right) e^{- i m \Omega t},
\end{align}

To implement the exterior multipolar expansion efficiently in our code, we use only scalar spherical harmonics and their derivatives. Specifically, we apply Equation~\eqref{VSH_definition_simplified} in Equation~\eqref{eqnApp:MagFieldVecBout}, together with $\partial_{\phi}(Y_{lm}(\theta, \phi)) = i m Y_{lm}(\theta) e^{i m \phi}$ and the linear scaling relation, $g_{lm}^{B,{\rm out}} = - i \mu_{0} m \Omega f_{lm}^{D,{\rm out}}$, to write,
\begin{equation}
    \boldsymbol{B}_{\rm out}(r, \theta, \phi, t) = \sum_{l=1}^{\infty}  \sum_{m=-l}^{l} \left[ B_{r,{\rm out}} (r, \theta, \phi, t) \, \bsym{e}_{r} + B_{\theta,{\rm out}} (r, \theta, \phi, t) \, \bsym{e}_{\theta} + B_{\phi,{\rm out}} (r, \theta, \phi, t) \, \bsym{e}_{\phi}  \right],
\end{equation}
where,
\begin{align*}
    B_{r,{\rm out}} = & - \frac{\sqrt{l(l+1)}}{r} f_{lm}^{B,{\rm out}} Y_{lm}(\theta) e^{i m \phi} e^{- i m \Omega t}, \\
    B_{\theta,{\rm out}} = & - \frac{\partial_{r}(r f_{lm}^{B,{\rm out}})}{r \sqrt{l(l+1)}} \partial_{\theta}(Y_{lm}(\theta)) e^{i m \phi} e^{- i m \Omega t}
     + \frac{i \mu_{0} m \Omega f_{lm}^{D,{\rm out}}}{\sin{\theta} \sqrt{l(l+1)}} i m Y_{lm}(\theta) e^{i m \phi} e^{- i m \Omega t},\\
    B_{\phi,{\rm out}} = & -\frac{\partial_{r}(r f_{lm}^{B,{\rm out}})}{r \sin{\theta} \sqrt{l(l+1)}} i m Y_{lm}(\theta) e^{i m \phi} e^{- i m \Omega t}
     -  \frac{i \mu_{0} m \Omega f_{lm}^{D,{\rm out}}}{\sqrt{l(l+1)}} \partial_{\theta}(Y_{lm}(\theta)) e^{i m \phi} e^{- i m \Omega t}.
\end{align*}
With the convention of using only positive $m$ values (as described in Section~\ref{subsubsec:coeff_SVF}), for a single $l$ the expressions become,
\begin{equation}
    \bsym{B}^{l}_{\rm out}(r, \theta, \phi, t) = \sum_{m} \left[ B_{r,{\rm out}}^{l} (r, \theta, \phi, t) \, \bsym{e}_{r} + B_{\theta,{\rm out}}^{l} (r, \theta, \phi, t) \, \bsym{e}_{\theta} + B_{\phi,{\rm out}}^{l} (r, \theta, \phi, t) \, \bsym{e}_{\phi}  \right],
    \label{eqn:FinalFormFriezaField}
\end{equation}
where,
\begin{align*}
    \label{Br_l}
    B_{r,{\rm out}}^{l} = & \sum_{m=0}^{l} \left[ - \frac{\sqrt{l(l+1)}}{r} f_{lm}^{B,{\rm out}} Y_{lm}(\theta) e^{i m \phi} \right] e^{- i m \Omega t}, \\
B_{\theta,{\rm out}}^{l} (1) = & \sum_{m=0}^{l} \left[ - \frac{\partial_{r}(r f_{lm}^{B,{\rm out}})}{r \sqrt{l(l+1)}} \partial_{\theta}(Y_{lm}(\theta)) e^{i m \phi}  \right] e^{- i m \Omega t}, \\
B_{\theta,{\rm out}}^{l} (2) = & \sum_{m=1}^{l-1} \left[ \frac{i \mu_{0} m \Omega f_{l-1,m}^{D,{\rm out}}}{\sin{\theta} \sqrt{(l-1)l}} i m Y_{l-1,m}(\theta) e^{i m \phi}  \right] e^{- i m \Omega t} + \sum_{m=1}^{l} \left[ \frac{i \mu_{0} m \Omega f_{l+1,m}^{D,{\rm out}}}{\sin{\theta} \sqrt{(l+1)(l+2)}} i m Y_{l+1,m}(\theta) e^{i m \phi} \right] e^{- i m \Omega t},  \\
B_{\phi,{\rm out}}^{l}(1) = & \sum_{m=0}^{l} \left[ -\frac{\partial_{r}(r f_{lm}^{B,{\rm out}})}{r \sin{\theta} \sqrt{l(l+1)}} i m Y_{lm}(\theta) e^{i m \phi} \right] e^{- i m \Omega t}, \\
B_{\phi,{\rm out}}^{l}(2) = & \sum_{m=1}^{l-1} \left[ - \frac{i \mu_{0} m \Omega f_{l-1,m}^{D,{\rm out}}}{\sqrt{(l-1)l}} \partial_{\theta}(Y_{l-1,m}(\theta)) e^{i m \phi} \right] e^{- i m \Omega t} +  \sum_{m=1}^{l} \left[ - \frac{i \mu_{0} m \Omega f_{l+1,m}^{D,{\rm out}}}{\sqrt{(l+1)(l+2)}} \partial_{\theta}(Y_{l+1,m}(\theta)) e^{i m \phi} \right] e^{- i m \Omega t}.
\end{align*}
where $B_{\theta,{\rm out}}^{l} = B_{\theta,{\rm out}}^{l}(1) + B_{\theta,{\rm out}}^{l}(2)$ and $ B_{\phi,{\rm out}}^{l} = B_{\phi,{\rm out}}^{l}(1) + B_{\phi,{\rm out}}^{l}(2)$, each split according to the distinct summation limits over $m$. As discussed in Section~\ref{sec:mf_expansion} the $f_{lm}^{D}$ terms couple multipole order $l$ to orders $l+1$ and $l-1$, resulting in modified summation ranges over $m$, with only the $l+1$ contribution present for the dipolar case.

For the interior magnetic field, the expression becomes
\begin{align}
    \bsym{B}^{l}_{\rm in}(r, \theta, \phi, t) = \sum_{m} \left[ B_{r,{\rm in}}^{l} (r, \theta, \phi, t) \, \bsym{e}_{r} + B_{\theta,{\rm in}}^{l} (1) (r, \theta, \phi, t) \, \bsym{e}_{\theta} + B_{\phi,{\rm in}}^{l} (1) (r, \theta, \phi, t) \, \bsym{e}_{\phi}  \right],
\end{align}
with the corresponding $f_{lm}^{B,{\rm in}}$.

To evaluate the full field expansion up to $l = 3$, the coefficients $a_{lm}^{B,{\rm in}}$, $a_{lm}^{B,{\rm out}}$, and $a_{lm}^{D,{\rm out}}$ and the corresponding calculations (Section~\ref{sec:mf_expansion},~\ref{sec:application_SVM2F}) are listed in Table~\ref{tab:coeffs}.

\begin{table}[h!]
\centering
\small
\setlength{\tabcolsep}{15pt} \renewcommand{\arraystretch}{2.0} \caption{Expansion coefficients $q_{lm}$, $a_{lm}^{B,\mathrm{in}}$, $a_{lm}^{B,\mathrm{out}}$, and $a_{lm}^{D,\mathrm{out}}$, for dipole ($l=1$), quadrupole ($l=2$), and octupole ($l=3$) components for each $m$. Relevant equations used to compute these quantities are indicated. All expressions in the last three columns include a factor of $e^{i m \phi_{lm}}$, which is omitted for brevity.}
\label{tab:coeffs}
\begin{tabular}{c c c c c}
\hline
\hline
$m$ &
$q_{lm}$ &
$a_{lm}^{B,\mathrm{in}}$ &
$a_{lm}^{B,\mathrm{out}}$ &
$a_{lm}^{D,\mathrm{out}}$\\
\hline
\textbf{Equation} &
\ref{eq:qlm_param}-\ref{eq:Clm_general} &
\ref{eq:abin_param} &
\ref{eqn:almOutAndIn_m0}, \ref{eqn:almOutAndIn} &
\ref{eq:bcD_lplus1_0}-\ref{eq:Jlm_For_almD_out} \\
\hline
\multicolumn{5}{l}{\textbf{Dipole ($l=1$)}} \\
0 &
$\sqrt{\frac{8\pi}{3}} \cos\chi_D$ &
$- B_D R_\star^3 q_{10}$ & 
$- B_D R_\star^3 q_{10}$ &
$a^D_{20} = \sqrt{\frac{1}{5}}\, \epsilon_0 \Omega B_D R_\star^5 q_{10}$ \\
1 & 
$\sqrt{\frac{16\pi}{3}} \sin\chi_D$ &
$B_D R_\star^3 q_{11}$ & 
$q_{11} \frac{B_D R_\star}{h^{(1)}_1 (kR_\star)}$ &
$a^D_{21} = \sqrt{\frac{3}{5}}\, \frac{\epsilon_0 \Omega B_D R_\star^2 q_{11}}{\partial_r (r h^{(1)}_2 (kr)) |_{r = R_\star}}$ \\
\hline
\multicolumn{5}{l}{\textbf{Quadrupole ($l=2$)}} \\
0 &
$\sqrt{\frac{4\pi}{3}} \cos\chi_{Q1}$ &
$B_Q R_\star^4 q_{20}$ &
$q_{20} B_Q R_\star^4$ &
$\begin{aligned} 
    a_{10}^D &= \frac{2}{\sqrt{5}} \epsilon_{0} \Omega B_Q R_\star^4 q_{20} \\
    a_{30}^D &= -2 \sqrt{\frac{2}{35}} \epsilon_{0} \Omega B_Q R_\star^6 q_{20} 
\end{aligned}$ \\
1 &
$\sqrt{\frac{8\pi}{3}} \sin\chi_{Q1}\cos\chi_{Q2}$ &
$B_Q R_\star^4 q_{21}$ &
$q_{21} \frac{B_Q R_\star}{h_2^{(1)} (k R_\star)}$ &
$\begin{aligned}
    a_{11}^D &= -\sqrt{\frac{3}{5}} \frac{\epsilon_{0} \Omega B_Q R_\star^2 q_{21}}{\partial_r (r h^{(1)}_1 (kr)) |_{r = R_\star}} \\
    a_{31}^D &= \frac{8}{\sqrt{35}} \frac{\epsilon_{0} \Omega B_Q R_\star^2 q_{21}}{\partial_r (r h^{(1)}_3 (kr)) |_{r = R_\star}}
\end{aligned}$ \\
2 &
$\sqrt{\frac{8\pi}{3}} \sin\chi_{Q1}\sin\chi_{Q2}$ & 
$B_Q R_\star^4 q_{22}$ &
$q_{22} \frac{B_Q R_\star}{h_2^{(1)} (2 k R_\star)}$ &
$a_{32}^D = 2 \sqrt{\frac{2}{7}} \frac{\epsilon_{0} \Omega B_Q R_\star^2 q_{22}}{\partial_r (r h^{(1)}_3 (2kr)) |_{r = R_\star}}$ \\
\hline
\multicolumn{5}{l}{\textbf{Octupole ($l=3$)}} \\
0 &
$\frac{2\sqrt{2\pi}}{3} \cos\chi_{O1}$ &
$B_O R_\star^5 q_{30}$ &
$q_{30} B_O R_\star^5$ &
$\begin{aligned}
    a_{20}^D &= 3 \sqrt{\frac{2}{35}} \epsilon_{0} \Omega B_O R_\star^5 q_{30} \\
    a_{40}^D &= -\sqrt{\frac{5}{21}} \epsilon_{0} \Omega B_O R_\star^7 q_{30}
\end{aligned}$ \\
1 &
$\frac{4\sqrt{\pi}}{3} \sin\chi_{O1}\cos\chi_{O2}$ & 
$B_O R_\star^5 q_{31}$ &
$q_{31} \frac{B_O R_\star}{h_3^{(1)} (k R_\star)}$ &
$\begin{aligned} 
    a_{21}^D &= - \frac{8}{\sqrt{35}} \frac{\epsilon_{0} \Omega B_O R_\star^2 q_{31}}{\partial_r (r h^{(1)}_2 (kr)) |_{r = R_\star}} \\ a_{41}^D &= \frac{5}{\sqrt{7}} \frac{\epsilon_{0} \Omega B_O R_\star^2 q_{31}}{\partial_r (r h^{(1)}_4 (kr)) |_{r = R_\star}} 
\end{aligned}$ \\
2 & 
$\frac{4\sqrt{\pi}}{3} \sin\chi_{O1}\sin\chi_{O2}\cos\chi_{O3}$ & $B_O R_\star^5 q_{32}$ &
$q_{32} \frac{B_O R_\star}{h_3^{(1)} (2 k R_\star)}$ &
$\begin{aligned} 
    a_{22}^D &= - 2 \sqrt{\frac{2}{7}} \frac{\epsilon_{0} \Omega B_O R_\star^2 q_{32}}{\partial_r (r h^{(1)}_2 (2kr)) |_{r = R_\star}} \\
    a_{42}^D &= 2 \sqrt{\frac{5}{7}} \frac{\epsilon_{0} \Omega B_O R_\star^2 q_{32}}{\partial_r (r h^{(1)}_4 (2kr)) |_{r = R_\star}} 
\end{aligned}$ \\
3 & 
$\frac{4\sqrt{\pi}}{3} \sin\chi_{O1}\sin\chi_{O2}\sin\chi_{O3}$ &
$B_O R_\star^5 q_{33}$ &
$q_{33} \frac{B_O R_\star}{h_3^{(1)} (3 k R_\star)}$ &
$a_{43}^D = \sqrt{\frac{5}{3}} \frac{\epsilon_{0} \Omega B_O R_\star^2 q_{33}}{\partial_r (r h^{(1)}_4 (3kr)) |_{r = R_\star}}$ \\
\hline
\end{tabular}
\end{table}

\section{Octupole field polar caps} \label{app:PolarCaps_l3}
\begin{figure}[htbp]
    \centering
    \includegraphics[width=0.8\linewidth]{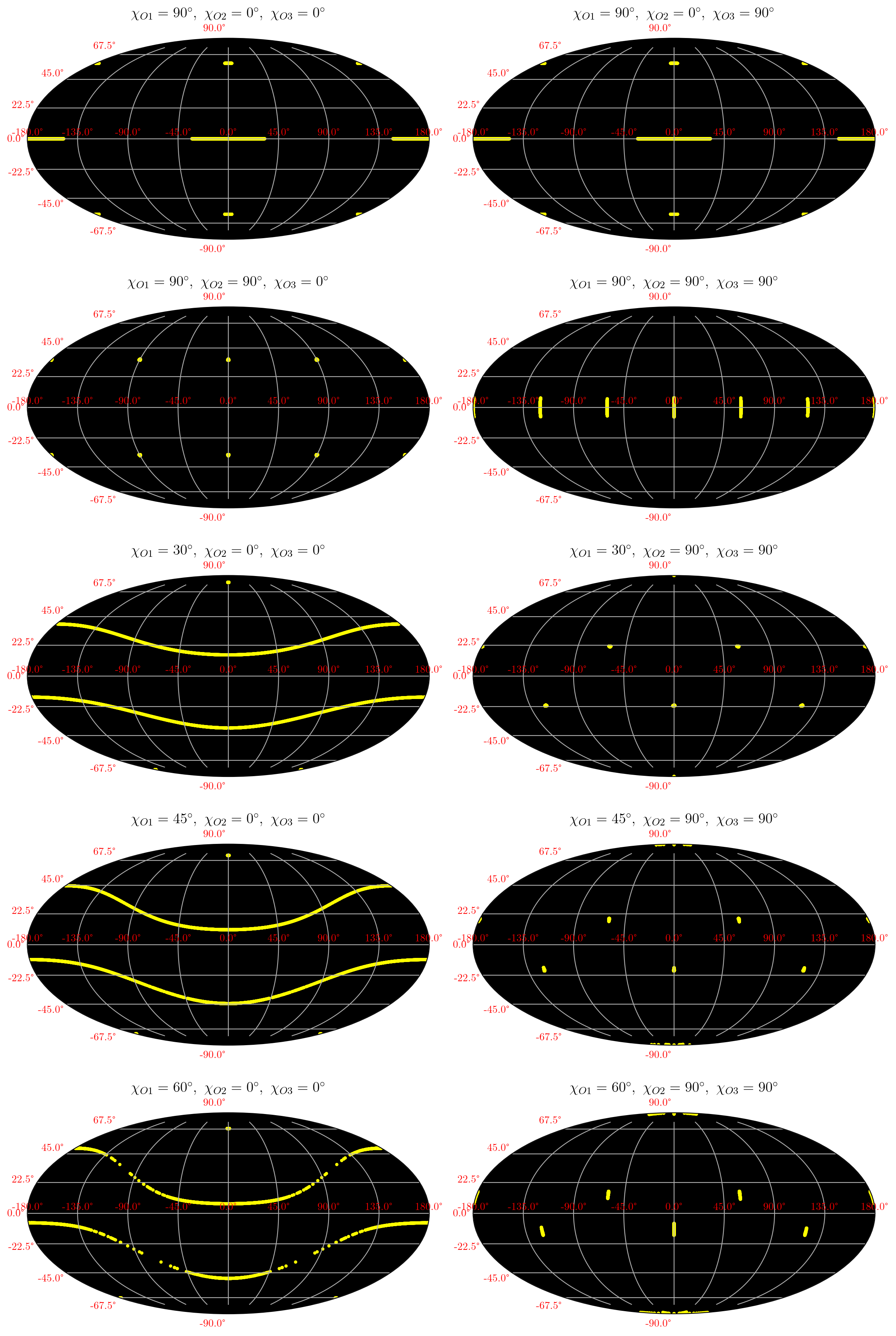}
    \caption{Polar caps for a hypothetical scenario of octupole-only magnetic field for J$0030$ shown in yellow. Several  configurations for $\{\chi_{O1}, \chi_{O2}, \chi_{O3}\}$ varying between $0^\circ$ and $90^\circ$ on a Mollweide projection are presented.}
    \label{fig:PC_l3_AppendixPlot}
\end{figure}

Although a star having solely an octupole magnetic field is unlikely to be realized in nature, we present their polar caps for several orientations to have benchmark cases as challenging numerical tests. We show ten configurations in Figure~\ref{fig:PC_l3_AppendixPlot} for J$0030$ with the angular octupole parameters, $\{\chi_{O1}, \chi_{O2}, \chi_{O3}\}$, varying between $0^\circ$ and $90^\circ$. The boundary conditions on the radial distance are $r = (R_{\star}, 13 R_{LC})$ along with a constraint that the cylindrical radius around the star is $\leq R_{LC}$, and the grid resolution is $10,000 \times 10,000$. The polar caps are clearly tiny as is expected for an octupole-only field, and the point size for all plots is amplified for better visualization. We note that even for such fine grid, the case of $\{ 60^\circ, 0^\circ, 0^\circ \}$ shows a discontinuous pattern -- this should converge towards more continuous streaks if the resolution is increased. We verified this saturation behavior for the similar cases where $\chi_{O1}$ is $30^\circ$ and $45^\circ$.

\section{MCMC execution details} \label{app:MethodsTables}
The MCMC execution details are summarized in Table~\ref{tab:MCMCexecutionDetails}.
\begin{table}[ht!]
\centering
\caption{MCMC execution details summarizing, for each truncation model, the total number of samples explored during each run, the number of CPUs utilized, the wall-clock time (in real hours), the values of, $a_{\mathrm{scale}}$, and the associated figures. \label{tab:MCMCexecutionDetails}}
\normalsize
\setlength{\tabcolsep}{0.03\textwidth}
\begin{tabular}{c c c c c l}
\hline
\hline
\textbf{Truncation} & \textbf{Samples explored} & \textbf{CPUs} & \textbf{Time (hours)} & $a_{\mathrm{scale}}$ & \textbf{Relevant figures} \\
\hline
\multicolumn{6}{c}{\emph{Run 1}: Parallel MCMC with NN model.} \\
\UQ &
$2.8 \times 10^{10}$ & 4000 & $\approx 132$ & Varying & 
Figures~\ref{fig:corner_l12},~\ref{fig:lightCurvesEnsemble},~\ref{fig:l12HotspotsAll},~\ref{fig:ReducedChiSquared_DQ_DQO} \\
\UO &
$2.4 \times 10^{10}$ & 4000 & $\approx 109$ & 12 &
Figures~\ref{fig:corner_l123},~\ref{fig:lightCurvesEnsemble},~\ref{fig:l123HotspotsAll},~\ref{fig:ReducedChiSquared_DQ_DQO} \\
\hline
\multicolumn{6}{c}{\emph{Run 2}: Parallel MCMC with physical model.} \\
\multicolumn{6}{c}{Continues Run 1. Grid resolution: $600 \times 600$.} \\
\UQ &
$1.2 \times 10^7$ & 4000 & $\approx 59$ & 12 & -- \\
\UO &
$1.2 \times 10^7$ & 4000 & $\approx 72$ & 12 & -- \\
\hline
\multicolumn{6}{c}{\emph{Run 3}: Serial chains exploration with physical model to determine best-fit.} \\
\multicolumn{6}{c}{Explores peak log-likelihood regions from Run 2. Grid resolution: $1,000 \times 1,000$.} \\
\UQ &
$4 \times 10^4$ & 40 & $\approx 32.5$ & -- &
Figures~\ref{fig:BestFitLightCurves},~\ref{fig:BestFitFieldLines1},~\ref{fig:BestFitSurfaceMagneticField} \\
\UO &
$4 \times 10^4$ & 40 & $\approx 42.5$ & -- &
Figures~\ref{fig:BestFitLightCurves},~\ref{fig:BestFitFieldLines2},~\ref{fig:BestFitSurfaceMagneticField} \\
\hline
\end{tabular}
\vspace{2pt}
{\scriptsize
\textit{Note 1.} Figures~\ref{fig:l12HotspotsAll},~\ref{fig:l123HotspotsAll} use posterior samples drawn from the NN model run but the hotspots are generated using physical SVM2F module.\\
}
{\scriptsize
\textit{Note 2.} As the runtime varies slightly depending on the CPU architecture used, the reported acceleration factors were evaluated using runs executed on identical CPU types to ensure consistency.
}
\end{table}
\bibliography{bibliography}{}
\bibliographystyle{aasjournalv7}
\end{document}